\documentclass[useAMS,usenatbib,twocolumn]{mn2e}
\pdfoutput=1
\usepackage{graphicx}
\usepackage[usenames]{color}
\usepackage{wasysym}

\newcommand{\um}{\ensuremath{\mu\mbox{m}}}
\newcommand{\Deg}{\ensuremath{^\circ}}

\newcommand{\ntempl}{12}

\newcommand{\refsec}[1]{Section~\ref{#1}}
\newcommand{\reffig}[1]{Fig.~\ref{#1}}

\newcommand{\kms}{km\,s$^{-1}$}

\newcommand{\FeH}{[\ensuremath{\mathrm{Fe}/\mathrm{H}}]}
\newcommand{\aFe}{[\ensuremath{\alpha/\mathrm{Fe}}]}
\newcommand{\mgb}{\ensuremath{\mathrm{Mg}\,b}}
\newcommand{\MgbFe}{(\mgb,$\langle \mathrm{Fe} \rangle$)}
\newcommand{\oiii}{[O\,\textsc{iii}]}
\newcommand{\nni}{[N\,\textsc{i}]}

\title[Two component nature of NGC\,7217]
{Re-growth of stellar disks in mature galaxies:\\
The two component nature of NGC\,7217 revisited with VIRUS-W
$^{\dagger,\diamond}$}

\author[M. H. Fabricius et al.]{Maximilian~H.~Fabricius,$^{1,2}$\thanks{E-mail:mxhf@mpe.mpg.de}
Lodovico~Coccato,$^{3,4}$
Ralf~Bender,$^{1,2}$
Niv~Drory,$^{5}$ 
\newauthor
Claus~G\"ossl,$^{2}$
Martin~Landriau,$^{5}$
Roberto~P.~Saglia,$^{1}$
Jens~Thomas,$^{1}$
\newauthor
Michael~J.~Williams$^{1,6}$
\\
$^{1}${Max Planck Institute for Extraterrestrial Physics,
Giessenbachstra\ss e, 85748 Garching, Germany}\\ 
$^{2}${University Observatory Munich, Scheinerstra\ss e 1, 81679
Munich, Germany}\\
$^{3}${European Southern Observatory, Karl-Schwarzschild-Stra\ss e 2, 
D-85748 Garching bei Muenchen, Germany}\\
$^{4}${ICG, University of Portsmouth, Dennis Sciama Building, Burnaby Road,
Portsmouth, PO1 3FX, United Kingdom.}\\
$^{5}${McDonald Observatory, The University of Texas at Austin, 
2515 Speedway, Stop C1402, Austin, Texas 78712-1206, USA}\\
$^{6}${Department of Astronomy, Columbia University, New York 10027, USA}
}

\begin{document}

\pagerange{\pageref{firstpage}--\pageref{lastpage}} \pubyear{2013}

\maketitle

\label{firstpage}

\begin{abstract} 
Previous studies have reported the existence of two
counter-rotating stellar disks in the early-type spiral galaxy
NGC\,7217. We have obtained high-resolution optical spectroscopic
data ($R \approx 9000$) with the new fiber-based Integral Field Unit
instrument VIRUS-W at the 2.7\,m telescope of the McDonald
Observatory in Texas. Our analysis confirms the existence of two
components. However, we find them to be co-rotating. The first
component is the more luminous ($\approx$\,77\% of the total light),
has the higher velocity dispersion ($\approx$\,170\,\kms) and
rotates relatively slowly (projected $v_\mathrm{max} = 50$\,\kms). The
lower luminosity second component, ($\approx$\,23\% of the total
light), has a low velocity dispersion ($\approx$\,20\,\kms) and
rotates quickly (projected $v_{max} = 150$\,\kms). The difference in
the kinematics of the two stellar components allows us to perform a
kinematic decomposition and to measure the strengths of their Mg and
Fe Lick indices separately. The rotational velocities and
dispersions of the less luminous and faster component are very
similar to those of the interstellar gas as measured from the \oiii\
emission. Morphological evidence of active star formation in this
component further suggests that NGC\,7217 may be in the process of
(re)growing a disk inside a more massive and higher dispersion
stellar halo. The kinematically cold and regular structure of the
gas disk in combination with the central almost dust-free morphology
allows us to compare the dynamical mass inside of the central
500\,pc with predictions from a stellar population analysis. We find
agreement between the two if a Kroupa stellar initial mass function
is assumed.

\vspace{.25 cm}

\noindent
{}$^{\dagger}${This paper includes data taken at The McDonald
Observatory of The University of Texas at Austin.}\\
{}$^{\diamond}${This paper contains data obtained at the Wendelstein Observatory
of the Ludwig-Maximilians University Munich.}
\end{abstract}


%\clearpage
%%%%%%%%%%%%%%%%%%%%%%%%%%%%%%%%%%%%%%%%%%%%%%%%%%%%%%%%%%%%%%%%%%%%%%%%%%%%%%%
\section{Introduction}
%%%%%%%%%%%%%%%%%%%%%%%%%%%%%%%%%%%%%%%%%%%%%%%%%%%%%%%%%%%%%%%%%%%%%%%%%%%%%%%

Photometric studies have been decomposing galaxies into multiple stellar
components for a long time now \citep[e.g.][]{de-Vaucouleurs1959}. The
technique has become common practice in the attempt to disentangle the
formation histories of galaxies.  The problem of a kinematic decomposition,
i.e.\ the detection of genuinely separate components in the line-of-sight
velocity distributions (LOSVDs) --- especially in later type galaxies ---
places higher demands on the data quality, both in terms of spectral resolution
and signal-to-noise ratio. Nevertheless, disk-like structures have been
detected in elliptical galaxies
\citep{Bender1988a,Franx1988,Davies1988,Jedrzejewski1989,Scorza1990,
Scorza1995} and spectroscopic surveys now provide statistics on the occurrence
of kinematic subcomponents in early-type \citep{Krajnovic2011}, S0s
\citep{Kuijken1996}, and spiral galaxies \citep{Pizzella2004}.

\begin{figure*}
\begin{center}
\includegraphics[width=\textwidth, bb=0 30 750 610]{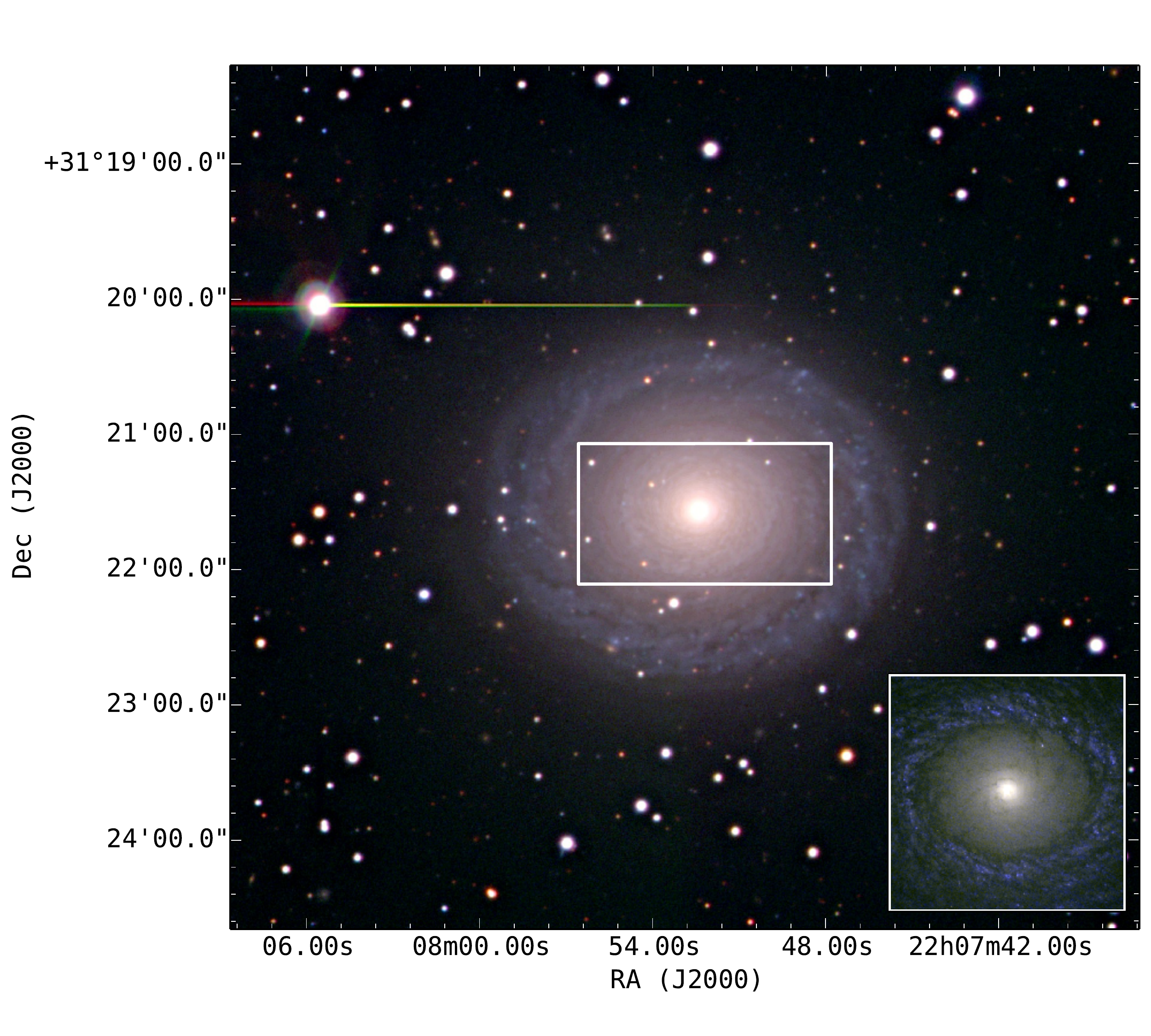}
% finderchart.py
\end{center}
\caption{A \textit{gri} composite of NGC\,7217. The white box outlines
    105\arcsec$\times$55\arcsec~field of view of the VIRUS-W IFU.\ We
    obtained the images with the Wendelstein Wide Field Imager at the
    new 2\,m Fraunhofer Telescope \citep{Gossl2012, Hopp2012} on the 25
    and the 27 of October 2013. The inset in the lower right shows an
    HST F450W/F336W false color composite of the central
25\arcsec$\times$25\arcsec\ (Program 11128; PI David Fisher).}
\label{fig:finderchart}
\end{figure*}

We have constructed a new, high-spectral resolution Integral Field Unit
spectrograph called VIRUS-W \citep{Fabricius2008b,Fabricius2012b}, that
is designed specifically to study the stellar LOSVDs of nearby disk
galaxies. It offers a spectral resolution of $R \approx 9000$
($\sigma_{inst}=15$\,\kms) in the optical which allows us to resolve the
low velocity dispersions of a few tens of \kms\ typically found
in disky systems, and to study the fine structure of the corresponding
LOSVDs.

As a test case, we observed the early-type spiral galaxy NGC\,7217. Its
dynamical structure has been of particular interest because it hosts
multiple rings. There are two star-forming rings with diameters of
63\arcsec\ and 156\arcsec, and a third inner dust ring with a diameter
of 21\arcsec\ (\citealp{Buta1995}, hereafter B95). This inner nuclear
ring marks a significant change in morphology: the outer flocculent
spiral disappears completely and gives away to a smooth central light
distribution \citep[][see \reffig{fig:finderchart}]{Fisher2008}. The
ring locations seemingly correspond to resonances
\citep{Verdes-Montenegro1995} but NGC\,7217 does not host any obvious,
non-axisymmetric structure such as a stellar bar that could create
corresponding resonances. Given its relative isolation
\citep{Karachentseva1973} tidal effects caused by other galaxies can
also be ruled out as a source of the resonances. However, B95 carried
out an extensive photometric analysis of NGC\,7217 and through a Fourier
analysis find a weak perturbation to the axisymmetry which may be a
faded bar.  

Previous work has claimed that NGC\,7217 hosts a large-scale
counter-rotating stellar disk \citep{Merrifield1994,Silchenko2000,
Fabricius2012a}, a phenomenon that has been observed only in a handful
of systems so far. The prototypical example of this class of systems is
NGC\,4550, where its bimodal LOSVD reveals that $\approx$50\% of the
stars are on retrograde orbits \citep{Rubin1992, Rix1992, Emsellem2004}.
Only a few similar objects are known: \citet{Prada1996} claim that
NGC\,7331 has a counter-rotating bulge, and~\cite{Zeilinger2001}
describe stellar counter-rotation in NGC\,3521. NGC\,3593 hosts a
counter-rotating component that dominates the light in the central
regions \citep{Bertola1996}. Further examples include NGC\,4138
\citep{Jore1996, Haynes2000} and counter-rotation caused by interaction
in NGC\,5719 \citep{Vergani2007, Coccato2011a}.

It was hypothesized that the counter-rotation in NGC\,7217 may be the
result of a minor merger event or the cold accretion of gas
\citep{Merrifield1994, Pizzella2004}. The observed ring structure
\citep{Lovelace1997} has been attributed to both the putative minor
merger \citep{Silchenko2006} and instabilities created by
counter-rotation.

In an attempt to disentangle the two counter-rotating disks and to probe
our ability to detect and to study kinematic substructure in stellar
systems, we obtained observations of the central region of NGC\,7217,
covering one of its two stellar rings (see \reffig{fig:finderchart}). We
recover non-parametric LOSVDs with the proper treatment of nebular
emission. We do not confirm the existence of counter-rotation in this
galaxy. Rather we find a sub-dominant, kinematically cold and rapidly
rotating stellar disk embedded in a higher dispersion,
co-rotating essentially spherical stellar halo.

In the next section we will briefly describe the characteristics of the
new spectrograph that we use for this work. In
\refsec{sec:observations}, we will then describe the observations. In
\refsec{sec:data_reduction}, we discuss the basic data reduction, the
algorithm that we use for the recovery of the non-parametric LOSVDs, the
kinematic decomposition, and the method of direct spectral
decomposition that we use to derive abundances of the two stellar
components. In \refsec{sec:results}, we present kinematic maps for
the two stellar components and the ionized gas. We also present maps for
the line strength determinations of the two individual components.  In
\refsec{sec:tilted_ring} we analyse the gas velocity field through
a tilted ring model and in \refsec{sec:mass_to_light} we derive a
central mass-to-light ratio from the gas rotation and compare this value
to the prediction from a stellar population analysis.  We discuss the
implications of our findings in \refsec{sec:discussion}, and
summarize in \refsec{sec:conclusions}. 

%%%%%%%%%%%%%%%%%%%%%%%%%%%%%%%%%%%%%%%%%%%%%%%%%%%%%%%%%%%%%%%%%%%%%%%%%%%%%%%
\section{The Instrument}
%%%%%%%%%%%%%%%%%%%%%%%%%%%%%%%%%%%%%%%%%%%%%%%%%%%%%%%%%%%%%%%%%%%%%%%%%%%%%%%

The observations of NGC\,7217 were carried out using the VIRUS-W
Integral Field Unit (IFU) spectrograph at the 2.7\,m Harlan J. Smith
Telescope of the McDonald Observatory in Texas. Its design is based on
the Mitchell Spectrograph \citep[][formerly VIRUS-P]{Hill2008b}, which
is the prototype for the HETDEX survey instrument \citep{Hill2010}.  Its
IFU consists of 267 optical fibers that are arranged in a dense pack
scheme in a rectangular field of view. Each fiber has a diameter of
150\,$\um$ or 3.2\arcsec\ on sky. The total field of view covers
105\arcsec$\times$55\arcsec\ with a fill factor of 1/3, such that three
dithered exposures are needed to cover the field completely. The IFU is
attached to the bent Cassegrain port of the 2.7\,m which delivers a beam
of $f/8$ that is converted to $f/3.65$ by the Mitchell Spectrograph's
focal reducer. An iris shutter and an exchangeable optical filter for
order separation and stray light reduction are placed in front of the
focal reducer. A 25\,m long fiber bundle guides the light to the
instrument that is mounted on an optical bench inside the telescope
control room.

The collimator of VIRUS-W accepts $f/3.22$ which is slightly faster than
the nominal $f/3.65$ with which the fibers are fed. This allows for
slight focal ratio degradation in the fibers. Two Volume Phase
Holographic gratings offer two modes of spectral resolution.
The higher resolution kinematics mode used in this work covers the
spectral range of 4850\,\AA--5480\,\AA\ with a resolution of $R \approx
8700$ ($\sigma_\mathrm{inst} = 15$\,\kms) and a linear dispersion of
0.19\,\AA\,pixel$^{-1}$. We use a SDSS \textit{g}-band filter in this
mode.

The refractive camera of VIRUS-W is a significant deviation from the
prototype design. It eliminates the central obscuration and increases
the total efficiency of the instrument. Tests on standard stars have
shown that, including atmosphere and telescope optics, the throughput
peaks at 27\%. A Marconi (today e2v) \mbox{CCD44-82} back side illuminated
2k$\times$4k 15\,$\um$ detector is used to record the spectra. At the
operating temperature of $-130\,\Deg$C it has an average dark current of
4.1\,$e^{-}$\,pixel$^{-1}$\,h$^{-1}$. In the adopted science mode the
camera electronics reads out at 100\,kHz over two amplifiers with a gain
of 1.61 ADU/$e^{-}$ and a read noise of 2.55\,$e^{-}$. 

%%%%%%%%%%%%%%%%%%%%%%%%%%%%%%%%%%%%%%%%%%%%%%%%%%%%%%%%%%%%%%%%%%%%%%%%%%%%%%%
\section{Observations}
%%%%%%%%%%%%%%%%%%%%%%%%%%%%%%%%%%%%%%%%%%%%%%%%%%%%%%%%%%%%%%%%%%%%%%%%%%%%%%%
\label{sec:observations}

The observations were carried out in the nights of 2011 August 6, 7 and
9. We observed one field centered on the galaxy and dithered the
observation by small offsets to fill the gaps between the fibers in the
IFU.\ We obtained three 1800\,s exposures in each dither position that
we bracketed and interleaved with 600\,s sky nods. The total on-object
integration time per fiber is 1.5\,h. The FWHM of the seeing varied
between 1.0\arcsec\ and 1.3\arcsec\ during the observations. We took
bias frames and simultaneous Hg and Ne arc lamp exposures for spectral
calibration. We also recorded dome flats to trace the fiber positions on
the detector and to compensate for fiber-to-fiber variation of the
throughput. 

Since the commissioning of VIRUS-W we have built up a library of
spectra of stars to serve as kinematic template spectra (see
Table~\ref{tab:vwtemplates}) and to calibrate our data against the Lick
system \citep{Worthey1994}. 

\begin{table}
\begin{center}
\caption{Previously observed stellar templates.}
\begin{tabular}{lllll}
\hline
Identifier     &     type       & \FeH\ &  date of obs. & Flag\\
\hline
HR\,2600 & K2III  & -0.35$^{(1)}$  & 2010 Dec 02  &K, S, L\\
HR\,3369 & G9III  & 0.16$^{(1)}$   & 2010 Nov 15  &K, S  \\
HR\,3418 & K2III  & 0.11$^{(1)}$   & 2010 Dec 02  &K, S, L\\
HR\,3427 & K0III  & 0.16$^{(1)}$   & 2010 Dec 02  &K, S, L\\
HR\,3428 & K0III  & 0.23$^{(2)}$   & 2010 Dec 03  &K, S  L\\
HR\,3905 & K0III  & 0.23$^{(1)}$   & 2010 Dec 02  &K, S, L\\
HR\,4435 & G9IV   & -0.40$^{(3)}$  & 2010 Dec 03  &K, S, L\\ 
HR\,6770 & G8III  & -0.05$^{(1)}$  & 2010 Dec 03  &K, S, L\\ 
HR\,6817 & K1III  & -0.06$^{(1)}$  & 2010 Dec 03  &K, S, L\\
HR\,7148 & K1III  & -0.09$^{(1)}$  & 2011 May 26  &K, S  \\
HR\,7176 & K1III  & 0.17$^{(1)}$   & 2011 May 27  &K, S, L\\
HR\,7576 & K3III  & 0.42$^{(1)}$   & 2010 Dec 04  &K, S  L\\
HR\,8165 & K1III  & -0.09$^{(1)}$  & 2011 May 25  &K, S, L\\
HD\,13791&  G9III & ---             & 2012 Feb 24   & S   \\
HD\,64606 & G8V   & 0.97$^{(1)}$   & 2010 Dec 03   & S   \\
HD\,74377 & K3V   & -0.07$^{(1)}$  & 2010 Dec 03   & S   \\
HD\,101501& G8V   & -0.13$^{(1)}$  & 2010 Dec 03   & S, L\\
HD\,107685& F6V   & -0.06$^{(3)}$  & 2012 Feb 24   & S, L\\
HD\,108154& F6    & -0.06$^{(3)}$  & 2012 Feb 24   & S, L\\
HD\,114762& F9V   & -0.68$^{(4)}$  & 2012 Feb 24   & S, L\\
HD\,161817& A2VI  & -0.95$^{(1)}$  & 2012 Feb 24   & S, L\\
\hline
\end{tabular}
\end{center}
\begin{minipage}{0.48\textwidth}
Notes: These stellar templates were previously observed with VIRUS-W,
using the instrumental setup used for NGC\,7217. Col. 1: Identifier.
Col. 2: Spectral type.  Col. 3: Metallicity: (1) from \citet{miles}, (2)
from \citet{Koleva+12}, (3) from \citet{Anderson+12}, and (4) from
\citet{Arnadottir+10}.  Col. 4: Date of Observation.  Col. 5: Flag that
specifies what we used the star for: K = used for kinematic
measurement; S = used for spectroscopic decomposition; L = Lick
spectrophotometric standard, used to compute the offset to the Lick
system. 
\end{minipage}
\label{tab:vwtemplates}
\end{table}
%
%%%%%%%%%%%%%%%%%%%%%%%%%%%%%%%%%%%%%%%%%%%%%%%%%%%%%%%%%%%%%%%%%%%%%%%%%%%%%%%
\section{Data Reduction and Kinematic Extraction}
%%%%%%%%%%%%%%%%%%%%%%%%%%%%%%%%%%%%%%%%%%%%%%%%%%%%%%%%%%%%%%%%%%%%%%%%%%%%%%%
\label{sec:data_reduction}
%%%%%%%%%%%%%%%%%%%%%%%%%%%%%%%%%%%%%%%%%%%%%%%%%%%%%%%%%%%%%%%%%%%%%%%%%%%%%%%
\subsection{Basic reduction}
%%%%%%%%%%%%%%%%%%%%%%%%%%%%%%%%%%%%%%%%%%%%%%%%%%%%%%%%%%%%%%%%%%%%%%%%%%%%%%%

The basic data reduction is carried out using the \texttt{fitstools}
package by \citep{Gossl2002}. The generation of master bias, arc, and
flat frames follows standard procedures. We use a slightly modified
version of the Cure pipeline that was developed by our group for the
HETDEX experiment \citep{Hill2004} for the wavelength calibration and
the spectral extraction. Cure first traces the fiber positions on the
masters of the dome flat frames. It then extracts the positions of the
spectral line peaks along these traces. The distortion and the spectral
dispersion are modelled by a two-dimensional 7th degree Chebyshev
polynomial. The corresponding model translates between pixel positions
on the detector and fiber-wavelength pairs. Cure also calculates
corresponding inverse and cross transformations.

We use 24 lines for the wavelength calibration. The standard
deviation of fitted line positions to the model prediction is
0.1\,pixel or 0.02\,\AA~or 1.2\,\kms.

With the models in place, Cure extracts spectra from the science frames
by \textit{walking} along the previously determined trace positions. We
use an extraction aperture that is 7 pixels wide. Tests have shown that this
results in less than 1\,\% aperture loss and less than 0.1\,\% crosstalk
between neighbouring spectra of equal signal level. The extraction is
directly carried out in $\ln(\lambda)$-space with a spectral step width
of 10\,\kms, which closely matches the physical pixel size of the
detector. The signal is distributed in a flux-conserving manner from
detector pixels to spectral elements according to two-dimensional
overlap in area. The dome flat spectra are extracted in the same way as
the science data.

The final reduction steps are carried out using a dedicated pipeline. At
each wavelength we divide the signal of all fibers in the flat frames by
their average signal at that wavelength. The resulting frame is used to
correct the science and the sky frames for differences in the
fiber-to-fiber transmission, vignetting of the spectrograph camera at
the ends of the spectral range, and sensitivity variations of the
detector.

For each science exposure, we average the surrounding sky spectra while
rejecting spurious events. In order to increase the signal-to-noise
ratio we average the sky signal of 20 neighbouring fibers in a moving
window approach while further rejecting outliers through kappa-sigma
clipping. The sky is then scaled according to the exposure time and
subtracted from the science data.

The three flat-fielded and sky-subtracted science frames for each dither
position are then averaged while further rejecting outliers. Finally, we
combine the per-fiber spectra into a common datacube.  The astrometry of
the IFU is tied to the guider images and the sky positions of the
individual fiber apertures are derived in as described in
\citep{Adams2011a}. We create a pixel grid that covers the whole field
of view and set the pixel size to 1.6\arcsec$\times$1.6\arcsec. We
assign fluxes from fibers to pixels by calculating the overlap between
the individual fiber apertures and the spatial extent of each pixel.
This is done for all wavelength steps of the extracted spectra and
results in a three-dimensional datacube with the dimensions RA, dec\
and $\ln(\lambda)$.

%%%%%%%%%%%%%%%%%%%%%%%%%%%%%%%%%%%%%%%%%%%%%%%%%%%%%%%%%%%%%%%%%%%%%%%%%%%%%%%
\subsection{Recovery of non-parametric LOSVDs}
%%%%%%%%%%%%%%%%%%%%%%%%%%%%%%%%%%%%%%%%%%%%%%%%%%%%%%%%%%%%%%%%%%%%%%%%%%%%%%%
\label{sec:kin_extract}
\begin{figure*}
\begin{center}
\includegraphics[width=\textwidth]{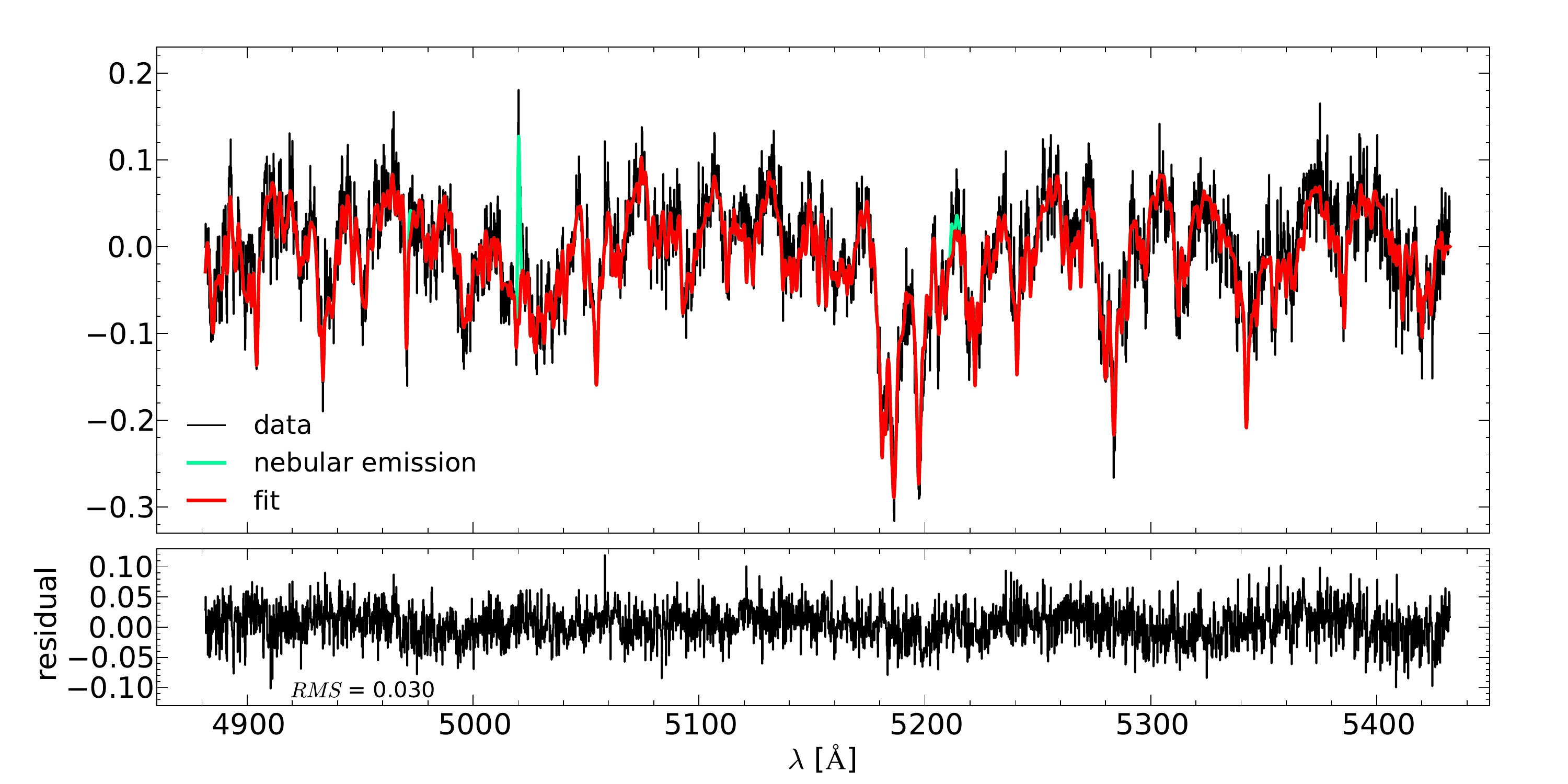}
% plot_fit.py
\end{center}
\caption{Example of a kinematic fit to bin 130. In the upper panel, we
    plot the actual recorded spectrum in black, the corresponding
    best-fitting stellar model spectrum in red, and the best-fitting gas
    model in green. The lower panel shows the residuals.
}
\label{fig:impl_fit}
\end{figure*}

A number of algorithms exist today which extract line-of-sight stellar
velocity distributions from galaxy spectra. The Fourier Correlation
Method (FCQ) by~\cite{Bender1990} recovers full line-of-sight velocity
distributions. It is relatively stable against mismatches of the
observed stellar templates and the intrinsic galaxy spectrum due to the
method of the deconvolution of the peak of the cross-correlation
function of the galaxy and the template spectrum. However, as it
operates in Fourier space, masking of spectral regions is problematic,
and potential contamination by nebular emission lines has to be treated
outside of FCQ \citep{Saglia2010, Fabricius2012b}. The Maximum Penalized
Likelihood Method (MPL) by \citep{Gebhardt2000b} overcomes the template
mismatch problem by simultaneously fitting a linear combination of a set
of different stellar templates. The relative weights of the different
templates enter the optimisation routine as additional free parameters.

Similarly, the more recent Penalized Pixel-Fitting
\citep[][pPXF]{Cappellari2004} operates in pixel space and uses a set
of multiple template spectra. GANDALF \citep{Sarzi2006} extends pPXF to
model nebular emission lines and to include them in the fitting of the
stellar kinematics. The latter two methods extract parametric velocity
distributions by means of modelling them by a Gauss-Hermite expansion
\citep{Gerhard1993}.

In the case of NGC\,7217 we expected very non-Gaussian LOSVDs and
considered an a priori parametric treatment problematic. We therefore
implemented a new algorithm that is based on MPL and recovers a
non-parametric LOSVD, but extends MPL to treat nebular emission. After
the removal of the continuum, the modelling of the stellar kinematics
operates completely analogously to MPL.\ A linear combination of all
template spectra is convolved with a trial LOSVD.\ A standard least
square fitting routine \citep[MINPACK \texttt{lmdif}, ][]{More1980}
minimizes the residuals between the galaxy spectrum and the
broadened template by varying the amplitude of the individual
velocity channels of the LOSVD and by adjusting the template
weights. As in MPL we regularize by calculating the sum of the
squared second derivative of the LOSVD.\ We then normalize the sum
by dividing it by the number of velocity channels and multiply the
result with a smoothing parameter before adding it to the residuals.
A small smoothing factor will result in little regularisation and
the fitted LOSVDs will tend to show noise-induced oscillations
especially in their wings. If a large smoothing factor is chosen,
then the routine is forced to produce a very smooth LOSVD at the
cost of increasing the residuals. We chose a degree of
regularisation that does not significantly affect the RMS of the
difference between the observed and best-fitting
model spectra. The statistical, relative variation of the RMS
value is given by $1/\sqrt{2 n}$, where $n$ is the number of pixels
in the spectrum. For our 3200-pixel spectra, the expected variation
is 1.25\%. We find that the use of a smoothing value of 5 never
results in an increase of the RMS by more than 1\% and use this
value for the fit of all our spectra (see
Appendix~\ref{apx:smoothing}).  

Our fitted spectral range includes several emission lines. For an
initial kinematic fit to the stellar continuum, the algorithm masks the
regions of nebular emission based on the rest frame wavelength of the
emission line, the systemic velocity and a predefined velocity range.
The best-fitting model spectrum is then subtracted from the observed
spectrum and the gas emission lines are fitted following a very similar
methodology to GANDALF.\ Each emission line is modelled by a
Gaussian with a central wavelength and a dispersion. The velocity and
the dispersion are derived from a subset or a single bright line, while
for the fainter lines only the amplitude is fitted.

In a second step, all parameters --- i.e.\ all velocity channels of the
LOSVD, all template weights and all parameters of the model for the gas
emission --- are refitted on the entire spectrum. The best fitting
parameters from the first step are used as initial guesses. Both steps
include one iteration of removal of spurious pixels through kappa-sigma
clipping.

We use a Voronoi tessellation to bin our data using a Python version of
the algorithm by~\cite{Cappellari2003} kindly provided by Eric
Emsellem. We fit the spectral range from 4865\,\AA--5415\,\AA\ and
model and remove the continuum with a 7th degree polynomial. We further
use an initial set of \ntempl~templates, including K and G giants with a
range of different metallicities (see Table~\ref{tab:vwtemplates}). We
first fit a representative subset of all galaxy spectra and successively
remove templates that are given very small weights from the list. Our
final set contains four template spectra (HR\,2600, HR\,6770, HR\,6817,
and HR\,7576). We fit 110 velocity channels and use a Gaussian with
a velocity dispersion of 500\kms\ as initial guess for the
LOSVD.\

The fitted spectral range contains the \oiii$\lambda\lambda$
4959,5007\,\AA\ and \nni$\lambda\lambda$ 5198,5200\,\AA\ emission
lines (see \reffig{fig:impl_fit}).  All lines are included in the fit
while the velocity and the dispersion are only derived from the brighter
\oiii\ line at 5007\,\AA.\

From the kinematic fit we obtain a median $S/N$ of 30 per spectral pixel
for all Voronoi bins. We determine errors to all parameters through the
generation of 30 Monte Carlo realisations of synthetic spectra with
artificial noise, based on the best fitting set of parameters for each
spectrum.

%%%%%%%%%%%%%%%%%%%%%%%%%%%%%%%%%%%%%%%%%%%%%%%%%%%%%%%%%%%%%%%%%%%%%%%%%%%%%%%
\subsection{Kinematic double-Gaussian decomposition}
%%%%%%%%%%%%%%%%%%%%%%%%%%%%%%%%%%%%%%%%%%%%%%%%%%%%%%%%%%%%%%%%%%%%%%%%%%%%%%%
\label{sec:kindecomp}
\begin{figure}
\begin{center}
\includegraphics[width=0.45\textwidth]{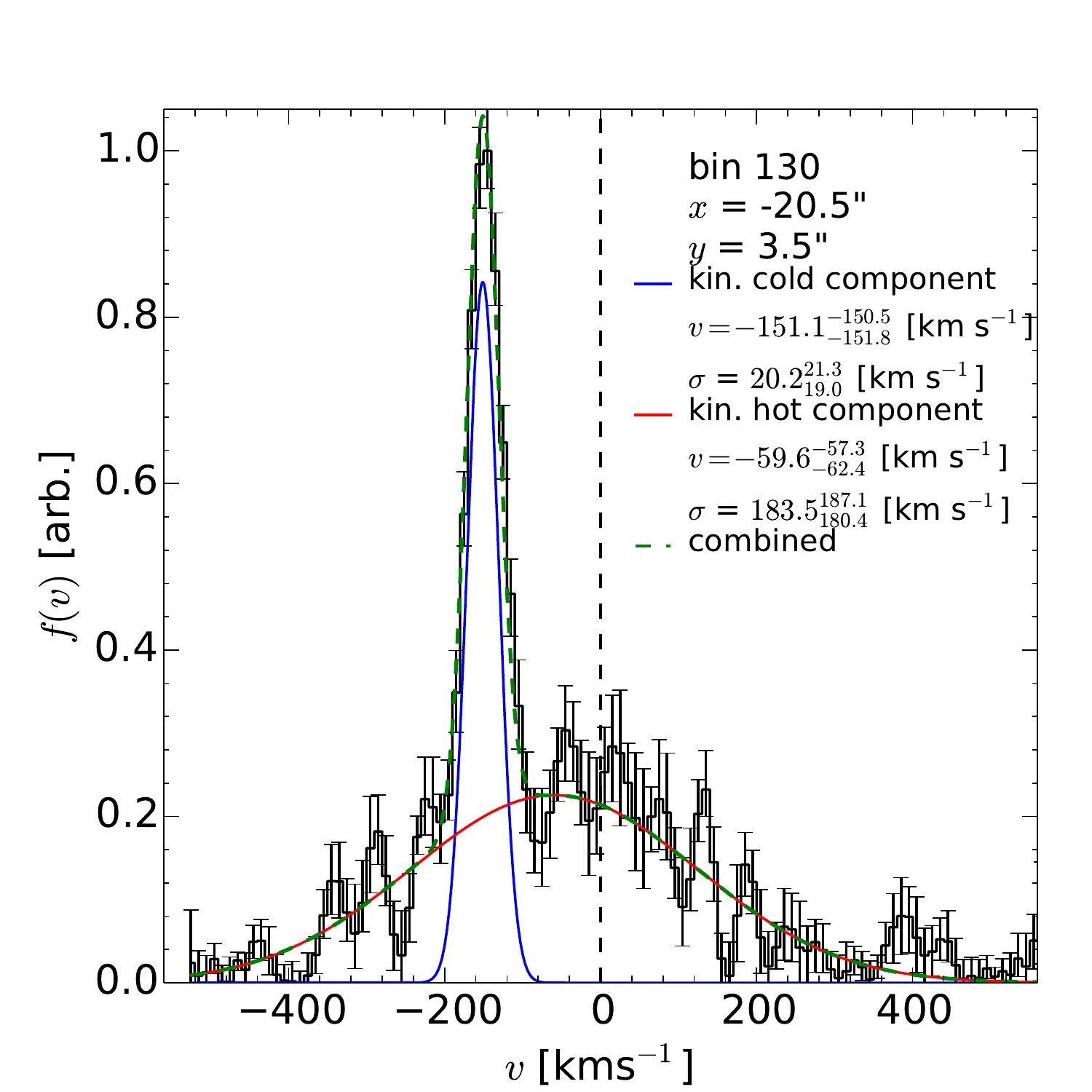}
\includegraphics[width=0.45\textwidth]{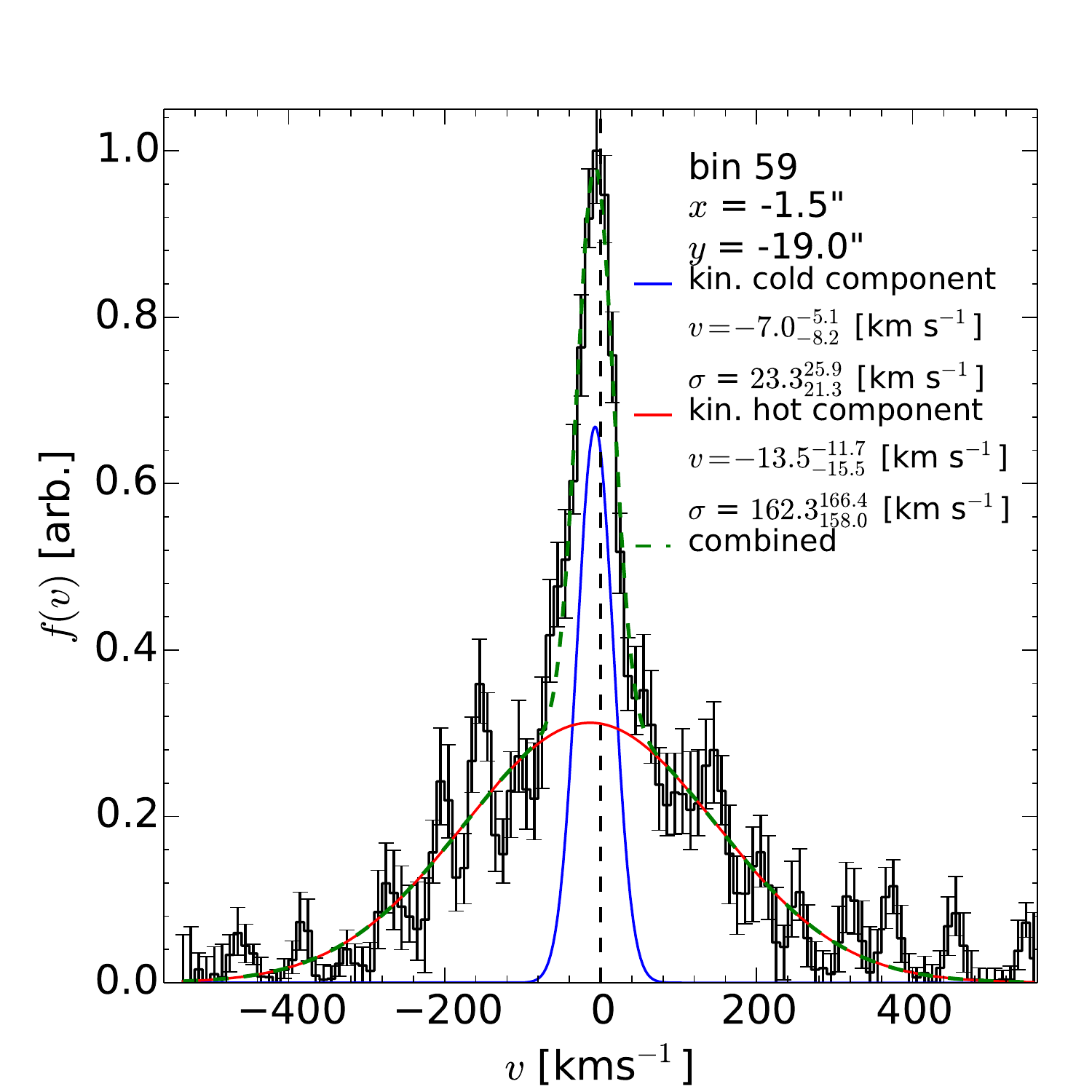}
% plot_losvd2.py
\end{center}
\caption{Examples of two recovered line-of-sight velocity distributions
    and their double-Gaussian decompositions. The upper panel shows the
    distribution for bin 130, which is located along the major axis at a
    radius of about 20\arcsec\ (compare \reffig{fig:impl_fit}). The
    red and blue lines show our best fitting double-Gaussian
    decomposition of the LOSVD.\ The vertical dashed line shows the
    systemic velocity, which we set to zero. The lower panel shows the
    same for a position along the minor axis located about 19\arcsec\
    away from the nucleus.
}
\label{fig:losvds}
\end{figure}
A visual inspection of the recovered LOSVDs immediately reveals a
two-component nature (see \reffig{fig:losvds}): the majority of all
distributions shows a narrow $\sigma \approx 20$ \kms\ peak that
is superimposed on a significantly broader component. In a similar
approach to \citet{Scorza1995} we fit a double-Gaussian distribution
\begin{eqnarray}
	f(v) = \frac{A_{cold}}{\sigma_{cold} \sqrt{2 \pi}}e^{ \frac{-(v-\mu_{cold})^2}{2  \sigma_{cold}^2} } + 
		\frac{ A_{hot} }{\sigma_{hot} \sqrt{2 \pi}} e^{\frac{-(v-\mu_{hot})^2}{2  \sigma_{hot}^2}}
\end{eqnarray}
to the LOSVDs $f(v)$, where $A_{cold}$ and $A_{hot}$ are the relative
weights of each component, $\mu_{cold}$, and $\mu_{hot}$ are their mean
velocities and $\sigma_{cold}$ and $\sigma_{hot}$ are their velocity
dispersions. The labels \textit{hot} and \textit{cold} denote the higher
and the lower velocity dispersion component respectively. We optimize
the fit using the least square fitting routine \texttt{lmdif} of the
MINPACK package \citep{More1980}. The fit is degenerate as a priori it
is not clear which of the two Gaussians should represent which of the
two components. In practice we inspect every fit by eye and adjust the
starting parameters of the fit so as to yield smoothly varying velocity
fields. The fits are nevertheless very stable and, with the exception of
the actual assignment of the two Gaussian distributions to either of the
two components, we find little dependence on the starting values. We
repeat the fit to all 30 LOSVDs that result from the Monte Carlo
simulations to estimate errors on the best fitting parameters.

%%%%%%%%%%%%%%%%%%%%%%%%%%%%%%%%%%%%%%%%%%%%%%%%%%%%%%%%%%%%%%%%%%%%%%%%%%%%%%%
\subsection{Mg and Fe indices of the two kinematic components}
%%%%%%%%%%%%%%%%%%%%%%%%%%%%%%%%%%%%%%%%%%%%%%%%%%%%%%%%%%%%%%%%%%%%%%%%%%%%%%%
\label{sec:decomposition}

In the previous sections we presented evidence that there are two
distinct kinematic components in NGC\,7217. Here we study the properties
of their stellar populations to constrain their formation mechanisms. If
these two components formed at different epochs or from different gas
sources, we would also expect to find differences in their chemical
content.

We apply the spectral decomposition technique that was introduced in
\citet{Coccato2011a,Coccato2013}, which allows us to separate the
spectra of the two kinematically distinct components in the observed
galaxy spectrum. The algorithm is based on pPXF \citep{Cappellari2004}
but allows for two separate velocity distributions that are fit
simultaneously. The model spectra for the two kinematic components are
built independently from a linear combination of stellar templates that
are convolved with either of the two LOSVDs. In this way, the code
allows the two kinematic components to have different stellar population
properties. The LOSVDs are parametrized by Gaussian functions with
different velocities and velocity dispersions. After the optimisation,
the best-fitting linear combination of stellar templates represents the
stellar populations of the two kinematic components, and therefore
allows us to investigate their stellar populations separately. This
method is complementary to the technique of the recovery of the full
LOSVD and subsequent decomposition that we presented in the previous two
sections. In principle it therefore adds a completely independent
kinematic analysis as well as a chemical analysis.

The covered wavelength range allows us to measure the equivalent widths
of the \mgb, Fe5270, and Fe5335 Lick indices, as defined by
\citet{Worthey1994}. Because of the lack of Balmer absorption lines, we
cannot derive the ages of the two stellar components.
Nevertheless, the measurement of Mg and Fe, still provides important
information on the chemical composition of the different stellar
components.

We proceed as follows: the direct spectral decomposition requires a
larger signal-to-noise ratio than the previous method. We therefore
adopt a more aggressive spatial binning than that of
\refsec{sec:kin_extract}. Our Voronoi bins reach $S/N \geq 90$ per
spectral pixel and per bin.

In contrast to previous studies
\citep{Coccato2011a,Coccato2013,Johnston2012}, our data's relatively
small wavelength range introduces degeneracies to the simultaneous fit
of kinematics and stellar populations. In general, the kinematic moments
that we recover using either of the two methods are in excellent
agreement, independent of the initial guesses. However, the ambiguity in
the computation of the best-fitting templates is too high to reliably
constrain the line strength indices. In a similar approach to
\citet{Katkov2013}, we therefore use the kinematic information
determined in \refsec{sec:kindecomp} to constrain the velocities and
velocity dispersions of the two components. This lowers the number of
free parameters and removes much of the degeneracy, as demonstrated by a
series of Monte Carlo simulations (see \refsec{sec:indices_results}).

To perform the spectral decomposition, we use the stars in
Table~\ref{tab:vwtemplates} flagged with `S', which cover the \MgbFe\
plane. These stars provide a better fit to the galaxy spectra than the
stars from other stellar libraries in the literature \citep[e.g.
ELODIE][]{Prugniel2001}, probably due to subtle differences in the
spectral line spread functions of the libraries and our data.

The spectral decomposition code returns the spectra of the two best-fit
templates that are associated with the two kinematic components. We
broaden these spectra to match the 8.4\,\AA\ instrumental resolution of
the Lick System \citep{Worthey+97}, and we then measure the equivalent
width of the \mgb, Fe5270, and Fe5335 absorption line indices. The
systematic offset to the Lick System is computed and corrected using the
stars in our library in common with \citep{Worthey1994}, which are
marked by `L' in Table~\ref{tab:vwtemplates}. The measured offsets are
listed in Table\,\ref{tab:lick_offset}. The offsets are constant within
the range defined by the measurements. We do not apply any correction
for the \mgb\ index as the mean of the offsets is smaller than their
standard deviation.

\begin{table}
\caption{Offset to the Lick indices.}
\begin{center}
\begin{tabular}{l c c}
\hline
Index   &  Mean Offset          & Error      \\
        &  [\AA]                &    [\AA]   \\
        &      (1)              &   (2)      \\
\noalign{\smallskip}
\hline
\noalign{\smallskip}
Mg{\it b}     &        $-0.04$        &     0.25   \\
Fe5270  &    \  \  $ 0.47$      &     0.20   \\
Fe5335  &        $-0.92$        &     0.17   \\
\noalign{\smallskip}
\hline
\end{tabular}
\end{center}
\begin{minipage}{0.45\textwidth}
Notes-- 
Col. 1: Mean offset; offsets are defined as the difference
between values \citep{Worthey1994} and our measurements.
Col. 2: Standard deviation of the measured offsets.
\end{minipage}
\label{tab:lick_offset}
\end{table}

%%%%%%%%%%%%%%%%%%%%%%%%%%%%%%%%%%%%%%%%%%%%%%%%%%%%%%%%%%%%%%%%%%%%%%%%%%%%%%%
\section{Results}
%%%%%%%%%%%%%%%%%%%%%%%%%%%%%%%%%%%%%%%%%%%%%%%%%%%%%%%%%%%%%%%%%%%%%%%%%%%%%%%
\label{sec:results}
%%%%%%%%%%%%%%%%%%%%%%%%%%%%%%%%%%%%%%%%%%%%%%%%%%%%%%%%%%%%%%%%%%%%%%%%%%%%%%%
\subsection{Stellar and ionized-gas kinematics}
%%%%%%%%%%%%%%%%%%%%%%%%%%%%%%%%%%%%%%%%%%%%%%%%%%%%%%%%%%%%%%%%%%%%%%%%%%%%%%%
\label{sec:results_kin}
\begin{figure}
\begin{center}
\includegraphics[width=0.49\textwidth]{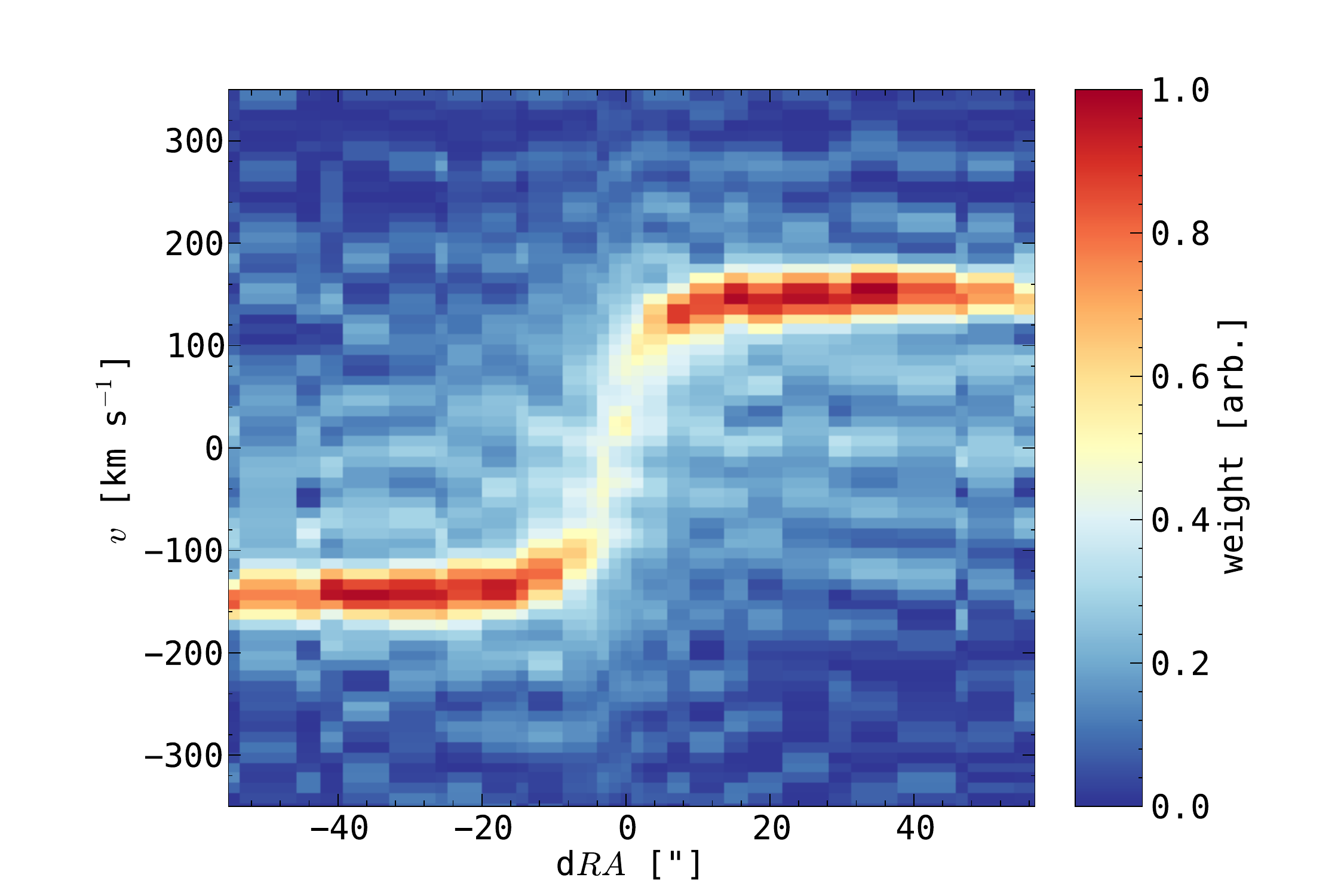}
\includegraphics[width=0.49\textwidth]{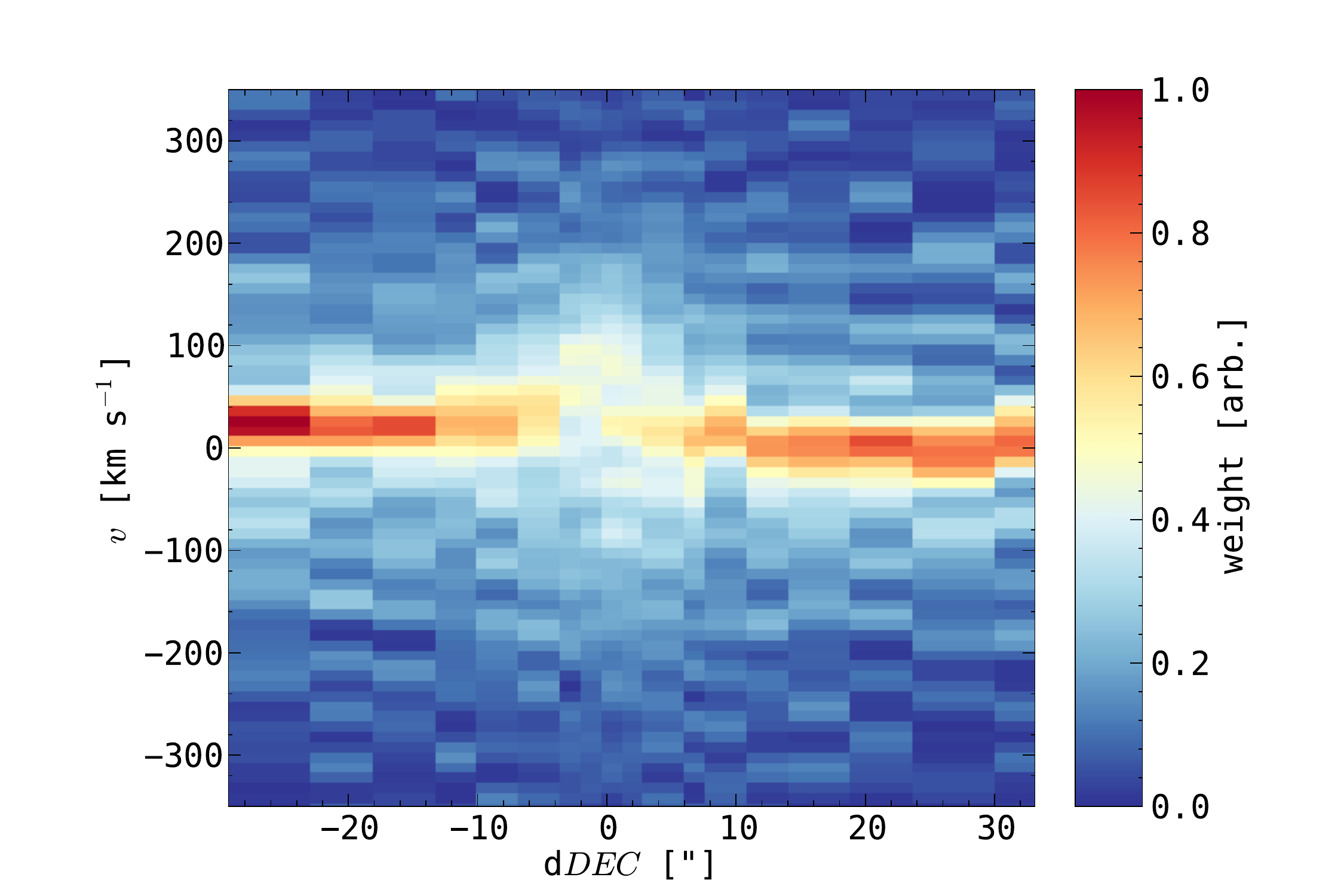}
% plot_bf.py
\end{center}
\caption{Recovered line-of-sight velocity distributions along cuts of
    constant dec (upper panel) and RA (lower panel) through the centre
    of the galaxy. We recover the LOSVD for each spaxel of our datacube
    as described in the text. The $y$-axis shows velocity channels with
    the systemic velocity subtracted. We restrict the plot to velocities
    of $\pm$ 350\,\kms.
}
\label{fig:bfs}
\end{figure}

\begin{figure*}
\begin{center}
\includegraphics[width=0.32\textwidth]{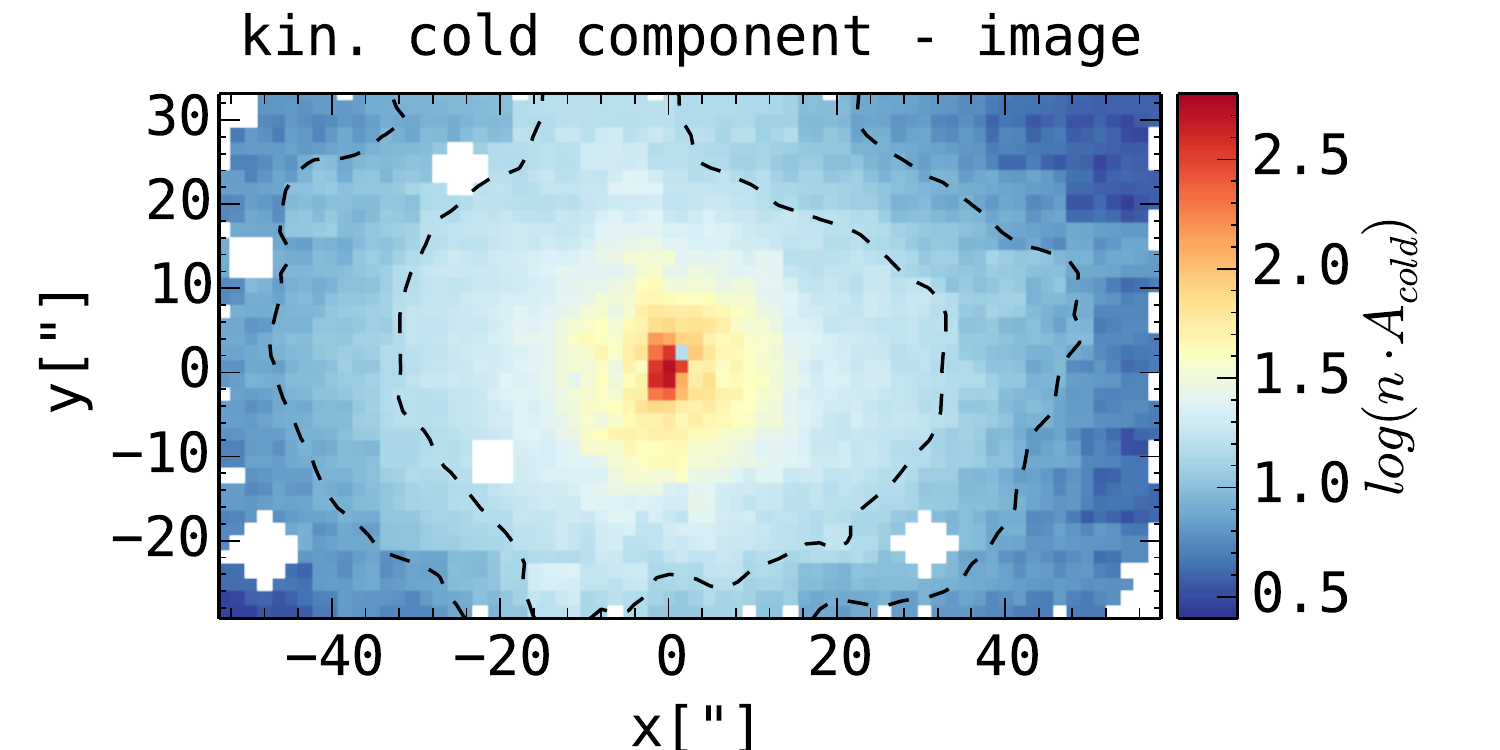}
\includegraphics[width=0.32\textwidth]{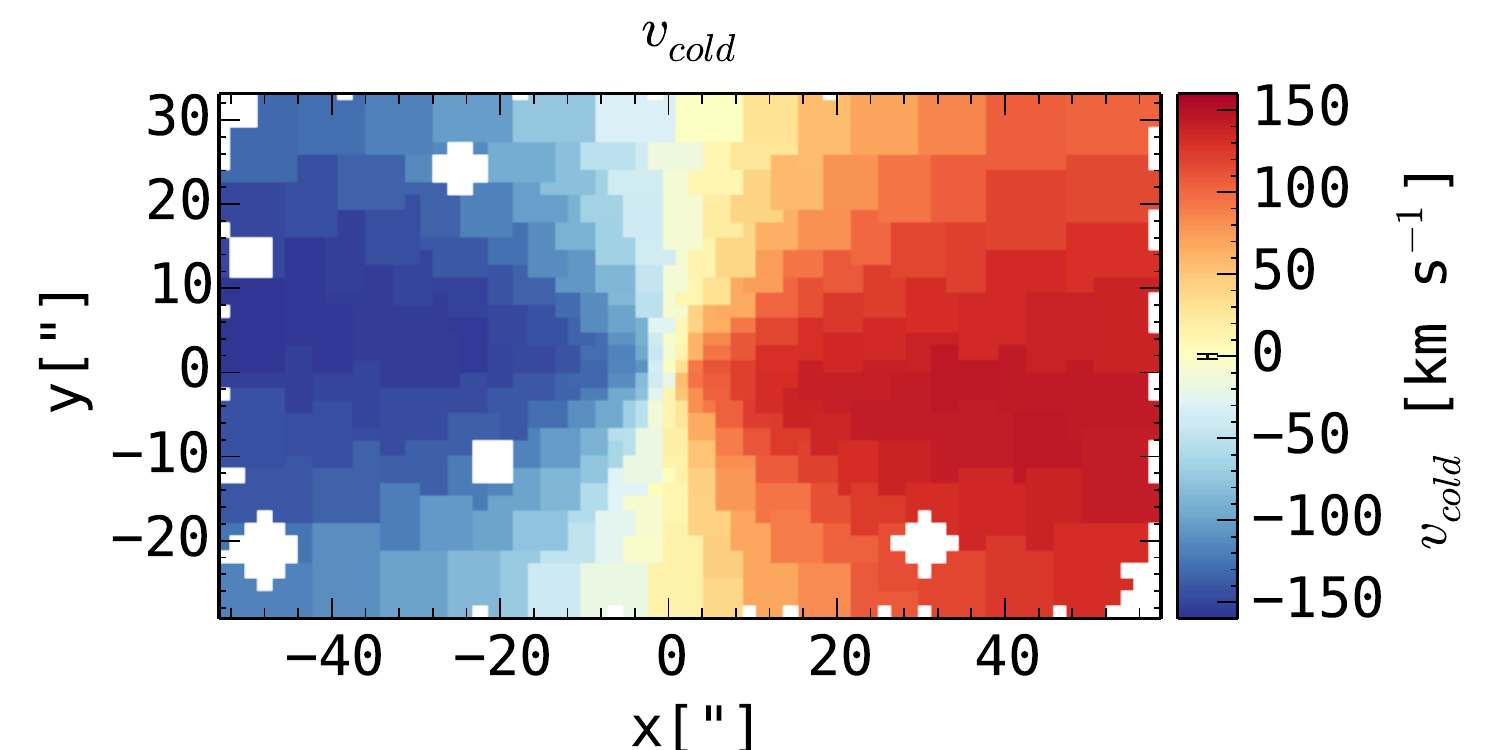}
\includegraphics[width=0.32\textwidth]{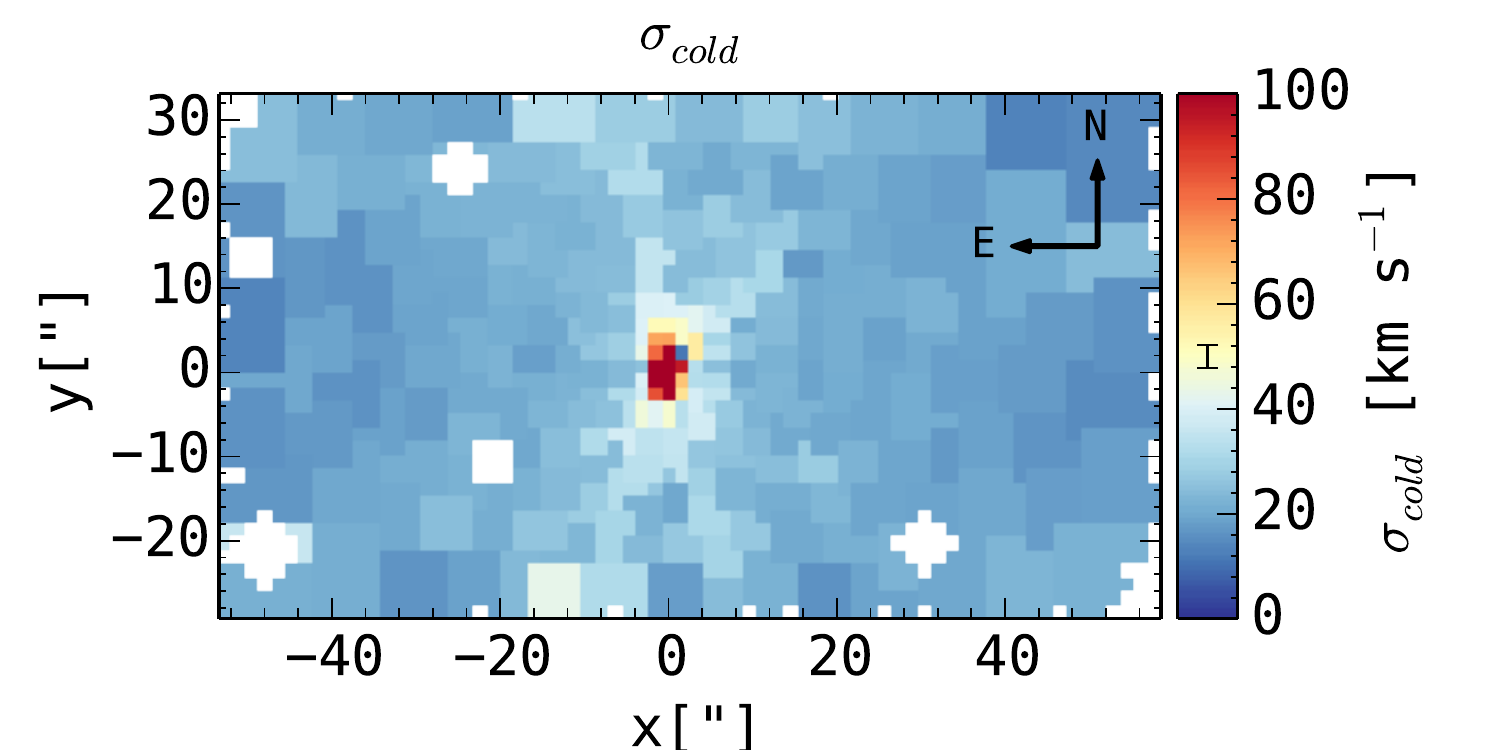}\\
\includegraphics[width=0.32\textwidth]{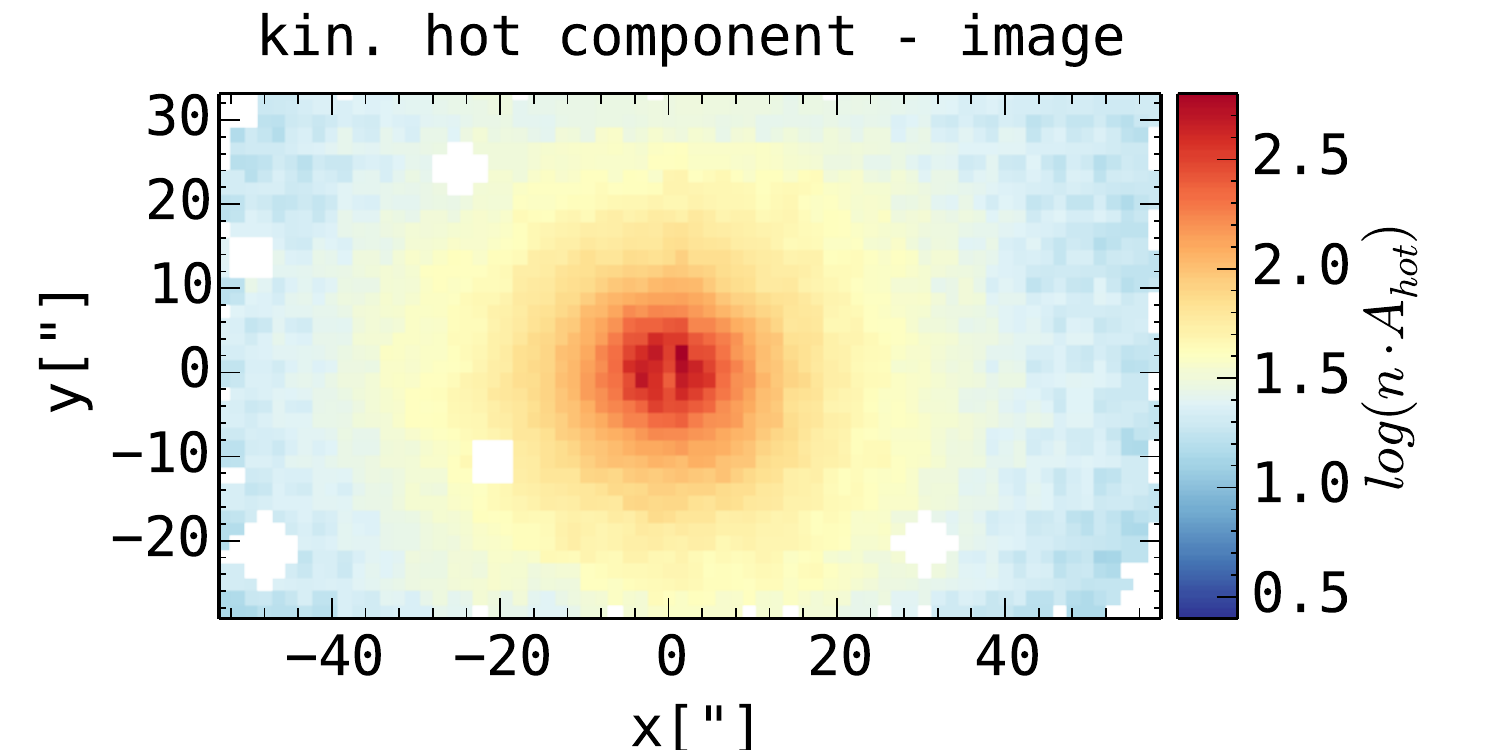}
\includegraphics[width=0.32\textwidth]{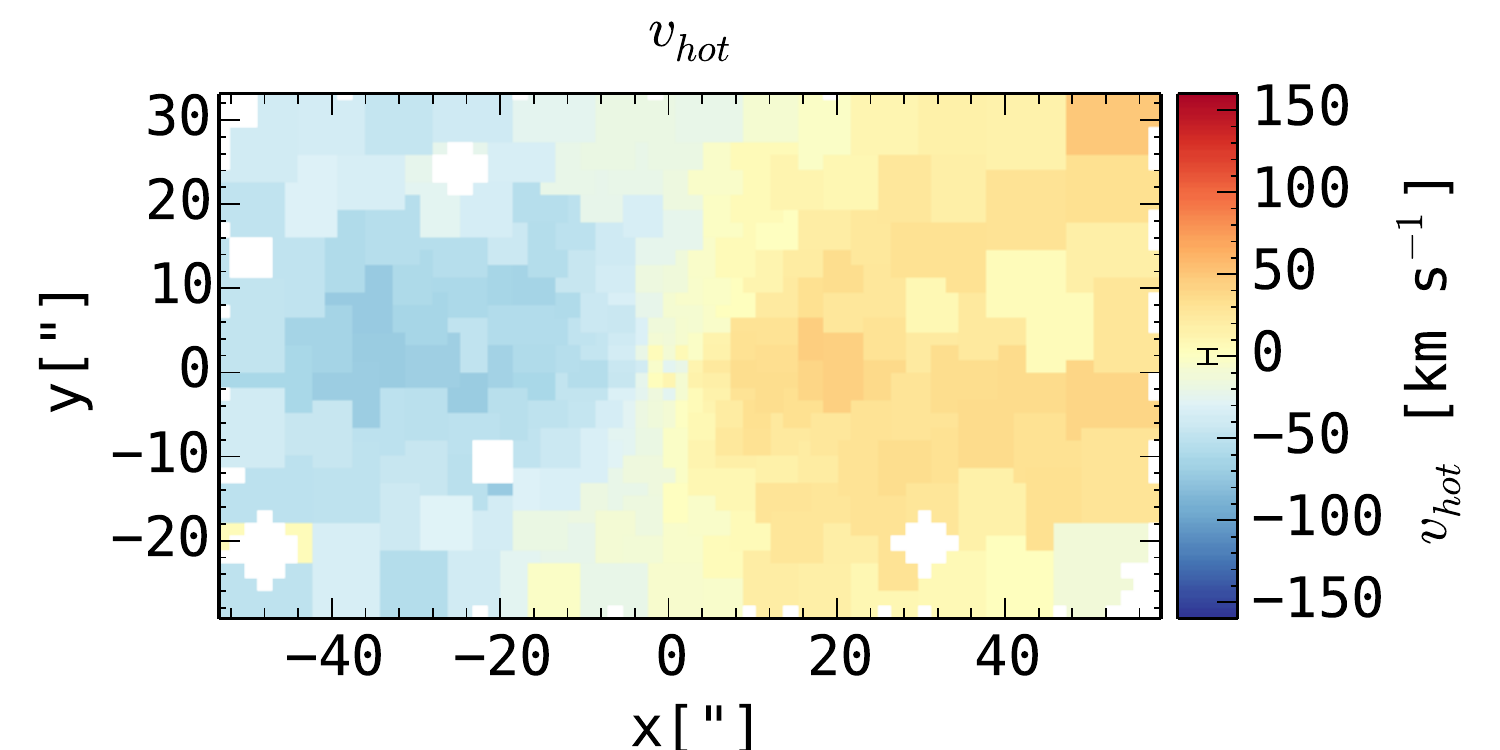}
\includegraphics[width=0.32\textwidth]{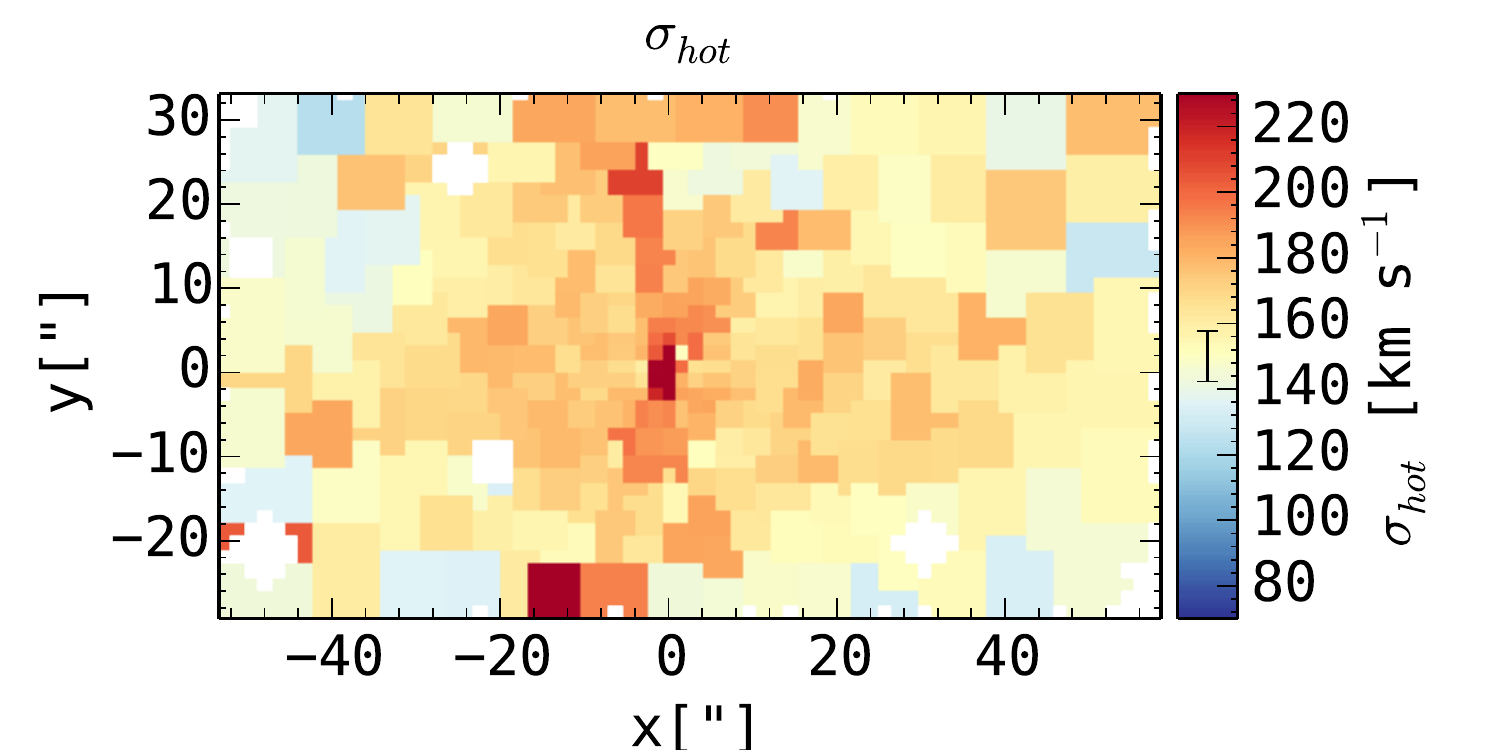}\\
\includegraphics[width=0.32\textwidth]{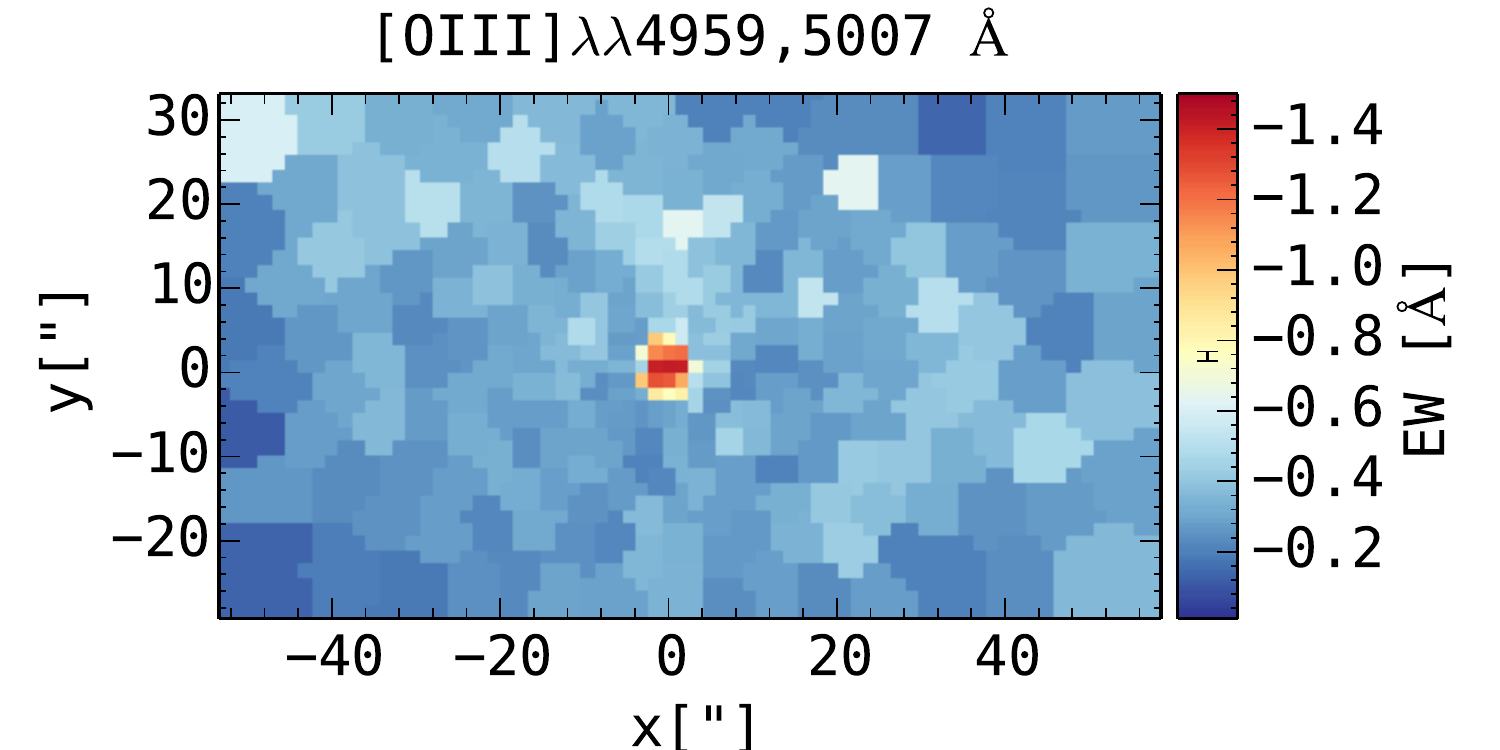}
\includegraphics[width=0.32\textwidth]{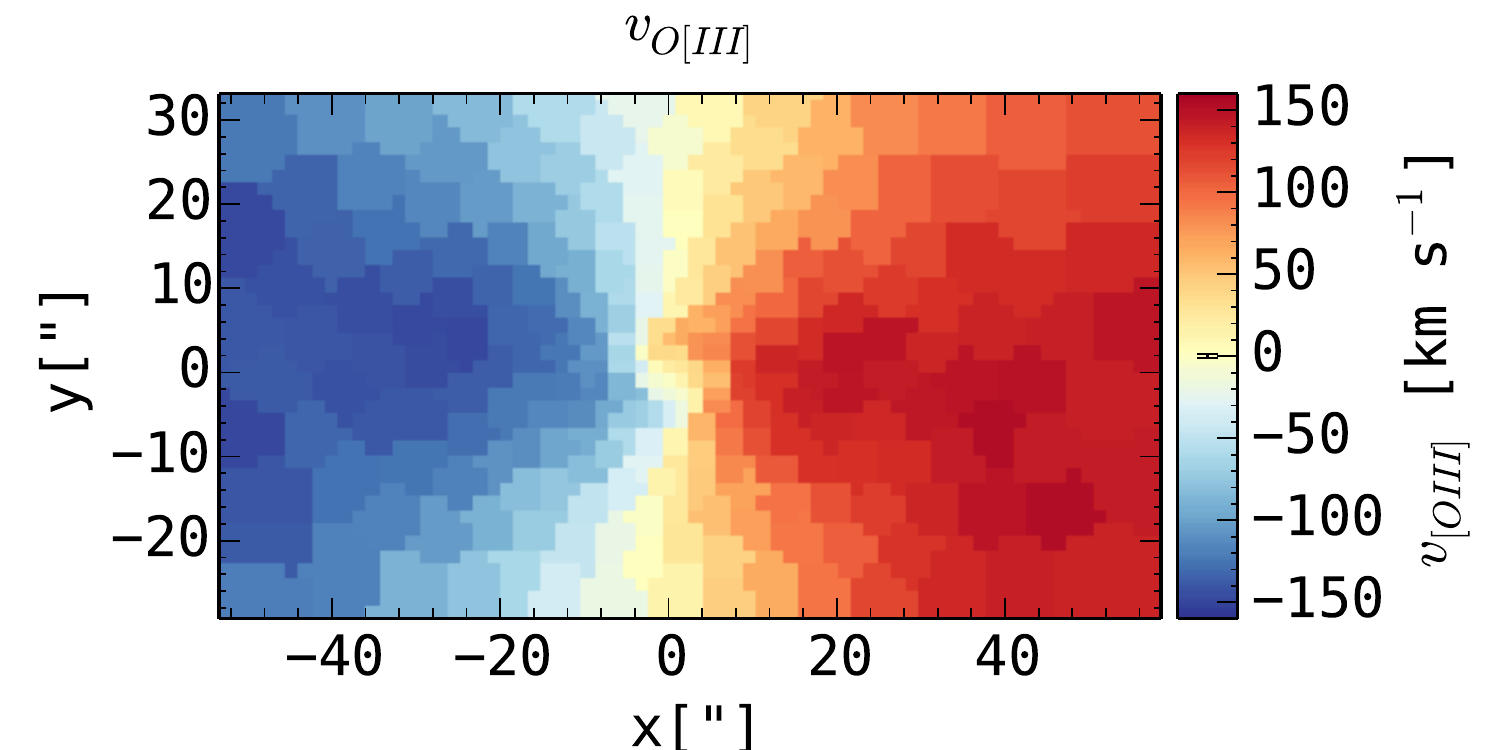}
\includegraphics[width=0.32\textwidth]{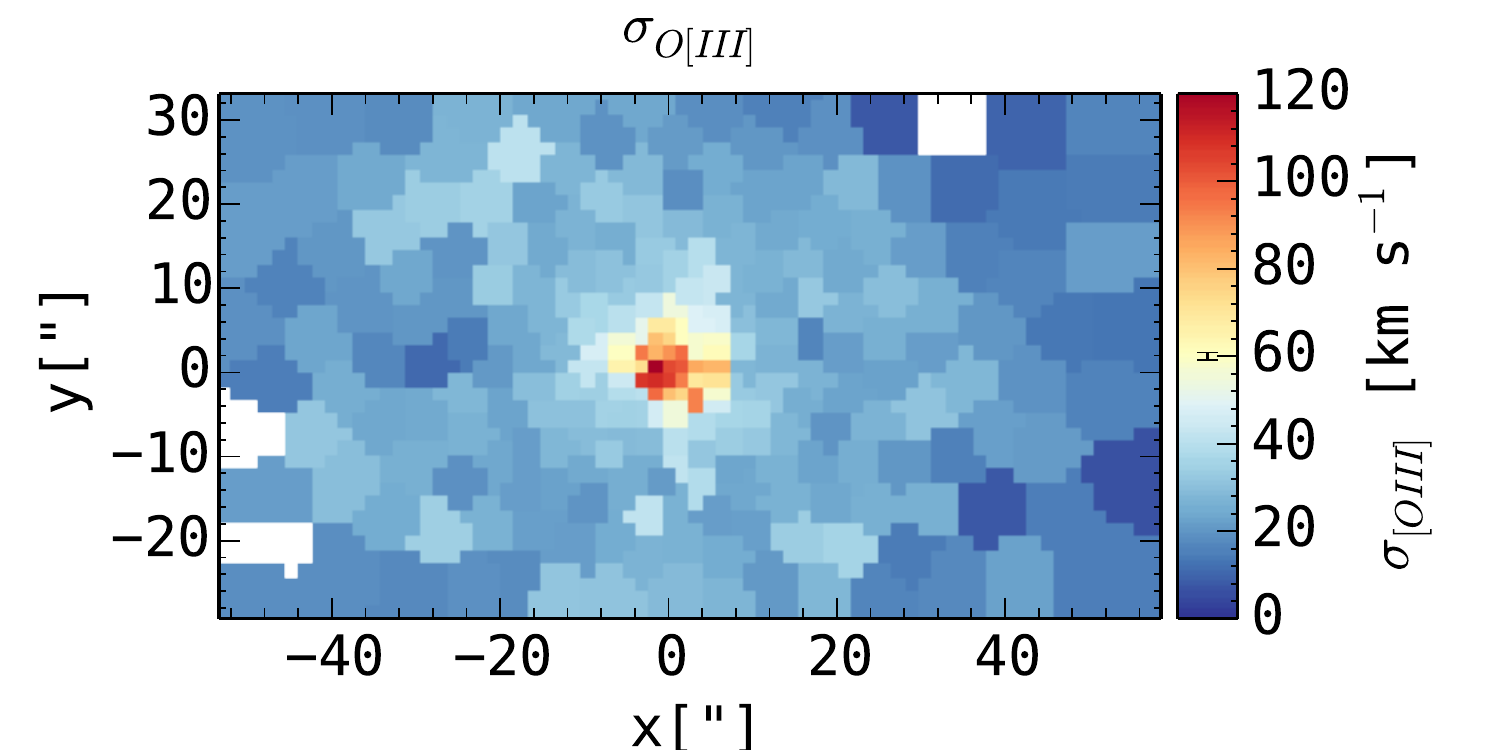}
% plot_maps2.py
\end{center}
\caption{Two-dimensional maps of the of the stellar kinematics
    obtained from the double-Gaussian decomposition and the gas
    kinematics. We decompose the recovered line-of-sight velocity
    distributions into two Gaussians (compare to \reffig{fig:losvds})
    and plot from left to right: the amplitude, the mean velocity and
    the velocity dispersion of the best fitting Gaussian model. The
    upper row plots the moments for the cold component while the middle
    row plots the hot component. The amplitudes for the two components
    have been normalized such that their integrated sum equals the
    wavelength collapsed datacube (see text). To highlight the
    flattening of the cold disk component, we added contours at two
    arbitrary values.  We caution the reader that the decomposition
    becomes degenerate inside of a radius of 3\arcsec, due to the
    relatively low weight of the kinematic cold component and the
    limited spatial resolution of VIRUS-W.  The bottom row shows the
    summed equivalent width of the \oiii$\lambda\lambda$
    4959,5007\,\AA\ emission lines and the velocity and velocity
    dispersion obtained by the kinematics extraction routine from the
    brighter line at 5007\,\AA.\ In all maps positive $y$-values point
    to north and positive $x$-values to west. 
}
\label{fig:kin_maps}
\end{figure*}

In this section we describe the results of the kinematic fit
(\refsec{sec:kin_extract} and \refsec{sec:kindecomp}) and the
double-Gaussian decomposition (\refsec{sec:decomposition}).

In \reffig{fig:bfs} we plot the full line-of-sight velocity
distributions along lines of constant dec and RA.\ The vertical axis is
velocity and the horizontal axis is the spatial position. The
constant dec cut (upper panel) falls close to the major axis because the
major axis of NGC\,7217 is closely aligned with the east-west axis. One
immediately identifies a rapidly rotating component with a relatively
low velocity dispersion. The rotation curve rises steeply in the centre
and flattens out at a radius of 20\arcsec. The rotation amplitude
reaches about 150\,\kms. The white areas in that panel show
another component with a much larger velocity dispersion. One may
compare this figure to \reffig{fig:losvds}, which shows individual
LOSVDs that correspond to vertical cuts in \reffig{fig:bfs}. There the
broader underlying component is even more easily identified. Similarly,
the plot along constant RA (lower panel) shows a narrow component close
to zero velocity and a much broader component underneath. The upper
panel shows no sign of a counter-rotating disk. Such a disk would show
up as an opposite s-shape structure. Similarly, \reffig{fig:losvds} does
not show an additional low-dispersion peak that would have to be located
at about 150\,\kms.

We therefore find no evidence for a counter-rotating disk. Rather we
find one low-dispersion disk that is embedded in a high-dispersion
component that shows little rotation.  

It is more difficult to estimate the amount of rotation of the hot
component from \reffig{fig:bfs}. We therefore turn to the results from
the decomposition shown in \reffig{fig:kin_maps}. From left to right,
this figure shows the amplitude, the mean velocity and the velocity
dispersion of the best-fitting Gaussian. The upper two rows show the
parameters for the low and high dispersion stellar components, while the
bottom row shows the parameters for gas visible in \oiii\ emission. The
fitted amplitudes (parameters $A_{hot}$ and $A_{cold}$ in equation 1)
correspond to line strength and therefore give little information on the
spatial extent of the two components. We therefore scale them by
enforcing the sum of the two amplitudes to equal the signal of the
collapsed datacube. We collapse the datacube in an emission line free
region redwards of the \mgb\ feature (rest frame 5230\,\AA--5400\,\AA).
The results are \emph{images} of the two components as they would appear
if they could be observed separately. It is these component images that
are shown on the left of \reffig{fig:kin_maps}. As the fiber size of
\mbox{VIRUS-W} is 3.2\arcsec\ and, because the low dispersion component
has relatively low weight at the nucleus, the decomposition becomes
unreliable inside a radius of 3\arcsec. We nevertheless plot the
innermost values for completeness. The component images suggest that the
light distribution of the cold component is flatter than that of the
high dispersion component. Also it appears more elliptical in shape and
aligned with the major axis position angle. The cold component typically
contributes 20\%--40\% of the light outside the central 8\arcsec. In
the next section we fit two-dimensional surface brightness models to
derive scale lengths and ellipticities of the two components.

The velocity fields of the two components show that they are actually
co-rotating. The rotational amplitude of the cold component reaches
150\,\kms\ while the dispersion stays mostly flat at 20\,\kms\
throughout the field of view. In contrast, the hot component reaches a
rotational amplitude of 58\,\kms\ and its dispersion rises from
$\approx$\,150\,\kms\ at the edges of the field of view to
$\approx$\,170\,\kms\ in the central regions. 

The kinemetry IDL routine \citep{Krajnovic2006} shows that the velocity
fields have best fitting position angles of $82.7 \pm 2.0$\Deg\ and
$80.6 \pm 3.8$\Deg\ for the cold and the hot components respectively
(three sigma errors).

%%%%%%%%%%%%%%%%%%%%%%%%%%%%%%%%%%%%%%%%
\subsection{Structural parameters}
%%%%%%%%%%%%%%%%%%%%%%%%%%%%%%%%%%%%%%%%
\begin{figure*}
\begin{center}
\includegraphics[width=0.32\textwidth]{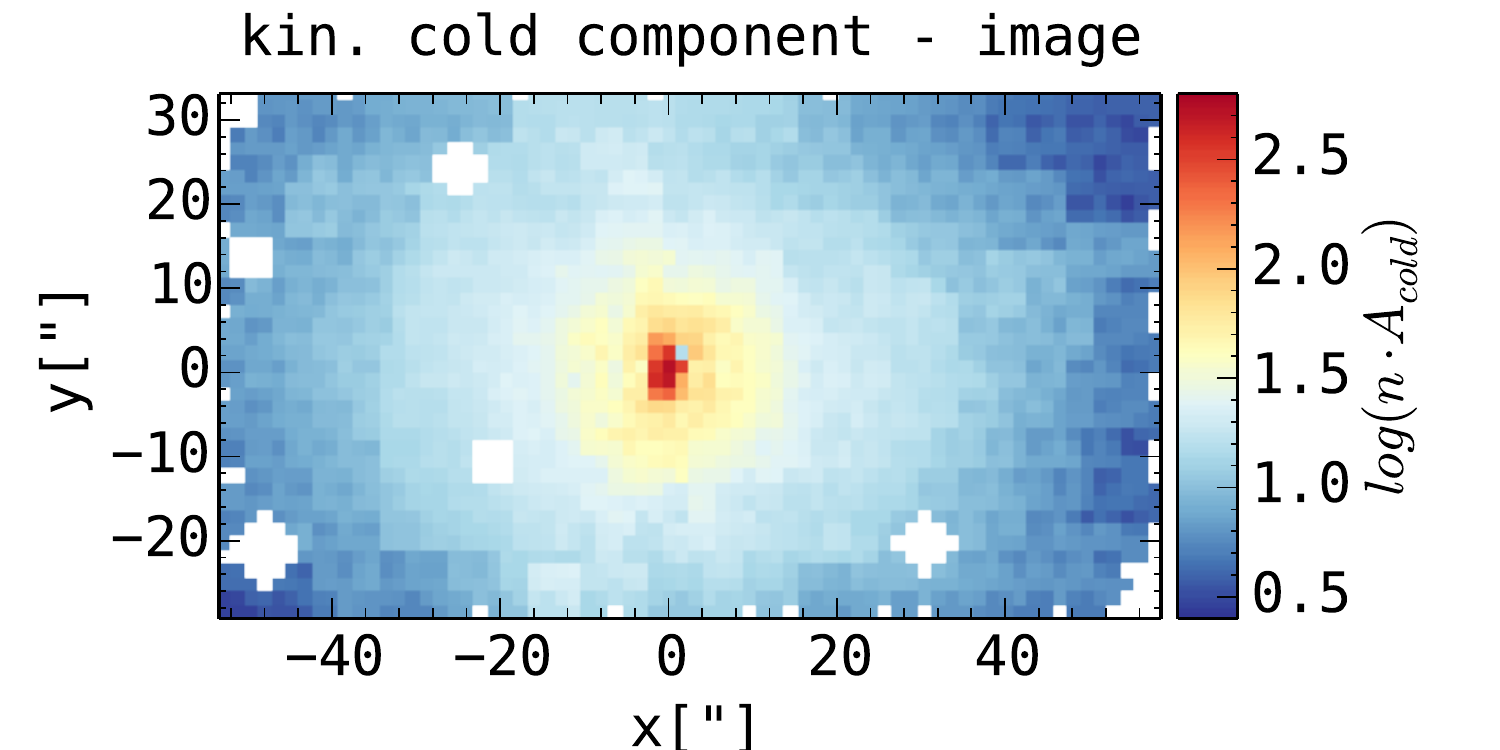}
\includegraphics[width=0.32\textwidth]{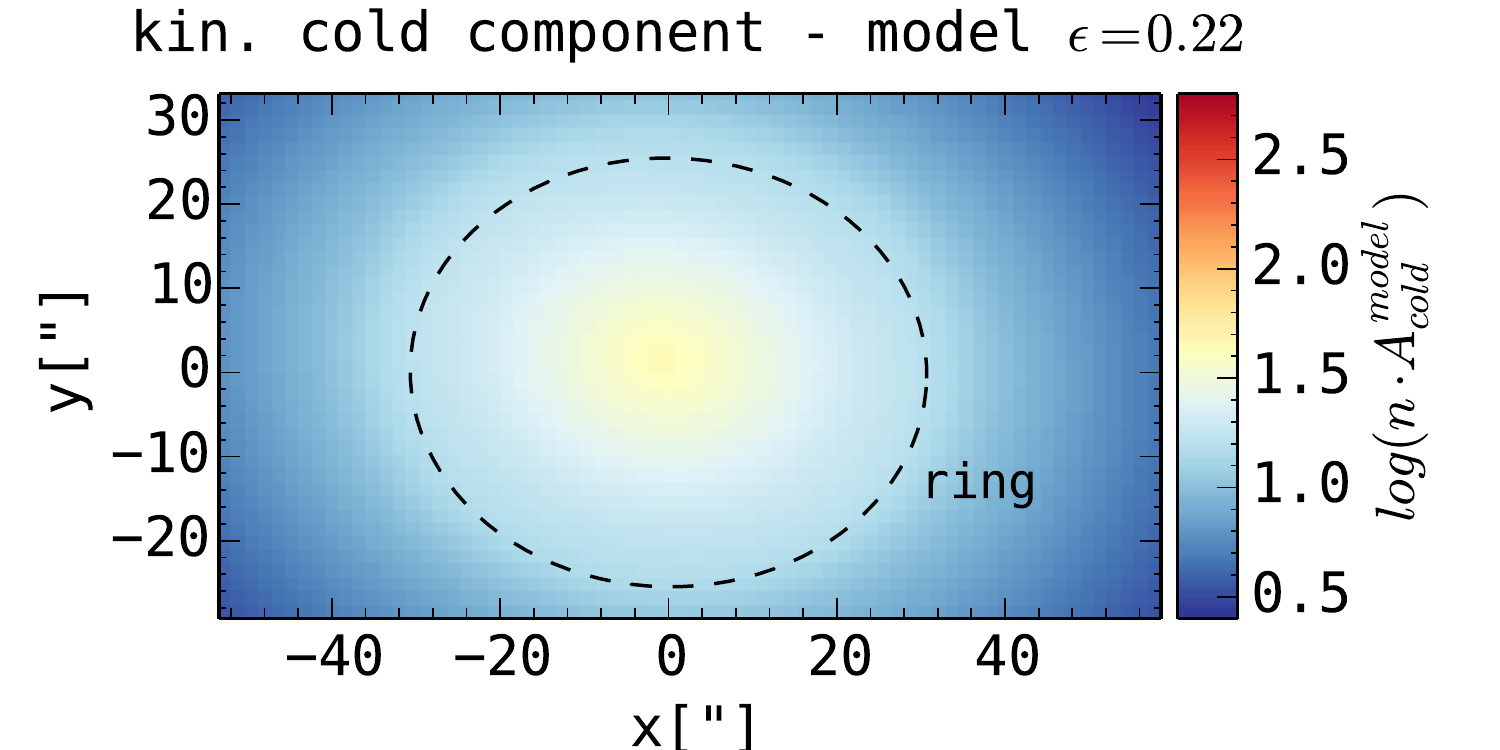}
\includegraphics[width=0.32\textwidth]{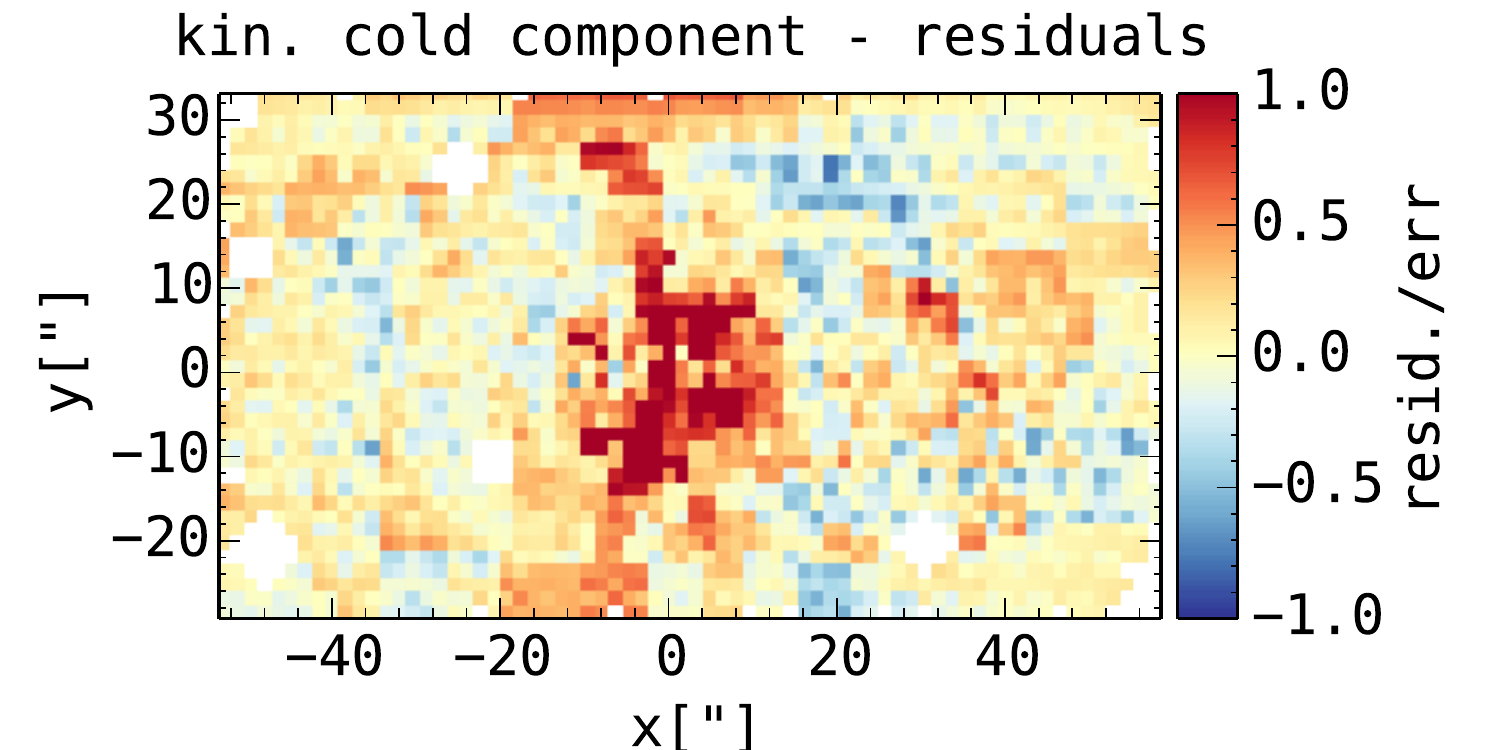}
\includegraphics[width=0.32\textwidth]{plots/nA_map_2}
\includegraphics[width=0.32\textwidth]{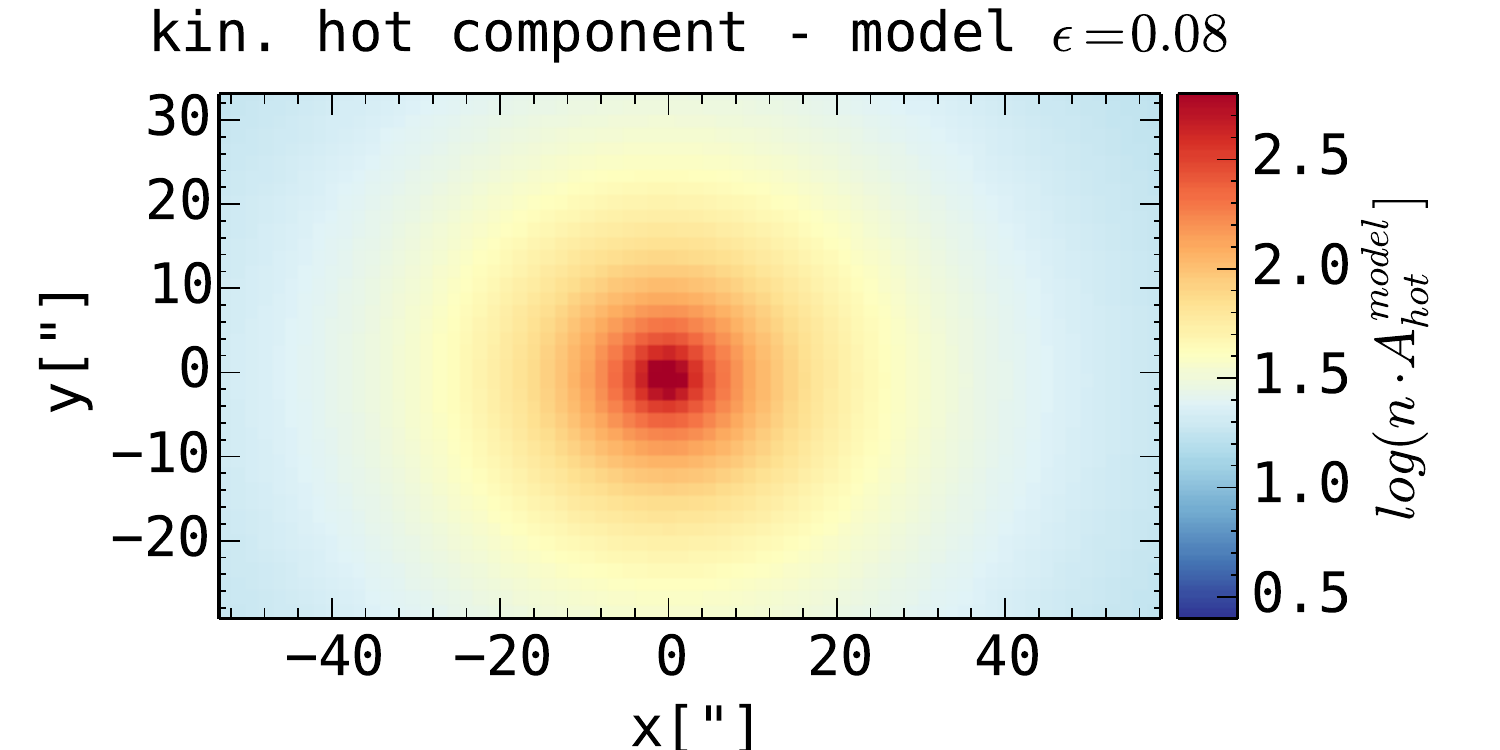}
\includegraphics[width=0.32\textwidth]{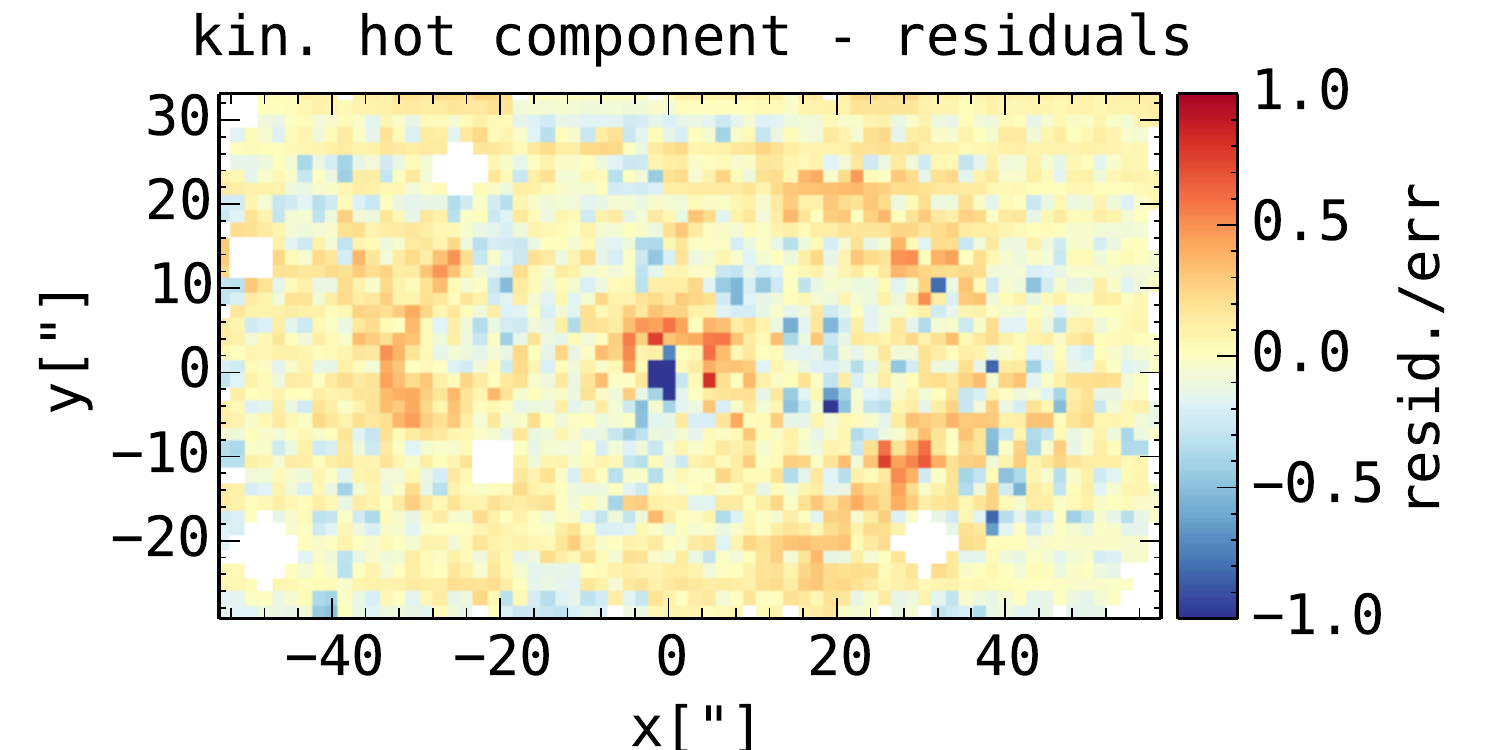}
\end{center}
\caption{Two dimensional models of the kinematic components in
    NGC\,7217. The \textit{images} of the two separate components are
    obtained through the double-Gaussian decomposition and proper
    scaling as described in the text. We reproduce the images of the two
    components from \reffig{fig:kin_maps} on the left. We fit an
    exponential disk model to the low dispersion component and also
    include a Gaussian ring to account for the inner stellar ring. We
    model the kinematically hot component with a de~Vaucouleurs'
    profile. The two best fitting models are shown in the middle column.
    The dashed line marks the location of the ring in the model for the
    cold component. The right columns shows the residual maps. The
    central $r > 15\arcsec$ of the cold component were excluded from
    the fit as they were poorly matched by an exponential light
    distribution.
}
\label{fig:phot_model}
\end{figure*}

In the previous section we reconstructed images of the two kinematic
components as they would appear if they could be observed separately.
For this the relative amplitudes from the double-Gaussian decomposition
are scaled, such that their sum equals the integrated flux in an
emission line free waveband. With these results we can now attempt to
derive structural parameters for the two individual components and to
compare them to the values that are derived from photometry. We use the
\texttt{Imfit}
package\footnote{http://www.mpe.mpg.de/$\sim$erwin/code/imfit/} by
\citet{Erwin2014} to fit two-dimensional models to respective images.

Following B95, we model the cold component with an exponential disk but
also include a Gaussian ring to account for the inner stellar ring. As
in B95, we adopt a de~Vaucouleurs' law for the surface brightness
distribution of the hot stellar component. We use the formal errors from
the double-Gaussian decomposition to create error images for
\texttt{Imfit} and mask the positions of foreground stars.

In \reffig{fig:phot_model} we reproduce the component images from
\reffig{fig:kin_maps} on the left, the best fitting model in the middle column,
and the residuals on the right. The central 20\arcsec\ of the model of the hot
component shows positive residuals. As the decomposition is degenerate in the
central region, we mask the corresponding area and constrain our fit to the to
data with $r > 15\arcsec$. We also fix the centre of both components to the
brightness centre that we determine from the collapsed datacube. We allow for a
variable flat sky background for the cold component.  We find that this is
necessary for the fit to converge to reasonable values. This may be due to
imperfect sky subtraction, which is perhaps not surprising given our reliance
on sky nods for background subtraction.

\begin{table}
\begin{center}
\caption{Structural parameters of the two kinematic components}
\begin{tabular}{lll}
\hline
Parameter     &     Value       & Uncertainty \\
%(1) & (2) & (3) & (4) \\
\hline
\multicolumn{3}{l}{Cold component --- exponential disk}\\
Position angle [\Deg]      & 82.1 & $\pm\ 2.5$   \\
ellipticity                & 0.23 & $\pm\  0.02$  \\
h [\arcsec]                & 43.19 & $\pm\  1.2$  \\
$D/T$                        & 0.221     &         \\
&&\\
\multicolumn{3}{l}{Cold component --- Ring}\\
Position angle [\Deg]            & 88.7 & $\pm\ 10.8$  \\
ellipticity                & 0.17 & $\pm\  0.07$  \\
radius [\arcsec]           & 30.7 & $\pm\  1.3$   \\
width $\sigma$ [\arcsec]   & 4.2  & $\pm\  1.0$   \\
$R/T$                        & 0.005      &        \\
&&\\
\multicolumn{3}{l}{Hot component --- de~Vaucouleurs}\\
Position angle [\Deg]            & 85.8 & $\pm\  7.9$  \\
ellipticity                & 0.08 & $\pm\  0.02$ \\
$r_e$ [\arcsec]            & 58.1 & $\pm\  11.9$ \\
$B/T$                        & 0.774      &        \\
\hline
\label{tab:struct_param}
\end{tabular}
\end{center}
\begin{minipage}{0.48\textwidth}
Notes-- $D/T$ is the ratio of the total model flux in the cold disk
component to the total flux of all components. $R/T$ is the
relative flux in the inner stellar ring model and $B/T$ is the
relative flux in the model of the hot stellar component.
\end{minipage}
\end{table}

We give our best fitting parameters in Table\,\ref{tab:struct_param}. We
cannot quote central surface brightnesses or effective surface
brightnesses as our data are not flux calibrated. 

%%%%%%%%%%%%%%%%%%%%%%%%%%%%%%%%%%%%%%%%%%%%%%%%%%%%%%%%%%%%%%%%%%%%%%%%%%%%%%%
\subsection{Direct decomposition \& line strength analysis}
%%%%%%%%%%%%%%%%%%%%%%%%%%%%%%%%%%%%%%%%%%%%%%%%%%%%%%%%%%%%%%%%%%%%%%%%%%%%%%%
\label{sec:indices_results}
\begin{figure*}
\begin{center}
\includegraphics[width=0.32\textwidth]{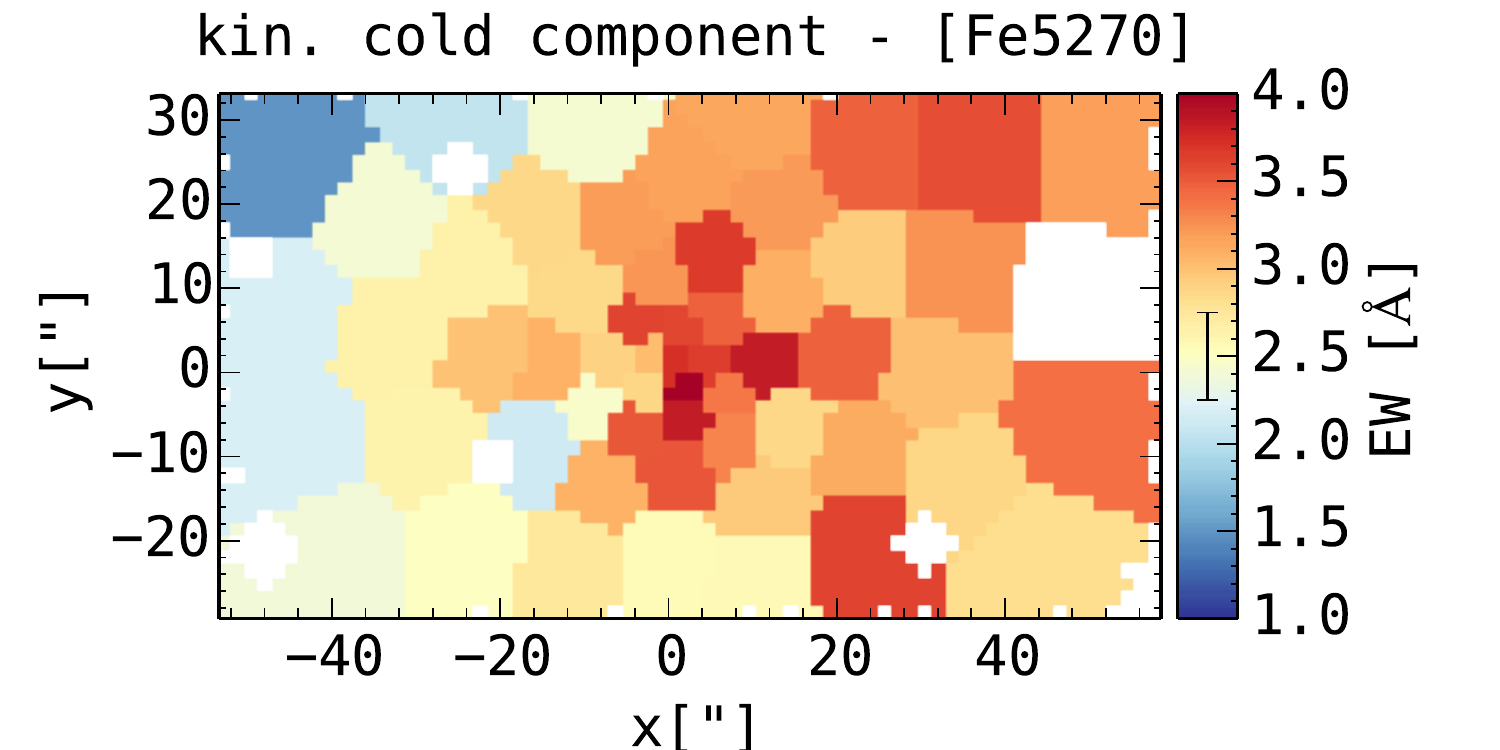}
\includegraphics[width=0.32\textwidth]{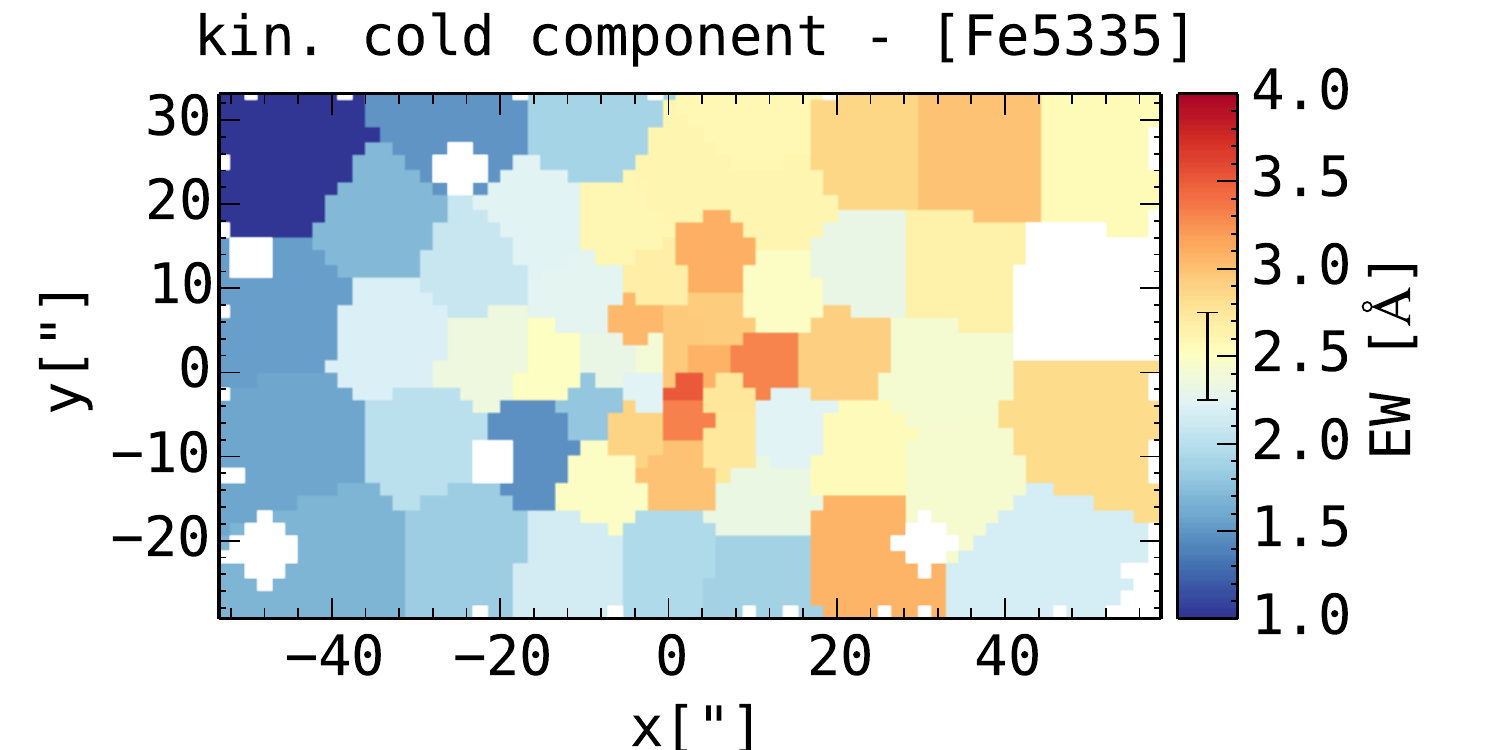}
\includegraphics[width=0.32\textwidth]{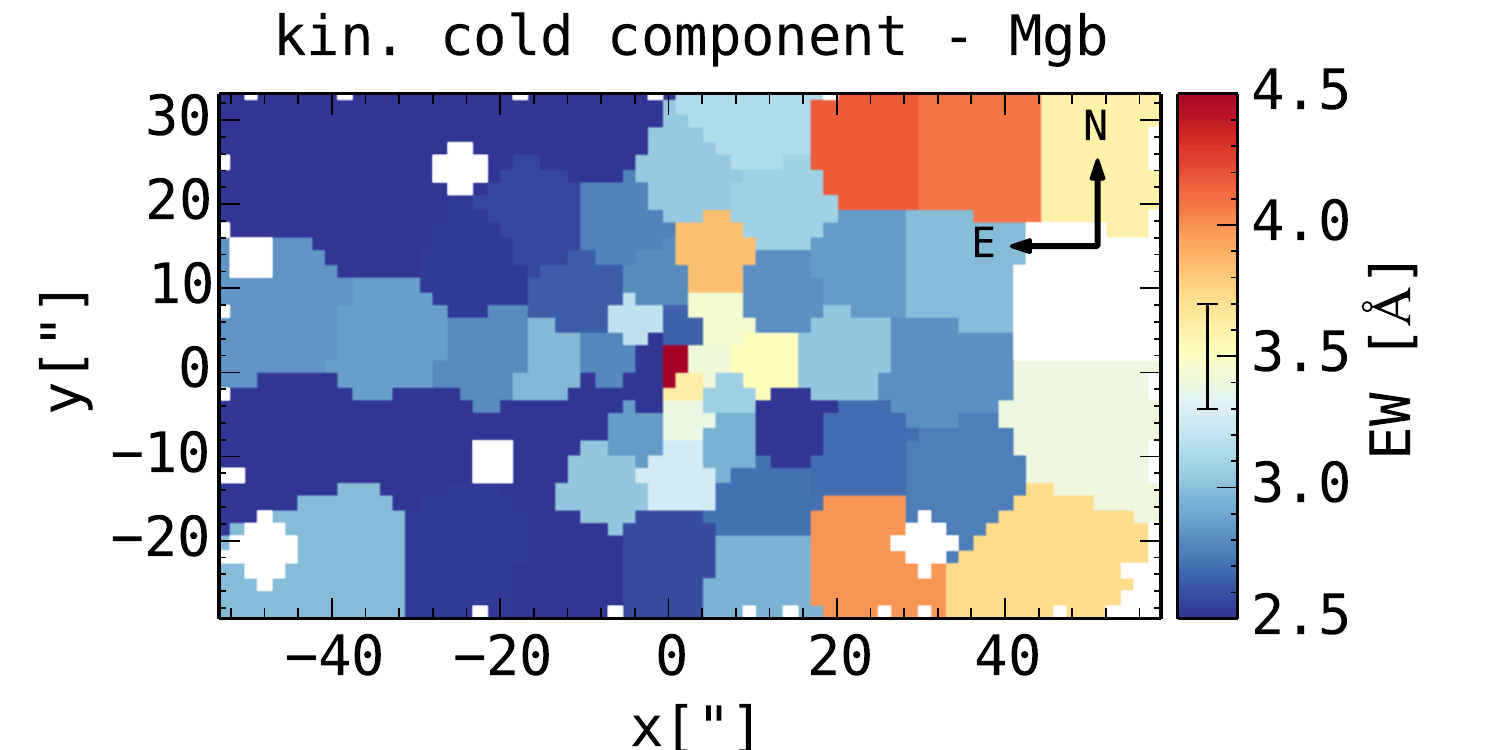}\\
\includegraphics[width=0.32\textwidth]{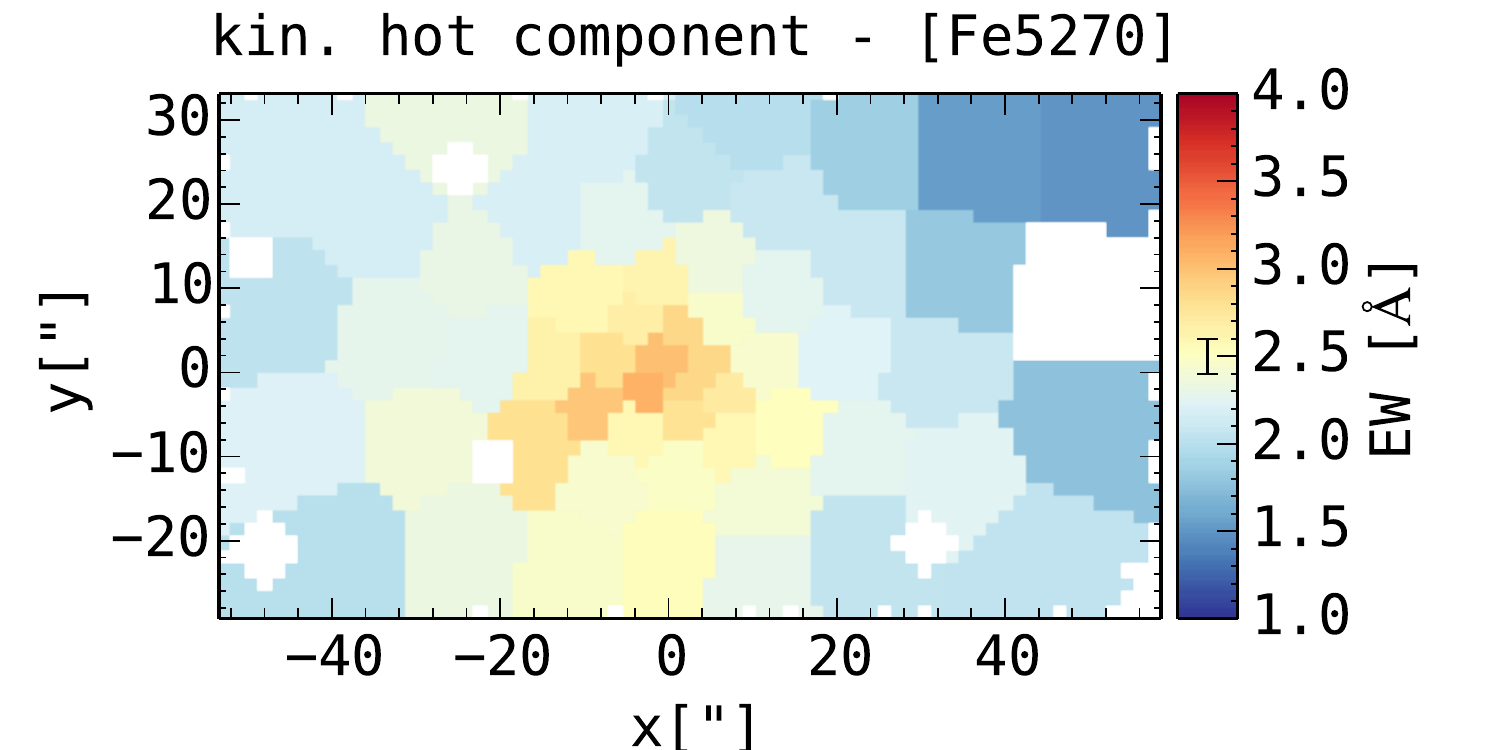}
\includegraphics[width=0.32\textwidth]{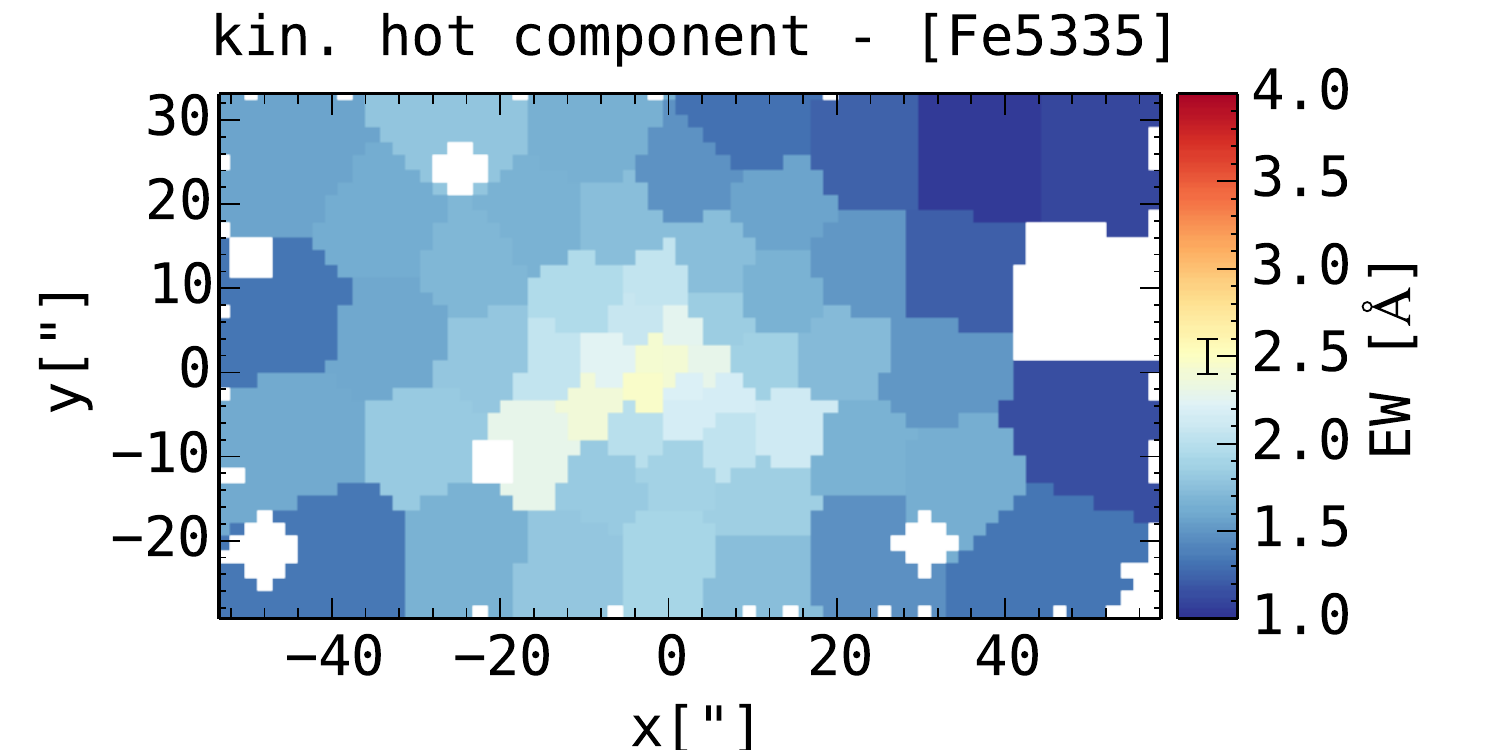}
\includegraphics[width=0.32\textwidth]{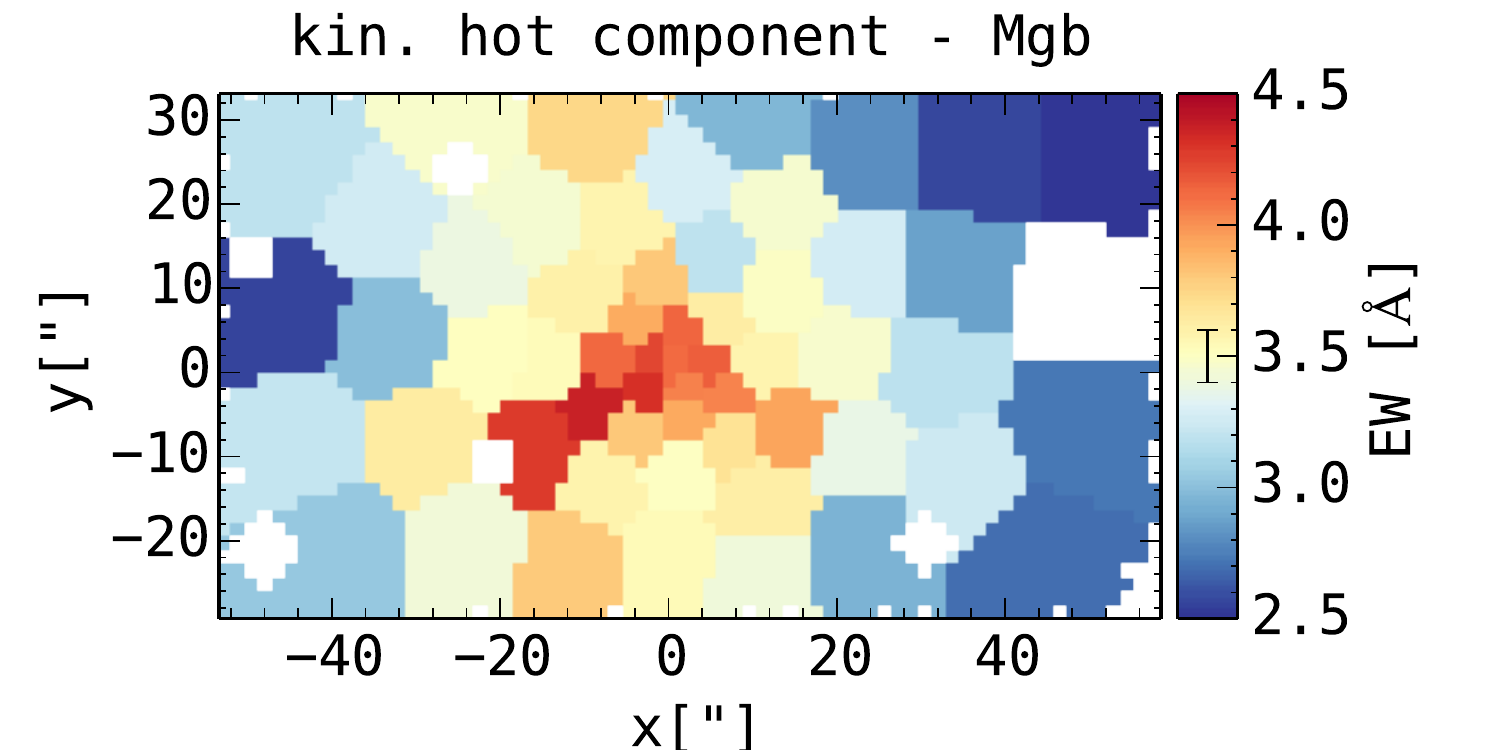}
% From Lodo & plot_maps.py
\end{center}
\caption{Two-dimensional maps of line strength indices, calibrated to
    the Lick system, for the cold (top panels) and hot stellar (top
    panels) components. The vertical bars in the color scale indicate
    the mean uncertainty as obtained from the MC simulations.} 
\label{fig:indices}
\end{figure*}

\begin{figure}
\begin{center}
\includegraphics[width=0.47\textwidth,angle=0,bb=60 360 370 645]{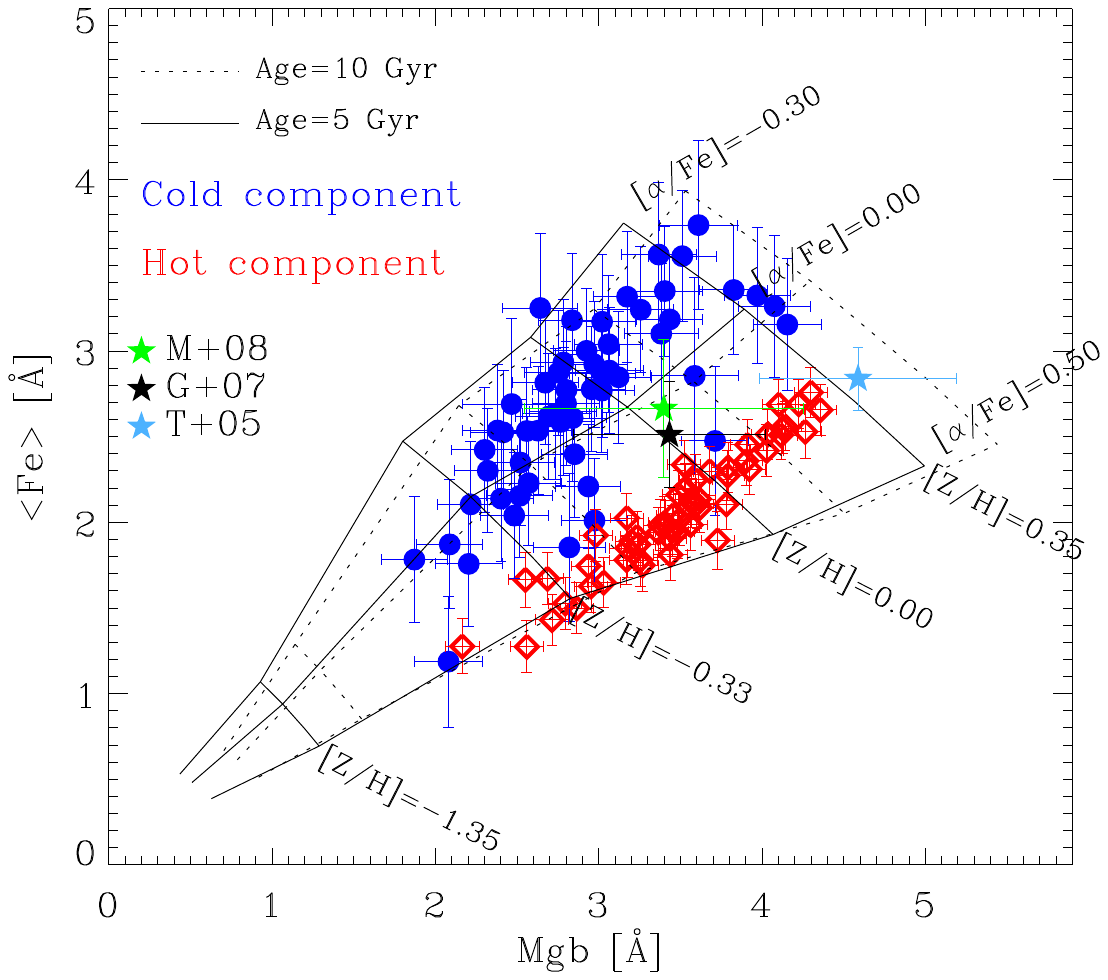}
% from Lodo
\end{center}
\caption{Mg\textit{b} and $\langle \mathrm{Fe} \rangle$ line strengths
    indices measured for the cold component (blue circles) and the hot
    component (red diamonds) in NGC\,7217. Predictions from single
    stellar population models \citep{Thomas+03} are also shown for
    comparison. The black, and green stars represent the mean values of
    the measurements by \citet{Gorgas+07} and \citet{Morelli+08} for a
    sample of bulges in spiral galaxies with the corresponding error
    bars representing the standard deviations of their measurements. The
    light blue star and its error bars represents the mean measurements
    and the standard deviation of a sample of early-type galaxies by
\citet{Thomas2005}.}
\label{fig:index_plane}
\end{figure}

In \reffig{fig:indices} we show the two-dimensional maps of the
equivalent width of \mgb, Fe5270, and Fe5335 for the two stellar
components. We use a set of Monte Carlo simulations that match the
observational setup (wavelength coverage, spectral resolution and
sampling, signal-to-noise) and the mean galaxy characteristics (velocity
separation between the two components, stellar velocity dispersion,
flux) to compute the uncertainty on the measured Lick indices. The
mean uncertainties from the simulations are about 0.2\,\AA\ and
0.1\,\AA\ for the indices of the cold and the hot component
respectively. In addition, these simulations show that the errors on the
Lick indices would have been $> 2$\AA, if we had not used the
constraints on the kinematics derived in \refsec{sec:kindecomp}. In
other words, the spectral decomposition code with the given instrumental
setup, needs independent kinematic measurements to remove the degeneracy
when recovering the best fitting stellar templates. The same results are
shown in \reffig{fig:index_plane}, where the measurements are plotted in
the $(\mgb, \langle \mathrm{Fe}
\rangle=(\mathrm{Fe5270}+\mathrm{Fe5335})/2$) plane and compared to the
predictions of single stellar populations models. The two components
have different line strengths: the cold component has significantly
lower values of \mgb, and significantly higher values of Fe5270 and
Fe5335 than the hot component.

We note a systematic trend in the measurements: the west side of the
galaxy has higher values than the east for all indices. This effect, of
unknown cause, is evident in the fainter cold component and negligible
in the brighter hot component, and explains the large scatter in the
measurements observed for the cold component (see
\reffig{fig:index_plane}). Nevertheless, this systematic trend is
smaller than the observed difference between the two kinematic
components in the \MgbFe\ plane, and does not invalidate our results.

\begin{figure}
\begin{center}
\includegraphics[width=0.46\textwidth,angle=0,bb=58 354 510 730]{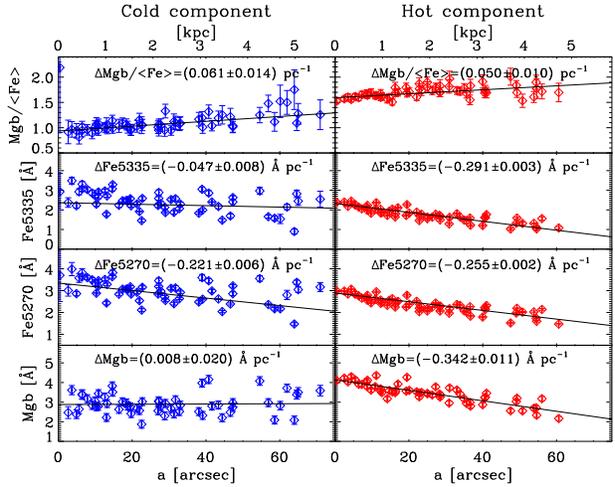}
\end{center}
\caption{Line strength indices of the cold component (left panels) and
    the hot component (right panels), as a function of radius. The black
    lines are linear fits to the data. The gradients of the fits are
shown in the panels.}
\label{fig:ind_profiles}
\end{figure}

The two components also show a radial gradient in the measured indices
(\reffig{fig:indices}): central regions have higher values of the Fe and
\mgb\ equivalent widths (see also \reffig{fig:ind_profiles}). Radial
gradients are more evident in the hot component than in the cold because
of the larger scatter and systematic trend present in the latter. The
corresponding gradients are shown in the figure.

%%%%%%%%%%%%%%%%%%%%%%%%%%%%%%%%%%%%%%%%%%%%%%%%%%%%%%%%%%%%%%%%%%%%%%%%%%%%%%%
\section{Ionized Gas Kinematics and Tilted Ring Analysis}
\label{sec:tilted_ring}
%%%%%%%%%%%%%%%%%%%%%%%%%%%%%%%%%%%%%%%%%%%%%%%%%%%%%%%%%%%%%%%%%%%%%%%%%%%%%%%
The velocity field of the \oiii\ emission shows a striking resemblance
to that of the cold component. The velocity dispersion of 20\kms\ to
30\kms\ is also similar to the values found in the cold component.
However, within a radius of 6\arcsec\ the velocity field of the gas shows
a strong twist. \citet{Silchenko2000} claim the existence of a nuclear
polar gas ring in NGC\,7217 that may be responsible for this twist. We
now investigate this scenario in detail.

We construct a tilted ring model to reproduce the ionized gas kinematics
under the assumption of purely circular motion. The model consists of 45
rings equally spaced in radius that are projected onto the sky plane.
The free parameters for each ring are: its relative weight, the observed
position angle, the inclination, and the circular velocity. We project
those rings onto the same pixel grid that we use for the binning of our
observed data and also compute mean velocities in the same Voronoi bins
as we use for the observations. We fit the model in two stages. First we
tie the position and inclination angles as well as the circular
velocities of all rings to each other which effectively results in a
simple thin disk model.  We then optimize these three free parameters
(the weight is fixed to one in the disk model) for minimal $\chi^2$.
This already generates a very good model for the gas disk velocities
outside of the central 15\arcsec\ with residuals that seem to be
dominated entirely by noise.  The best fitting parameters are
$\mathrm{PA} = 84.6 \pm 0.6 \Deg$, $i=33.9 \pm 1.2 \Deg$ and $v_c =
260.3 \pm 6.8$\,\kms.

\begin{figure}
\begin{center}
\includegraphics[width=0.32\textwidth]{plots/v_map_OIII} 
\includegraphics[width=0.32\textwidth]{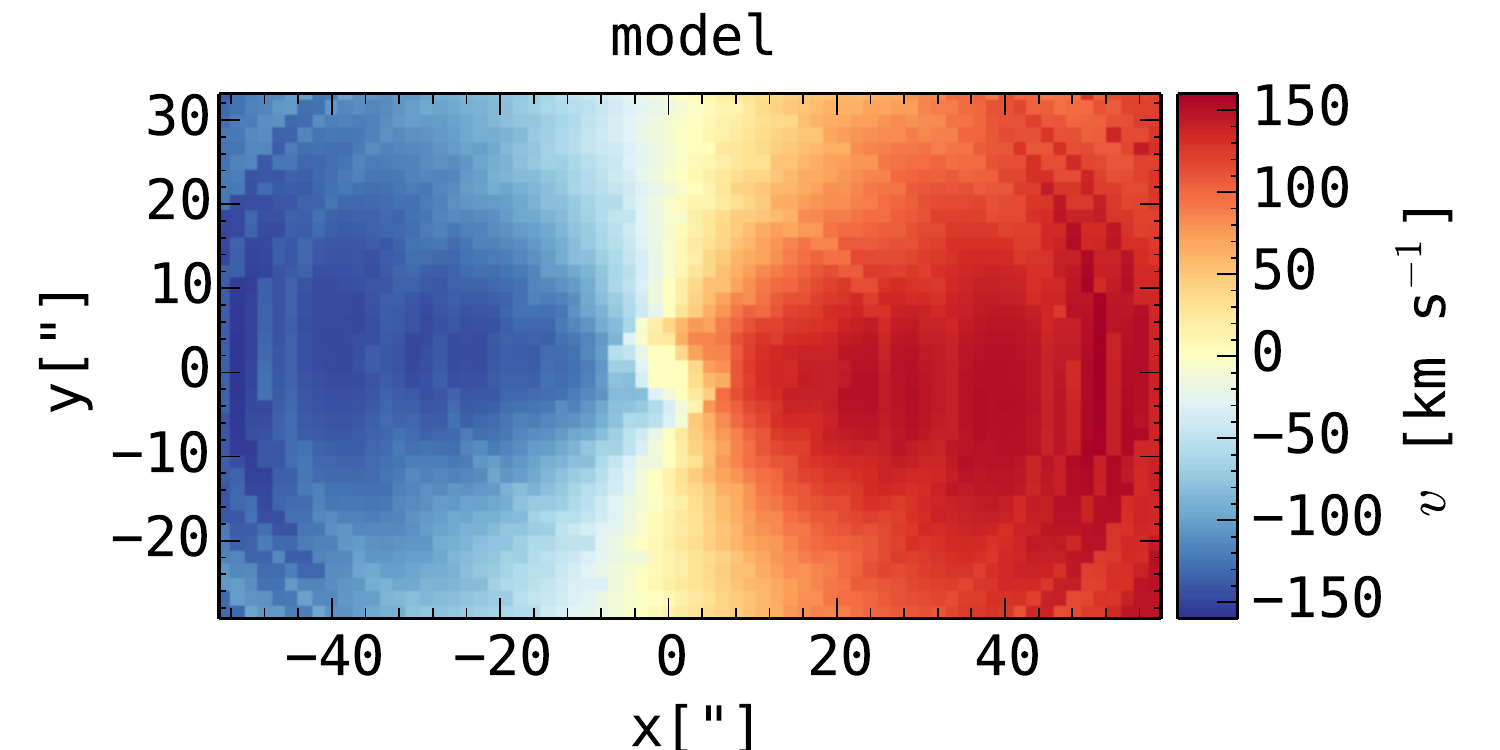} 
\includegraphics[width=0.32\textwidth]{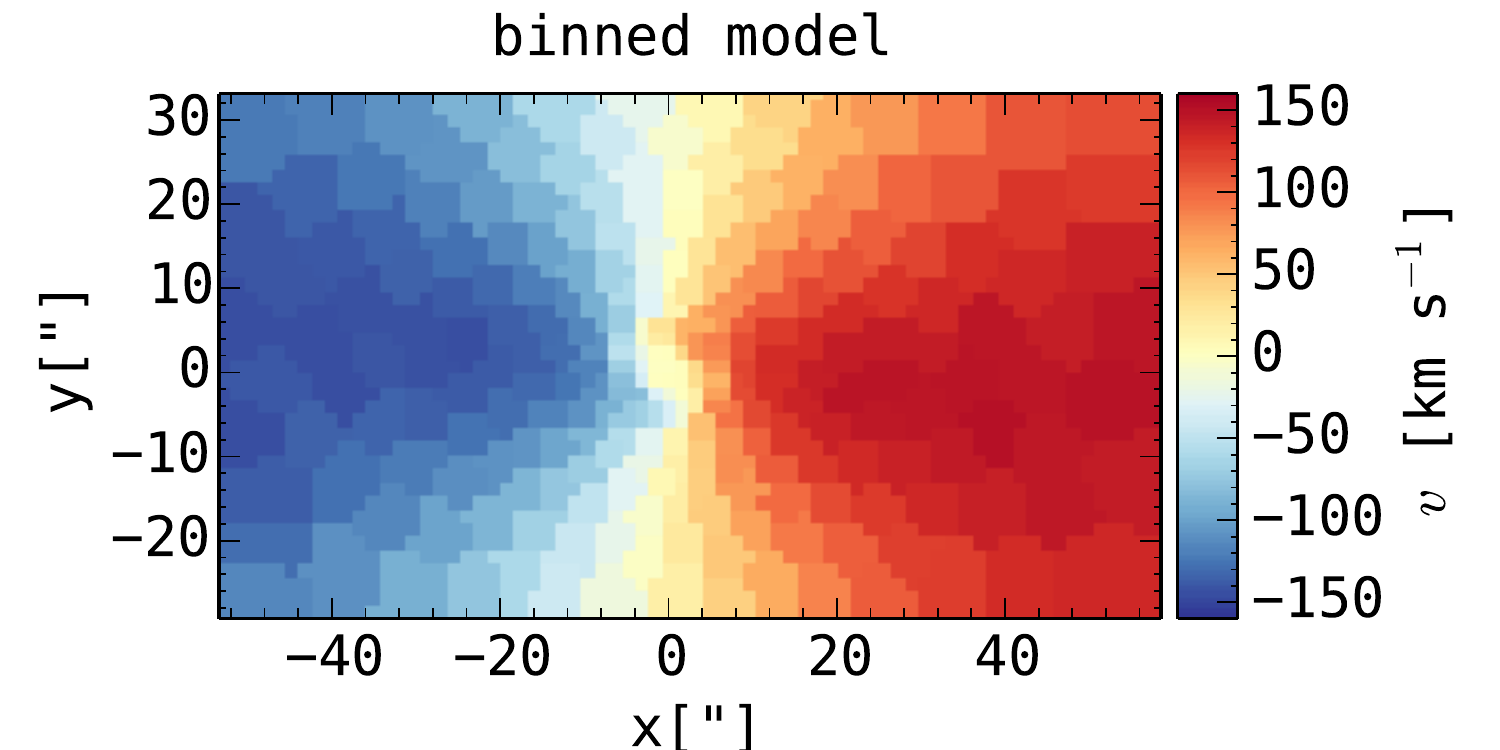} 
\includegraphics[width=0.32\textwidth]{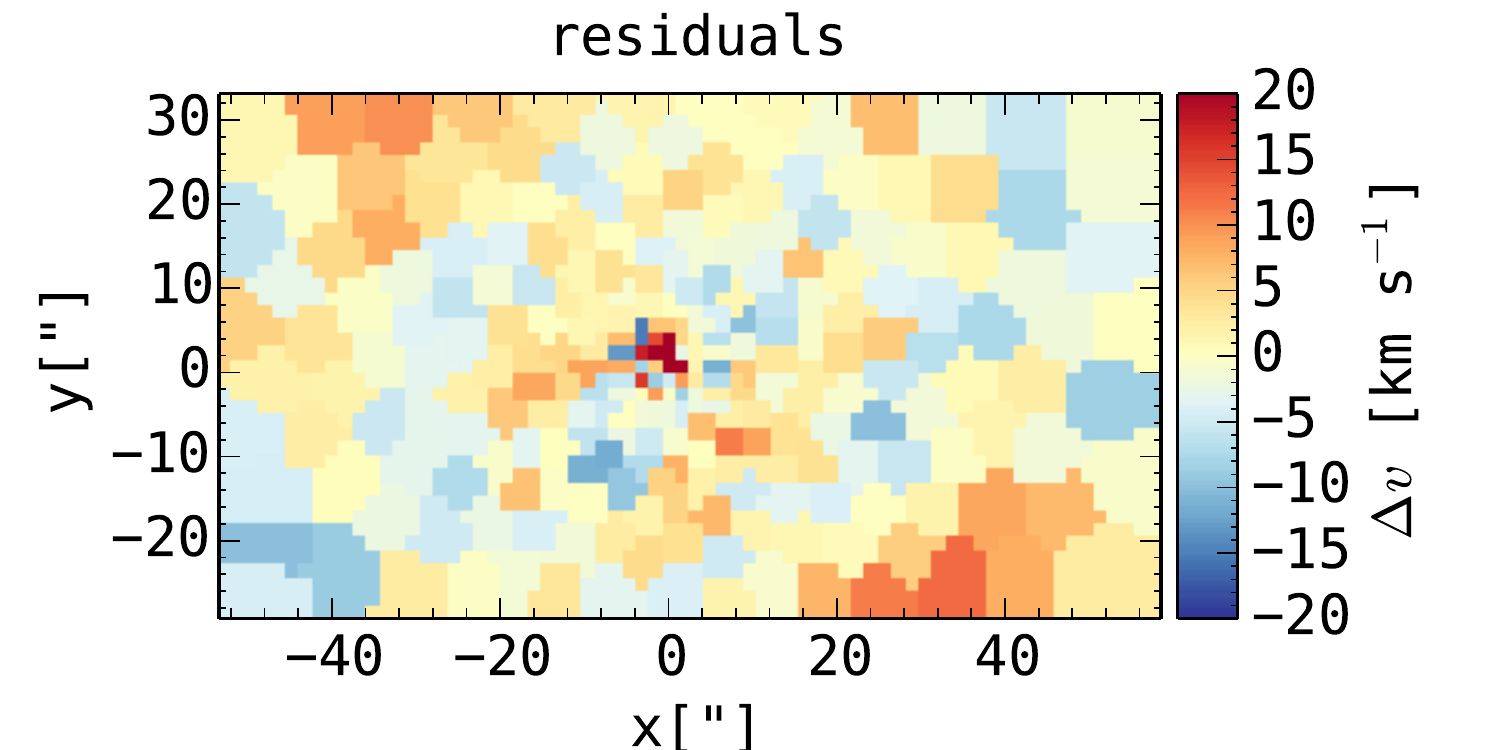} 
% ./tring/tring.py & plot_maps.py
\end{center}
\caption{Tilted ring model for the gas disk. The upper panel reproduces
    the mean line-of-sight velocity of the \oiii\ emission from
    \reffig{fig:kin_maps}.  We fit a series of 45 rings with a radial
    spacing of 1.6\arcsec. We sample each ring every 2\Deg\ and
    project the line-of-sight component of the circular velocity into
    the pixel grid. The fitted parameters are the relative weight of a
    ring, its position angle on the sky, its inclination angle and the
    circular velocity. The fit is carried out in two stages as described
    in the text. The best fitting parameters are plotted in
    \reffig{fig:tring_param}. The second panel shows
    our best fitting model. The third panel shows the model after
    the application of the same Voronoi binning scheme that we use for
    the data. The bottom panel shows the residuals.
} 
\label{fig:tring}
\end{figure}

In the second stage we now untie the relative alignment and the circular
velocities and rerun the fit --- again minimizing for $\chi^2$ --- while
using the best-fit parameters from the first stage as initial guesses.
In \reffig{fig:tring} we reproduce the gas velocity field from
\reffig{fig:kin_maps} and show plots of the best fitting model and a map
of the residuals. The third panel shows the Voronoi-binned velocity map
of the model. The central twist is reproduced well by the tilted ring
model. 

Using the parameters from the thin disk model, we derive the angular
separation $\phi$ of the rotation axis of the inner rings with respect
to the rotation axis of the outer disk. 
\begin{figure}
\begin{center}
\includegraphics[width=0.45\textwidth]{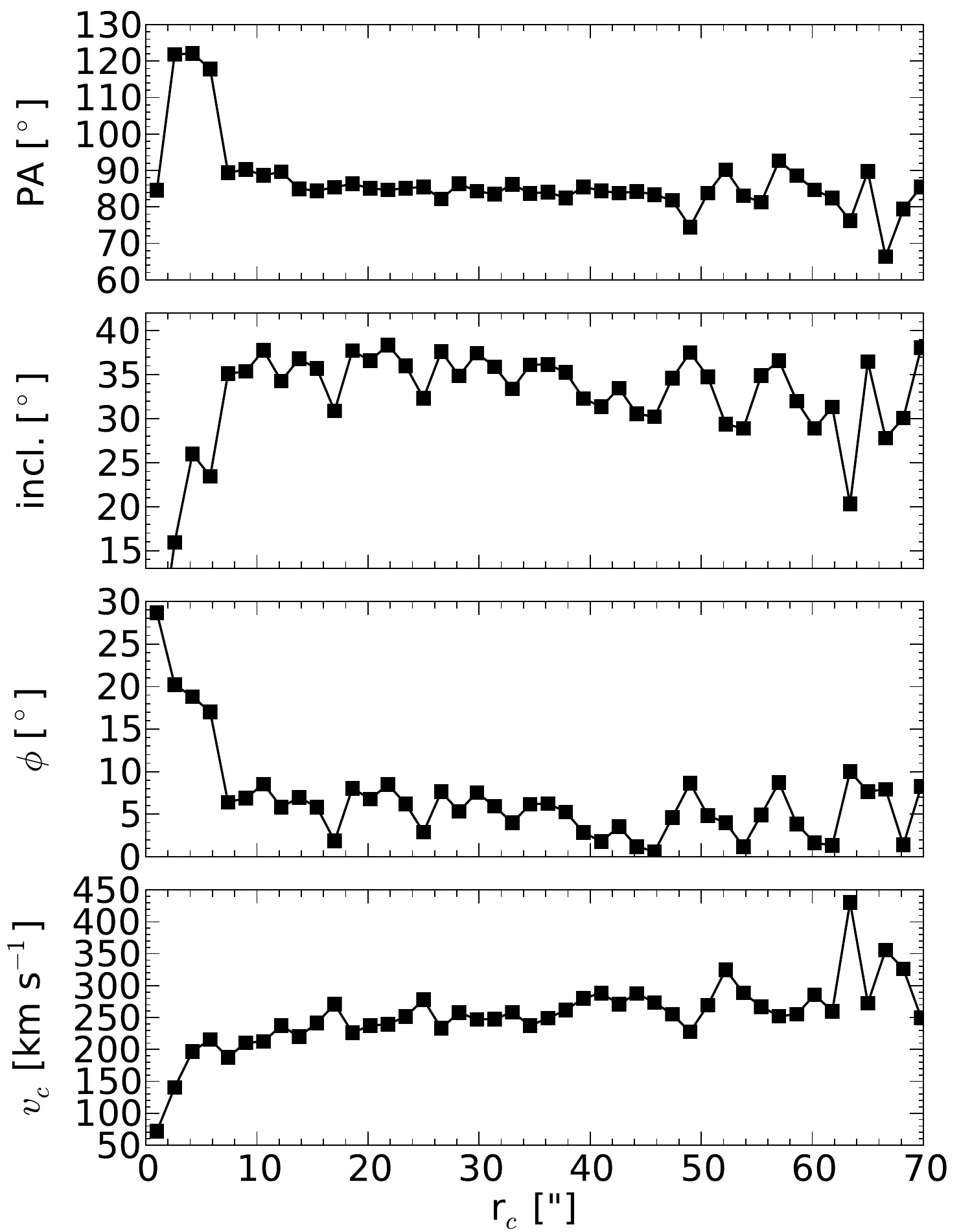} 
\end{center}
\caption{The best fitting parameters for the tilted ring model that is
    shown in \reffig{fig:tring}. The upper two panels plot the
    projected position angle on the plane of the sky and the inclination
    angle for all rings. The third panel shows the angular separation
    between the corresponding axis of rotation and the axis of the
    rotation of the best fitting global thin disk model that we derived
    in the first stage of our modelling. The lowermost panel shows the
    circular velocity as function of ring radius.  
} 
\label{fig:tring_param}
% ./tring/tring.py & plot_tring.py
\end{figure}
\reffig{fig:tring_param} shows the parameters that we derive for the
titled rings as a function of the ring radius. From top to bottom, the
panels show the projected position angle, the inclination, the angular
separation and the circular velocity. From our data the maximum angular
separation of the inner rings with respect to the outer disk is
30\Deg. This is significantly lower than the 90\Deg\ that would
imply a polar ring, and it resembles more an inner warped gaseous
structure, as observed, for example, in the spiral galaxy NGC\,2855
\citep{Coccato+07}. 

If the central regions of NGC\,7217 do host a non-axisymmetric
structure, as suggested by B95, then an alternative explanation to the
observed twist may lie in non-circular motions of the gas due to a
corresponding non-axisymmetry of the potential
(e.g.\,\citealt{deZeeuw+89, Athanassoula1992}). 

\begin{figure}
\begin{center}
\includegraphics[width=0.45\textwidth]{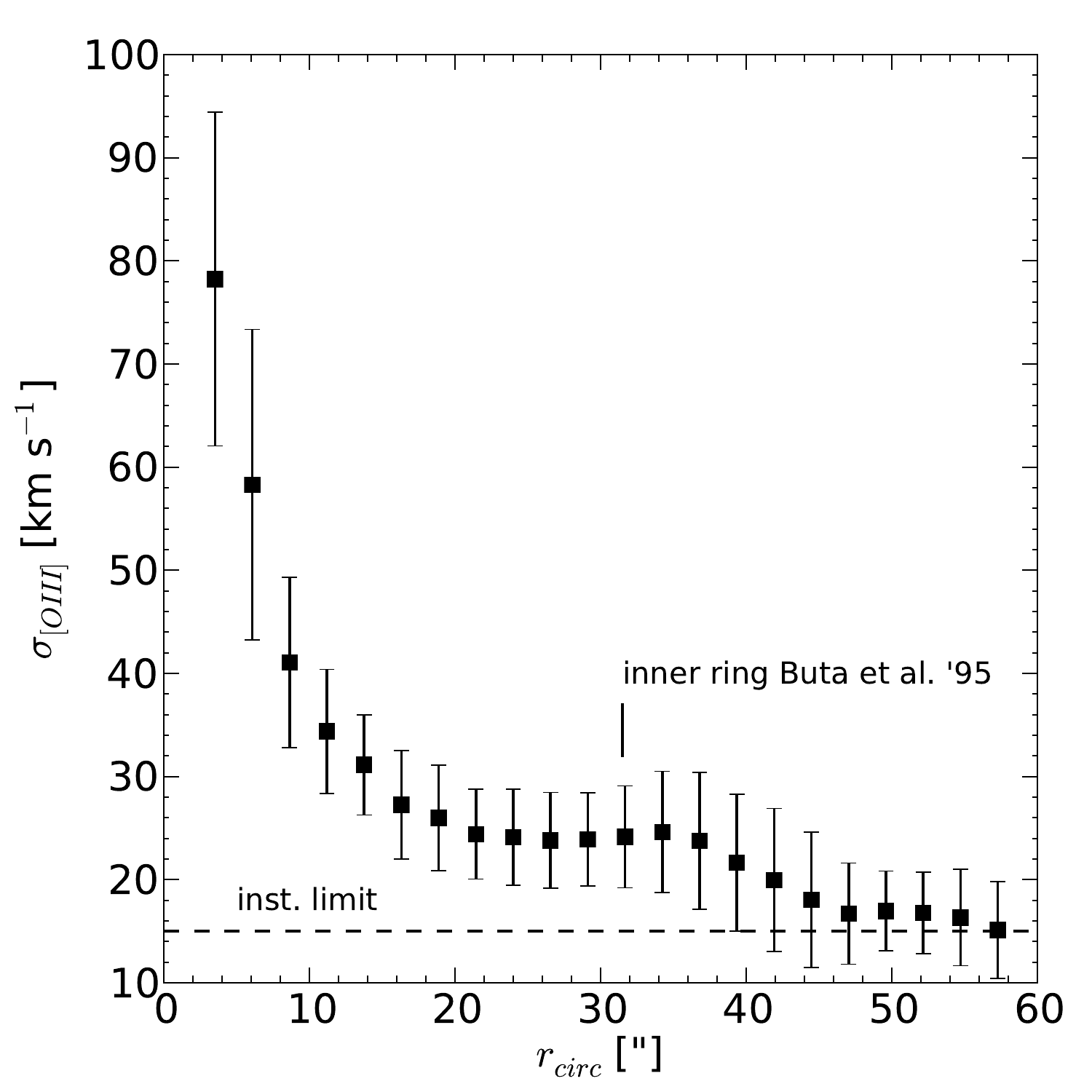}
% plot_sgasofr.py 
\end{center}
\caption{Velocity dispersions of the \oiii\ emission line as
    a function of radius. We derive the mean dispersion in elliptical
    annuli. The ellipticity of the annuli is set to 0.2. The horizontal axis
    plots the length of the semi-major axis. The error bars show the
    variation of the gas dispersion within each annulus. These 
    dispersions have been corrected for the instrumental resolution by
    subtracting 15\,\kms\ in quadrature. The horizontal dashed
    line indicates the resolution limit.
} 
\label{fig:vdisgas}
\end{figure}

We find a slight increase in the gas dispersion at the location of the
inner ring. This is best seen in \reffig{fig:vdisgas} where we plot the
averaged gas dispersion in elliptical annuli with an ellipticity fixed
to the value of 0.2 and a position angle of 85\Deg. B95 gives a radius
of 32\arcsec\ for the inner ring. The dispersion profile does show a
shelf or a hump at about the same location and falls off to $\approx
15$\,\kms\ --- which is just the instrumental resolution limit --- at a
radius of 45\arcsec. Inside the ring, the dispersion rises
continuously to about 80\,\kms\ at 3\arcsec, which is within the radius
our measurements become unreliable due to their limited spatial
resolution. 

%%%%%%%%%%%%%%%%%%%%%%%%%%%%%%%%%%%%%%%%%%%%%%%%%%%%%%%%%%%%%%%%%%%%%%%%%%%%%%%
\section{Dynamical Mass To Light Ratio Compared to Stellar Population Analysis}
\label{sec:mass_to_light}
%%%%%%%%%%%%%%%%%%%%%%%%%%%%%%%%%%%%%%%%%%%%%%%%%%%%%%%%%%%%%%%%%%%%%%%%%%%%%%

In the last section we saw that the ionized gas in NGC\,7217 shows very
regular rotation outside of $r \approx 5$\arcsec\ with a deprojected
velocity of 200\,\kms. In \refsec{sec:results_kin} we measure velocity
dispersions $\approx$20\,\kms. The gas disk is therefore well
approximated by a cold and thin disk and we can directly derive the
total gravitating mass enclosed in a sphere of a given radius. On the
other hand, the inner dust ring ($r \approx 10.5$\arcsec) marks a strong
transition from the outer dusty spiral morphology to a relatively dust
free central region giving a relatively unperturbed view of the central
stellar population (\citealt{Fisher2008}, see also
\reffig{fig:finderchart}), which is presumably dominated by the
spheroid. 

We can therefore attempt to derive a stellar mass-to-light ratio from a
stellar population analysis, and compare this value to the dynamical
measurement. We obtained archival SDSS \textit{ugriz}-band images
\citep{York2000}, FUV and NUV GALEX \citep{Martin2005} data from
Mikulski Archive for Space Telescopes, and Spitzer IRAC channel 1 to 4
data \citep{Werner2004, Fisher2010}. We measure the mean flux in an
aperture of diameter 10\arcsec\ centered on the galaxy and estimate the
background in empty regions outside of the galaxy.  The 10\arcsec\
aperture stays clear of the dust ring even in the lower resolution GALEX
data. We correct all images for galactic foreground extinction using an
$E(B-V)$ of 0.0761 (\citealt{Schlafly2011} recalibration of
\citealt{Schlegel1998}). We then use the updated 2007 version of the
\citet{Bruzual2003} models for single-aged stellar populations to invert
the colours into stellar population parameters and, in particular, to
derive mass-to-light ratios. As the models are relatively coarsely
spaced in metallicity, we interpolate the predicted fluxes between
metallicity values linearly, adding three equally spaced metallicity
values between the published quantities. For each of the resulting
models we compute the $\chi^2$ between the predicted magnitudes and the
measured values assuming an error of $\pm 0.15$\,mag in the GALEX bands
\citep{Gil-de-Paz2007} and $\pm 0.05$\,mag in the other bands. The
best-fitting model is chosen based on the minimum $\chi^2$ value and the
errors are estimated by considering the maximum and minimum values of
all models that have $\Delta\chi^2 = 1$. The IRAC 4 magnitudes were
typically poorly fit which may be a consequence of the poorer resolution
in this band and an onset of the contribution from dust emission. We
exclude this band from the fit.

The GALEX channels are particularly important to break the
age/metallicity degeneracy.  Without those data models with $[Z/H] =
0.0$, 0.1 and 0.2 give similarly low $\chi^2$ vales with very little
constraint on the mass-to-light ratio. Adding the GALEX data however
results in a clear preference of $[Z/H] = 0.2$, yielding a
\textit{V}-band mass-to-light ratio of $\Upsilon^*_V = 3.9_{3.7}^{4.3}
M_{\astrosun}/L_{\astrosun}$ and an age of $7.5_{7.2}^{8.3}$\,Gyr if a
Kroupa Initial Mass Function (IMF;\ \citealt{Kroupa2001}) is assumed. For
a Salpeter IMF \citep{Salpeter55} we obtain values of $\Upsilon^*_V =
6.1_{5.9}^{7.2} M_{\astrosun}/L_{\astrosun}$ and an age of
$7.5_{7.2}^{10}$\,Gyr. We do have an independent estimate of the
metallicity from the line strength analysis in
\refsec{sec:indices_results}. By comparing \reffig{fig:ind_profiles} and
\reffig{fig:index_plane} one can see that for the central Mg and Fe line
strengths of the hot component, the adopted \citet{Thomas2003} models
predict metallicities between $[Z/H] = 0.15$ assuming a 10\,Gyr old
population and $[Z/H] = 0.3$ when assuming a 5\,Gyr old population. This
range of metallicities is compatible with the photometrically derived
value. Adding this constraint to our parameter inversion while dropping
the FUV/NUV bands results in essentially identical predictions of the
stellar mass-to-light ratio and ages. 

To determine the dynamical mass-to-light ratio, we evaluate the gas
rotation curve of our tilted ring model in the previous section at a
radius or 6\arcsec. This is slightly larger that the aperture size we
used for the stellar population analysis but puts the analysis slightly
more comfortably outside of the twist in the velocity field. Using the
Tully-Fisher based distance estimate of $D = 18.4$\,Mpc
\citep{Russell2002} and a velocity of $200 \pm 10$\,\kms\ we obtain a
dynamical mass of $5.0 \pm 0.5 \times 10^9\,M_{\astrosun}$ inside a
radius of 535\,pc.

In the central regions of a stellar system, any aperture photometry
necessarily integrates also over the whole light in a cylinder with the
radius of that aperture. In order to compute the true dynamical
mass-to-light ratio, one must find the actual fraction of light that is
contained inside of a sphere of the radius at which the rotational
velocity is determined. For this we deproject the analytical model of
the \textit{V}-band surface brightness profile of B95 for the spheroid.
The deprojection is carried out axisymmetrically using the code of
\cite{Magorrian1999}. Using the magnitude of the sun of $V = 4.83$
\cite{Binney1998} and an extinction of $A_V = 0.24$ \citep{Schlafly2011}
we derive a luminosity of $1.1 \times 10^9\,L_{\astrosun}$ or dynamical
mass-to-light ratio of $\Upsilon^*_{\mathrm{dyn}} = 4.5
M_{\astrosun}/L_{\astrosun}$. This value is marginally larger than the
mass-to-light ratio that we derive from the stellar population analysis
for a Kroupa IMF and smaller than the values expected for a Salpeter
IMF.\ Given the central velocity dispersion of 150--170\,\kms\
this is in excellent agreement with relations derived for early-type
galaxies \citep{Treu2010,Thomas2011,Cappellari2013}. 

%%%%%%%%%%%%%%%%%%%%%%%%%%%%%%%%%%%%%%%%%%%%%%%%%%%%%%%%%%%%%%%%%%%%%%%%%%%%%%%
\section{Discussion}
%%%%%%%%%%%%%%%%%%%%%%%%%%%%%%%%%%%%%%%%%%%%%%%%%%%%%%%%%%%%%%%%%%%%%%%%%%%%%%%
\label{sec:discussion}
%%%%%%%%%%%%%%%%%%%%%%%%%%%%%%%%%%%%%%%%%%%%%%%%%%%%%%%%%%%%%%%%%%%%%%%%%%%%%%%
\subsection{Stellar co vs.\, counter-rotation}
%%%%%%%%%%%%%%%%%%%%%%%%%%%%%%%%%%%%%%%%%%%%%%%%%%%%%%%%%%%%%%%%%%%%%%%%%%%%%%%
We do not confirm the existence of a counter-rotating disk that was
previously claimed. The shape of our derived LOSVDs and the kinematic
decompositions rather suggest that NGC\,7217 hosts a sub-dominant cold
stellar component co-rotating with a bright hot stellar component.

In \citet{Silchenko2000} the rotation curve along $\mathrm{PA} =
240$\Deg\ is shown. Their rotation curve, derived using the absorption
line feature [Na\,\textsc{i}]$\lambda\lambda$ 5890,5896\,\AA\ shows
counter-rotation within the central 40\arcsec~(their Fig. 6). The main
stellar component rotates slowly and appears asymmetric. It reaches a
maximum amplitude of $\approx 100$\,\kms, and then stays flat at $R >
30$\arcsec, and declines at $R < -30$\arcsec. The asymmetry in the
counter-rotating component is even larger and reaches an amplitude of
$\approx 300$\,\kms\ at $R < -10$\arcsec\ but only $\approx 200$\,\kms\
for $R > 10$\arcsec.

While they do not derive rotation curves, \citet{Merrifield1994}
originally found counter-rotation based on their detailed analysis of
individual velocity distributions. The authors did consider the
contamination by a very massive bulge component as explanation for their
observed tail in the absorption line profiles but rejected that
hypothesis based on the argument that such a hot component had not been
observed and would be atypical for a disk galaxy with spiral structure.
Through deep photometry B95 however showed the presence of such a halo
and argued that the relative luminosities are in the right order to
generate the observed profiles. Our kinematic confirmation of a hot halo
now further completes this picture.
 
However, the velocity distributions presented by \citet{Merrifield1994}
in their Fig.\,6 do show a distinctive secondary peak that cannot be
explained by a bulge contamination. We can only speculate at the reason
for this secondary peak: first, their data are long slit spectra which
prohibit two dimensional binning. Compared to their Fig.\ 3, the Voronoi
bins of our IFU data have significantly higher $S/N$ at the important
radial range of the two-component overlap. Second, their data have a
spectral resolution of $\sigma_{\mathrm{instr}} \approx 25$\,\kms\
(measured at 5200 \AA), and the authors rebinned their spectra in [$\ln
\lambda$] to have a velocity step of 28.5\,\kms\ per pixel. In
comparison, our data have a spectral resolution of 15\,\kms\ and we bin
our spectra in $\ln \lambda$ with a velocity step of 10\,\kms\ per
pixel. The resolution of our spectra therefore allows a cleaner
separation of the two components. Finally, \citet{Merrifield1994} do not
adopt any special treatment of emission lines, which might affect the
shape of their derived LOSVDs. 

The resolution argument also applies to the work of
\citet{Fabricius2012a}. There a similar double-Gaussian decomposition of
FCQ-derived LOSVDs suggested a dynamically hot, counter-rotating
component. However, the spectral resolution of that work was 39\,\kms,
which is significantly lower than the resolution used here.

\begin{figure*}
\begin{center}
\includegraphics[width=.8\textwidth,bb=25 25 475 406]{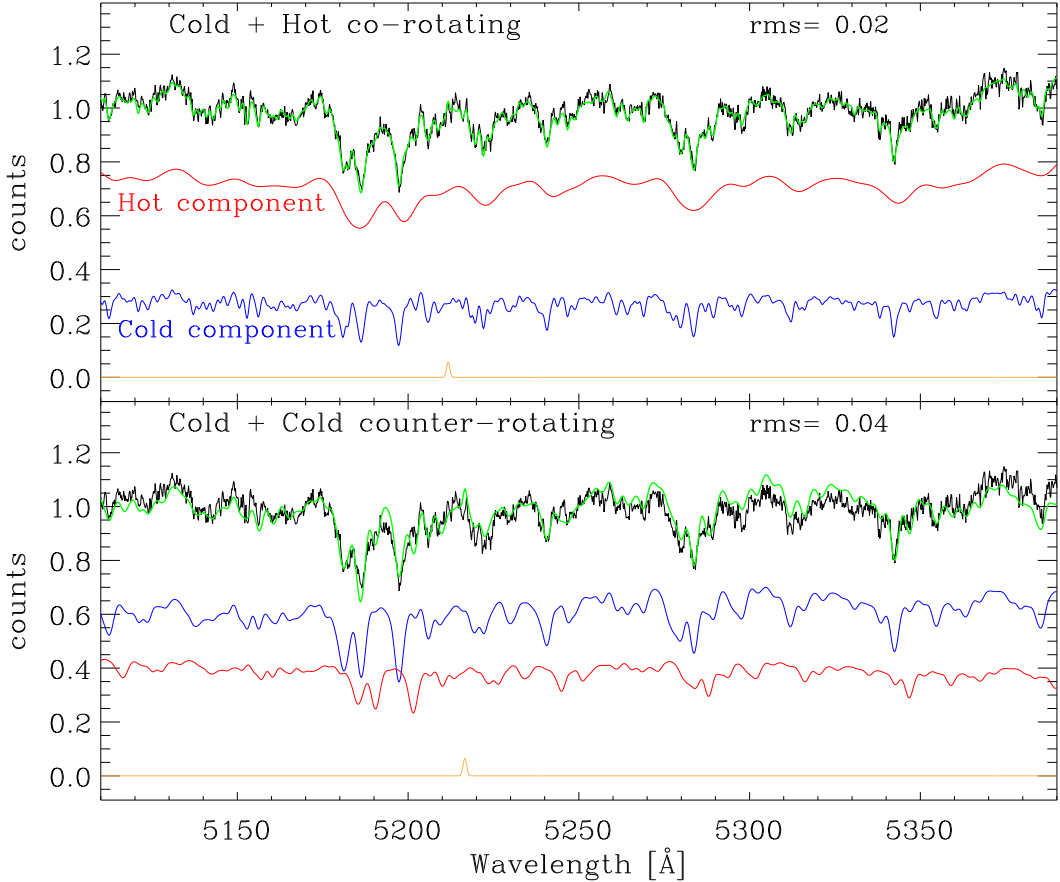}
\end{center}
\caption{Results of the spectral decomposition code. In both panels we
    show the actual data in black, while the best-fitting two components
    are plotted in red and blue. The green line shows the sum of those
    two components. The orange line shows the best-fitting model for the
    nebular emission. In the upper panel we show the actual best fit,
    while in the lower panel we enforce a counter-rotating disk
    scenario as suggested by \citet{Merrifield1994}.
}
\label{fig:enforced_counterrot}
\end{figure*}

We test that our results are not driven by systematics in the
non-parametric kinematic fit by cross-checking with the parametric
spectral decomposition technique of \citet{Coccato2011a}. We use
starting guesses that resemble two counter-rotating disks, as found by
\citet{Merrifield1994}. The best fit solution found by the spectral
decomposition code is still consistent with a cold disk co-rotating with
a hot stellar component, as found in \refsec{sec:results_kin}.
Further, our code allows us to force the fit to two counter-rotating
disks by locking their mean rotational velocities to be the opposite
with respect to the systemic velocity. In
\reffig{fig:enforced_counterrot} we compare the actual best-fitting
solution from the spectral decomposition (upper panel) with the enforced
counter-rotation solution. In that example the enforced counter-rotation
results in a RMS that is increased by a factor of two over the
non-constrained fit.

We stress the fact that both routines, i.e.\ the algorithm for the
recovery of non-parametric LOSVD with subsequent double-Gaussian
decomposition and the method of the direct spectral decomposition, are
completely independent. They both use their own treatment of the stellar
continuum and use different strategies for the derivation of the optimal
model spectrum.

%%%%%%%%%%%%%%%%%%%%%%%%%%%%%%%%%%%%%%%%%%%%%%%%%%%%%%%%%%%%%%%%%%%%%%%%%%%%%%%
\subsection{Structural parameters through kinematic decomposition}
\label{sec:kinematic_photometric_decomposition}
%%%%%%%%%%%%%%%%%%%%%%%%%%%%%%%%%%%%%%%%%%%%%%%%%%%%%%%%%%%%%%%%%%%%%%%%%%%%%%%

In Table~\ref{tab:struct_param} we list the structural parameters for
the two kinematic components. A kinematic double-Gaussian decomposition
cannot be expected to yield the same precision as a photometric
decomposition. But by comparing our values to the photometric data we
can conclude that our two components correspond to the stellar disk and
the spheroid that were inferred by photometric decomposition in B95.

Our derived values of the position angle are in reasonable agreement
with the value of 91\Deg\ from B95. In the radial range of $30\arcsec
\leq r \leq 120\arcsec$, our values for the ellipticity bracket their
values of about 0.3 (their Fig.\,6), which is expected given that they
measure the ellipticity of the superposition of the two components. For
$r > 140\arcsec$ B95, find an ellipticity of about 0.05 with some
variation, which is compatible with our findings.

\begin{figure}
\begin{center}
\includegraphics[width=0.45\textwidth]{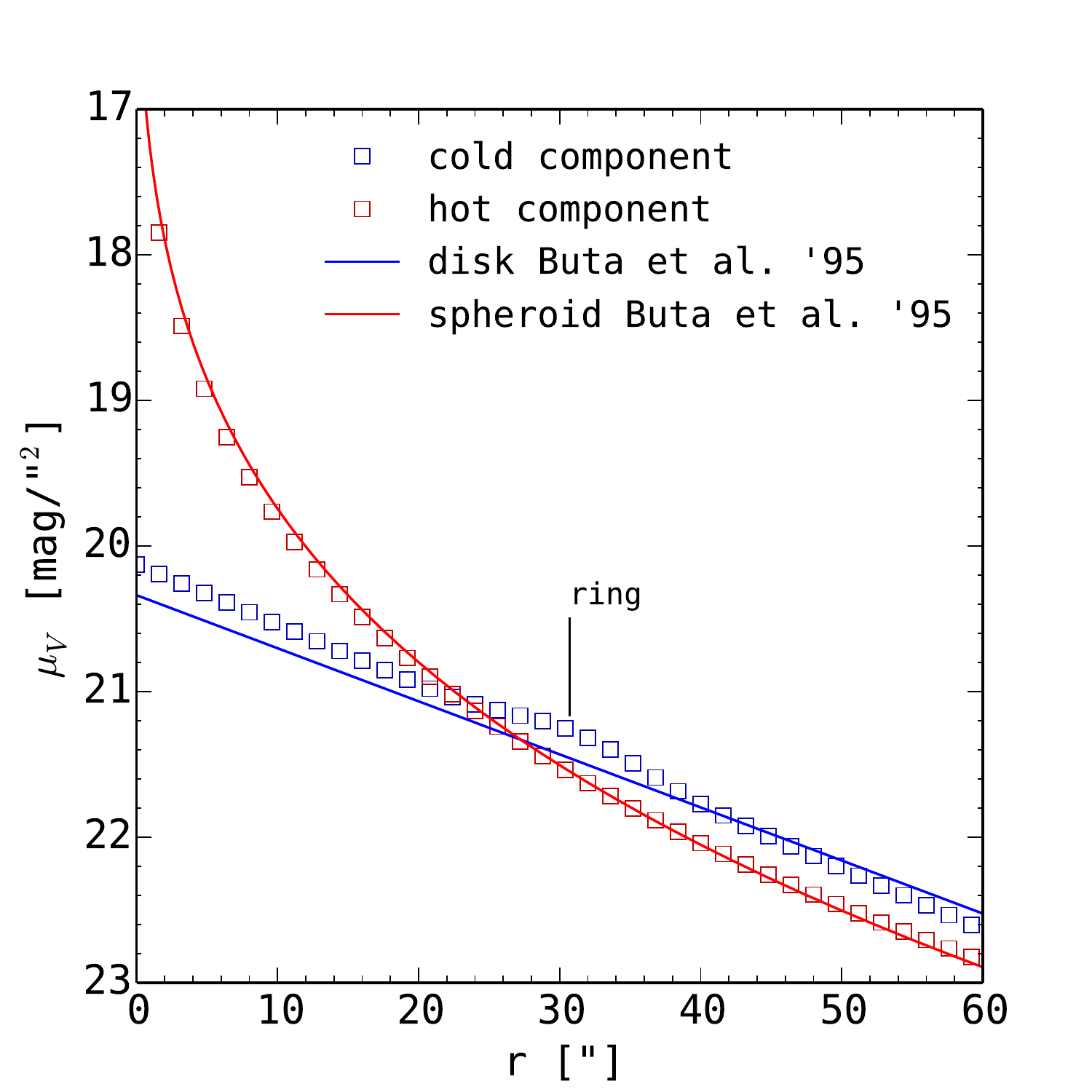}
\end{center}
\caption{Surface brightness models for the total stellar content in
NGC\,7217.  We fit two-dimensional models to the component images
shown in \reffig{fig:kin_maps} using the \texttt{Imfit} program by
Peter Erwin. Similarly to B95, we model the hot component with a
de~Vaucouleurs' surface brightness distribution and the cold
component with an exponential distribution. In the latter model we
include a Gaussian ring. The open squares represent one-dimensional
major-axis cuts through our models and the lines show the models
derived by B95. As our data are not flux calibrated we chose a zero
point to match our models against those of B95. We somewhat
arbitrarily choose the zero points for an optimal fit at 40\arcsec.
We note the need to apply zero points that differ by
$\approx$0.7\,mag to the two components.}
\label{fig:sb}
\end{figure}

For the spheroid, we find an effective radius of $58.1 \pm 11.9\arcsec$
which compares to the $V$-band value of $49.9 \pm 3.4\arcsec$ by B95.
For the scale length of the exponential disk we obtain $43.2 \pm
1.2\arcsec$ vs. B95's $29.7 \pm\ 2.0$\arcsec.\footnote{Following
\citet{Ciotti1991} we convert their $r_e$ to a scale length by $h =
r_e/1.678$.} The difference in the effective radii and scale radii are
easily explained by the much lower resolution and coverage of our
component images. The general shape of the radial distribution seems,
however, to be in acceptable agreement: in \reffig{fig:sb} we plot
one-dimensional surface brightness profiles that we derive from our
parameters and from the parameters of B95 as function of the major axis
radius. As our data are not flux-calibrated we scale our models for the
two components separately to match the B95 models at a radius of
40\arcsec. We find that we have to adopt different zero points to
match the two different components to the B95 data --- they differ from
each other by 0.7\,mag. However, as the relative weights in the
double-Gaussian decomposition are affected by differences in the
relative line strengths in a non-trivial fashion, the zero points cannot
be expected to be identical. Once the relative scaling has been
established at one radius, the relative fluxes of our decomposition are
very similar to those of the photometric decomposition.

We can therefore conclude with high confidence that the kinematic cold
and hot stellar components of NGC\,7217 found in this work are the
photometric disk and spheroid components found by B95. This demonstrates
that, given components that are sufficiently kinematically distinct, and
a high enough spectral resolution, it is possible to spectrally
decompose disk plus spheroid systems in a similar fashion to that common
in photometric studies.

%%%%%%%%%%%%%%%%%%%%%%%%%%%%%%%%%%%%%%%%%%%%%%%%%%%%%%%%%%%%%%%%%%%%%%%%%%%%%%%
\subsection{Stellar populations}
\label{sec:res_stellar_pop}
%%%%%%%%%%%%%%%%%%%%%%%%%%%%%%%%%%%%%%%%%%%%%%%%%%%%%%%%%%%%%%%%%%%%%%%%%%%%%%%

In \refsec{sec:indices_results}, we found evidence that the
decoupling of the two stellar components is not only in their
kinematics, but also in their chemical composition. 

Differences in the chemical composition of kinematically decoupled
stellar structures are observed in many classes of galaxies, and it is
interpreted as the indication that the two stellar components have
different origins. Observations range from large counter-rotating disks
\citep{Johnston2012,Coccato2011a,Johnston2012,Coccato2013},
to counter-rotating bulges \citep{Katkov2011}, to small kinematically
decoupled cores in early-type galaxies 
\citep[e.g.][]{Silchenko2002,SilChenko2005,McDermid2006}).

For the particular case of NGC\,7217, \citet{Jablonka1996} measured Mg
and Fe line strength indices, finding $\mathrm{Mg}_2=0.28$\,mag,
$\mathrm{Fe5270}=3.04$\,\AA, and $\mathrm{Fe5335}=2.45$\,\AA.\
\citet{SilChenko2011} performed a stellar template fit, and found age
and metallicity gradients in the central $40\arcsec$: the mean stellar
age decreases from 10--13\,Gyr in the centre to 5\,Gyr while the
metallicity decreases from nearly solar ($\FeH \approx -0.06$) to
subsolar ($\FeH \approx-0.3$).~\cite{Sarzi2007} found old ages
(5--10\,Gyr) in the central few arcseconds, consistent with
\citet{SilChenko2011}, but extremely low metallicity ($Z=0.05$ solar,
i.e. $\FeH \approx-1.3$), inconsistent with the Silch'enko~et~al.\
values. 
 
All these studies assumed a single stellar kinematic component when
measuring the stellar population properties, and did not attempt to
separate the kinematics. If we repeat the measurement on our spectra,
considering only one kinematic stellar component, we measure a
mean $\langle \mathrm{Fe} \rangle_\mathrm{sing.\ comp.} = 2.1 \pm 0.2$,
which is slightly lower than that found by \citet{Jablonka1996}
($\langle \mathrm{Fe} \rangle \approx 2.7$), and a mean
\mgb$_\mathrm{sing.\ comp.}=2.9\pm0.2$.

We do not measure $H\beta$ so we cannot constrain the age with our
measurements (see \refsec{sec:decomposition}). However, if we adopt the
age range 5\,Gyr $< \mathrm{Age} < 10$\,Gyr found by
\citet{SilChenko2011} and~\cite{Sarzi2007}, we derive a metallicity
range $-0.3 < \FeH < -0.2$, consistent with the range obtained by
\citet{SilChenko2011}, but not with that of~\cite{Sarzi2007}.

In \reffig{fig:index_plane} we compare the Lick indices for the two
stellar components with (i) the typical values observed in the bulges of
spiral galaxies given by \citet[32 galaxies]{Gorgas+07} and \citet[15
galaxies]{Morelli+08}\footnote{We only consider the measurements from
    \citet{Gorgas+07} obtained within half of the bulge effective
    radius. Measurements in \citet{Morelli+08} are performed in a region
    where the contribution to the total of the disk is equal or lower
the contribution of the bulge, as indicated by their photometric
bulge/disk decomposition.}, and (ii) the typical values observed in
early-type galaxies by \citet[124 galaxies]{Thomas2005}. We note that
the measurements for the hot stellar component, which we associate with
the bulge component of the galaxy
(\refsec{sec:kinematic_photometric_decomposition}), deviate from the
mean values observed in the bulges of spiral galaxies. However, 
the central values of \mgb\ and $\langle \mathrm{Fe} \rangle$ measured
for the same component (top-right part of the measurement distributions
in \reffig{fig:index_plane}) are consistent with those measured in the
central regions of early-type galaxies \citep{Thomas2005}.

%%%%%%%%%%%%%%%%%%%%%%%%%%%%%%%%%%%%%%%%%%%%%%%%%%%%%%%%%%%%%%%%%%%%%%%%%%%%%%%
\subsection{Implications for the formation of NGC\,7217}
\label{sec:formation_scenarios}
%%%%%%%%%%%%%%%%%%%%%%%%%%%%%%%%%%%%%%%%%%%%%%%%%%%%%%%%%%%%%%%%%%%%%%%%%%%%%%%

In the light of our findings it is interesting to revisit the question
of the formation process of this particular galaxy. With its massive
spheroid, it is likely that NGC\,7217 experienced a major merger in its
past, which may have been responsible for shutting down most of its star
formation activity. 

From SDSS images we derive a total extinction corrected $u-r$ color of
2.32\,mag 
%data/sdss/colors2.py
in a 9\arcmin\ diameter aperture.  This places NGC\,7217 as a whole
comfortably on the red sequence in a color magnitude diagram
\citep{Baldry2004}. If the relatively small contribution of the disk
were to be removed, then the spheroid would appear even redder (B95). In
fact, inside of the inner dust ring, where the spiral morphology
disappears completely, $u - r = 2.7$\,mag,
% sedfit/sedfit/regphot.sh see ext corrected mags in
% sedfit/sedfit/N7217.dat
placing NGC\,7217 on the red flank of the red sequence. Compared to
the correlations of effective radius, total stellar mass, luminosity,
S\'ersic index, and central velocity dispersion \citep{Kormendy2009,
Cappellari2013}, the spheroid would also appear as a normal early-type
galaxy.  

The resemblance of the spheroidal component of NGC\,7217 to an
elliptical galaxy is further strengthened by the line index analysis.
Indeed, the central values of the \mgb\ and $\langle \mathrm{Fe} \rangle$
line strengths of the hot component are in good agreement with the
values that \citet{Thomas2005} found in the central regions of
early-type galaxies (see \reffig{fig:index_plane}). We note the
importance of the spectroscopic decomposition in measuring the line
strengths of the disk and spheroid separately.  In fact, the results of
the single component analysis (\refsec{sec:res_stellar_pop}) would place
the bulge of NGC\,7217 in the \MgbFe\ plane closer to the bulges of
spiral galaxies than to early-type galaxies (\reffig{fig:index_plane}).

On the other hand, while its spheroid resembles in many respects a
merger-built elliptical galaxy, NGC\,7217 does host a stellar and a
gaseous disk. However, the disk contains only about 20\,\% of the total
light of the galaxy. Given its bluer colours it probably has a lower
mass-to-light ratio than the spheroid, making the difference in stellar
mass even more extreme. As such, NGC\,7217 bears a resemblance
to the Sombrero galaxy, which also looks in many respects like a
typical elliptical galaxy if its thin, dusty disk is masked out.
NGC\,7217 may just be a nearly face-on version of Sombrero.  

It is very unlikely that such a sub-dominant disk could have survived
the merger itself; it must have reformed later on. Its bluer colours
(B95), very low velocity dispersion and the sites of active star
formation show that the stellar disk must be younger and is still in the
process of growth. So what is the most likely formation scenario for the
disk component in NGC\,7217?

The disk may have formed as a consequence of a major merger. Such events can
result in the formation of a cold disk if gas is already in place in the
progenitors \citep{Steinmetz2002,Barnes2002,
Abadi2003,Springel2005,Robertson2006,Governato2007,Hopkins2009}. For example,
\citet{Governato2009} show that mergers can reform systems with bulge-to-disk
ratios very close to the value that B95 found photometrically for NGC\,7217.

Another possibility is that the gas that formed the disk was acquired later
through minor mergers or cold accretion from the intergalactic medium
\citep{Mazzuca2006, Eliche-Moral2010}. A factor seemingly pointing against
accretion is the fact that NGC\,7217 appears relatively isolated
\citep{Karachentseva1973}. This of course may not always have been the case
but, the near-perfect alignment of the rotation axis of the spheroid and the
gas disk strengthens that case against such a scenario. Misalignments between
ionized gas and stellar kinematics do seem to be a common phenomenon;
\citet{Davis2011} for instance finds misalignments in 42\% of their field
galaxies. In contrast, for NGC\,7217 we find, with the exception of $r \leq
6\arcsec$, close to perfect alignment between the rotation axis of the ionized
gas with the cold stellar component and the hot spheroidal component. If,
however, the spheroid is not round but rotationally flattened, then it could
force gas into alignment through gravitational torque. If we assume that the
spheroid is axisymmetric, then the projected ellipticity of 0.08 (see
\refsec{sec:kinematic_photometric_decomposition}) and an inclination of
33.9\Deg\ (see \refsec{sec:tilted_ring}), would correspond a substantial
intrinsic flattening of 0.7.  Correcting for the inclination, the maximum
rotational velocity of the spheroid in the field of view of VIRUS-W is
90\,\kms. A dispersion of 170\,\kms\ gives $v/\sigma = 0.8
\sqrt{\varepsilon/(1-\varepsilon)}$,
% see ell_test.py
indicating that the flattening is close to the value expected for an
isotropic rotator \citep{Binney1987, Kormendy1982b}. Therefore, even if the spheroid is
a perfectly isotropic system, its rotational flattening may have been
sufficient to eradicate any prior misalignment between its own rotation
axis and that of the gaseous disk.

It is possible that the gas is of internal origin, produced through, for
example, stellar mass loss. Stellar mass loss may return as
much as half of the stellar mass to the ISM over a Hubble time
\citep{Jungwiert2001, Lia2002, Pozzetti2007, Parriott2008, Bregman2009}.
The formation of the disk in NGC\,7127 would require that a reasonable
fraction of the gas ($\approx$20\% for NGC\,7217) can cool, recombine and reform a
disk.

\reffig{fig:index_plane} contains some important clues on the origin of
the disk. An obvious feature of that plot is the significant offset of
the stellar populations of the hot spheroid and the cold disk. They
form two completely separate sequences aligned roughly parallel to
the one-to-one line of the \MgbFe\ plane.  The sequences are separated
from each other by at least 1\,\AA\ in either $\langle \mathrm{Fe}
\rangle$ strength or \mgb\ strength or both. 

This is strikingly different to the cases of NGC\,5719
\cite{Coccato2011} or NGC\,3593 and NGC\,4550 \citep{Coccato2013}, where
the two populations form essentially one continuous sequence. Taken by
themselves, the spread in \mgb\ and $\langle \mathrm{Fe}\rangle$
probably reflects the evolution of a single stellar population if it was
left alone for a sufficient amount of time --- gradually building up
both $\alpha$-elements and other metals over time. The location of a
particular point along the sequence then reflects the different amounts
of self-enrichment. The fact that both sequences show a very similar but
modest increase of \aFe\ towards the right of the plot indicates that in
both cases the star formation was a prolonged process as a rapid burst
would have caused a sudden change in \aFe.

\reffig{fig:ind_profiles} reveals that the spread of both sequences
corresponds to negative radial \mgb\ and $\langle \mathrm{Fe} \rangle$
gradients. If the above picture is correct, larger radii correspond to a
lower degree of self-enrichment and probably lower ages, as would be
expected in an inside-out growth scenario.

The cause of the offset of the two sequences may be understood in
several ways: the analysis of the \mgb\ and Fe line strengths shows that
the cold component has, on average, larger Fe and similar Mg to the hot
component. This indicates that the disk must have experienced an
extended star formation history, allowing for enrichment of the ISM by
type Ia supernovae \citep[e.g.][]{Greggio1983, Thomas1998, Thomas2005}.
At first sight this extended star formation may have moved a disk population
off the bulge sequence and gradually towards smaller \aFe (the upper left of
\reffig{fig:index_plane}). However, the fact that the disk never reaches larger
metallicities than the spheroid seems to conflict with this scenario. Any
prolonged star formation should also have enhanced the metallicity beyond that
of the spheroid. The prolonged nature of this process makes it seem unlikely
that it could cause the strong, observed bimodality.  

Alternatively, through some event, the formation of the stars in the
disk may have restarted at much lower metallicities than were present at
the time in the spheroid. The disk population would then have developed
starting at this lower metallicity towards the upper right hand of
\reffig{fig:ind_profiles}. An inflow of relatively primordial gas
from some external reservoir would explain such a restart quite
naturally. Gas that is available in the spheroid would be depleted,
moving the sequence to the left of the plot. Any inflow of primordial
material would at the same time leave \aFe\ constant. The model lines of
constant \aFe\ then indicate that gas from the regions with the highest
self-enrichment was depleted significantly and brought from $[Z/H]
\approx 0.35$ down to values around $-0.33$ (following the 10\,Gyr model
lines).  This would mean a factor of about five to one for depletion if
the accreted material was truly primordial, and an even larger factor if
it was partly pre-enriched. Given this, it seems most likely that the
disk formation was triggered either through accretion from an external
reservoir or through a minor merger event.

While our IFU data do not cover the whole area inside the effective
radius, within the field of view the spheroid has a specific angular
momentum of $\lambda_r = 0.17$ which would put it marginally into the
regime of fast rotators \citep{Emsellem2007}. Sub-dominant disk
structures are very common in early type galaxies
\citep{Bender1989,Rix1990} and have been shown to exist in the majority
of fast rotators \citep{Krajnovic2013}. Aligned gas disks are also
observed frequently in early-type galaxies \citep{Davis2011}. NGC\,7217
seems to be one of the few cases found so far where the transformation
of the gas disk into a stellar disk can actually be observed. With its
total \mbox{H\,\textsc{i} + H$_2$} mass of $\approx 1.2 \times 10^9
M_{\astrosun}$ and its relatively low level of star formation of
$\approx$1\,$M_{\astrosun}\,\mathrm{yr}^{-1}$ (B95), this process could
be very long-lived. It may be that this is only possible because
NGC\,7217 happens to live in such relative isolation. It remains to be
seen whether denser environments are able to shut off star formation in
these low disk-to-total systems as they do in large-scale disks
\citep{Koopmann2004}.

One more definite conclusion can be drawn from our findings concerning
the origin of the ring structure:~\cite{Verdes-Montenegro1995} conducted
a detailed analysis of the ring structure and concluded that the ring
locations can be identified with the outer and inner Lindblad resonance
and the 4:1 ultra-harmonic resonance of a distortion in the axisymmetry.
But NGC\,7217 lacks obvious distortions such as a bar, an oval or a
strong spiral structure. We can now exclude the possibility that
instabilities caused by stellar counter-rotation \citep{Lovelace1997}
are responsible for the formation of the ring structure in NGC\,7217.
The counter-rotation has also been attributed to a previous minor merger
\citep{Comeron2010, Grouchy2010}, which could have led to the ring
structure. Based on our findings, the scenario proposed by B95 now seems
most likely: a weak non-axisymmetric distortion, discovered only through
a careful Fourier analysis, generates the resonance that results in the
rings.

%%%%%%%%%%%%%%%%%%%%%%%%%%%%%%%%%%%%%%%%%%%%%%%%%%%%%%%%%%%%%%%%%%%%%%%%%%%%%%%
\section{Conclusions}
%%%%%%%%%%%%%%%%%%%%%%%%%%%%%%%%%%%%%%%%%%%%%%%%%%%%%%%%%%%%%%%%%%%%%%%%%%%%%%%
\label{sec:conclusions}

Using the novel VIRUS-W spectrograph we have obtained moderately high
resolution $R \approx 9000$ optical Integral Field Unit observations
of NGC\,7217. Based on our high signal-to-noise data we are able to
revisit the kinematic structure of this galaxy and to test previous
claims of the existence of a counter-rotating stellar disk. Using a new
algorithm, we derive non-parametrized line-of-sight velocity
distributions and carry out double-Gaussian decompositions. We also use
the methodology introduced by \citet{Coccato2011a,Coccato2013} to
confirm our findings and to derive line strength indices for the two
stellar components.

\medskip \noindent Our main findings are: 

\begin{itemize} 

\item We confirm the existence of two dynamically distinct stellar
    components. In contrast to previous claims by
    \citet{Merrifield1994}, \citet{Silchenko2000}, and
    \citet{Fabricius2012a} we do not find them to be counter rotating.
    Rather we are able to decompose them into one hot, dominant, round,
    slowly-rotating component with a velocity dispersion of
    $\approx$170\,\kms, and a cold, co-rotating stellar disk with a
    velocity dispersion of $\approx$20\,\kms.

\item The velocity and velocity dispersion fields of the cold stellar
    disk are very similar to those of the gas as derived from the \oiii\
    emission lines. Together with the blue colours of the rings in this
    galaxy (B95) and the visible sites of active star formation, this
    supports a picture where the stellar disk is still in the process of
    regrowing.

\item The kinematic position angles of the cold stellar disk component
    and the hot spheroidal component are identical within our
    measurement errors, rendering an external origin of the gas unlikely.

\item We find an increase of velocity dispersion of the gas at the inner
    stellar ring. This may be a result of the resonant nature of the
    ring, but also a result of the ongoing star formation in that
    region.

\item The two components are clearly separated in \MgbFe\ space in the
    sense that the disk component shows larger equivalent widths in
    $\langle \mathrm{Fe} \rangle$ and lower equivalent widths in
    \mgb\ that cannot be explained by an age difference. This
    points to a different star formation history, with a shorter-lived
    period of star formation in the spheroid and a later and/or
    longer-lived star formation in the disk.

\item The Lick indices measured in the central regions of the hot
    component are more similar to those in the central regions of
    elliptical galaxies than to those in the bulges of spirals.

\item We confirm the existence of a misalignment of the gas velocity
    field in the central arcseconds. We attempt a tilted ring model, but
    cannot confirm a 90\Deg\ angular separation between the central
    disk and the outer disk rotation axes as would be necessary for a
    polar ring that was described by~\cite{Zasov1997, Silchenko2000}.
    The maximum angular separation we find is 30\Deg. While this may
    be a consequence of the limited spatial resolution of our data, we
    stress that a natural explanation for the central deviation from the
    global rotation could lie in the weak break in axisymmetry that was
    reported by B95 \citep{Athanassoula1992}.

\item The tilted ring analysis provides a rotation curve for the gas.
    This allows us to derive the total enclosed dynamical mass and a
    deprojected mass-to-light ratio of $\Upsilon^*_{dyn} = 4.5
    M_{\astrosun}/L_{\astrosun}$. Using GALXEX FUV/NUV, SDSS, and IRAC
    I1,2,3 bands we also conduct a stellar population analysis inside
    the central 10\arcsec\ and find that the predicted mass-to-light
    ratio is in reasonable agreement with the dynamical value if a
    Kroupa IMF is assumed.

\item The structural parameters of the two stellar components (scale
    lengths, ellipticities and position angles) are in good agreement
    with the values obtained from photometry (B95). This demonstrates
    that a kinematic decomposition is feasible for spheroid/disk
    systems. This method has the advantage of being completely model
    free; it does not rely on the extrapolation of model profiles. As
    such, it is fully complementary to photometric methods and allows us
    to test, for instance, assumptions where all the light that exceeds
    the inwards extrapolation of an outer exponential disk is attributed
    to the bulge of a system. It also enables us to probe the kinematic
    properties and to measure stellar population parameters beyond
    broadband colours. This extends the technique of the spectral
    decomposition \citep{Coccato2011a,Coccato2013} from the application
    to counter-rotating stellar disks to spheroid-disk systems, as long
    as the two components have sufficient separation in velocity
    dispersion. This makes it possible to probe the origins of
    individual components in a multicomponent system separately and
    unambiguously, rendering it a powerful tool to study the formation
    of such galaxies.

\end{itemize} 

We suggest that the main bulk of stars in NGC\,7217, i.e.\ the spheroidal
component, formed through a major merger. The merger remnant has
photometric and spectroscopic properties more similar to those of an
elliptical galaxy than to those of the bulges of spiral galaxies.
The disk component formed after the merger, presumably from 
relatively primordial gas acquired through minor mergers or cold
accretion from the intergalactic medium \citep{Mazzuca2006,
Eliche-Moral2010}, or an external reservoir as suggested by the
significant offset of the two stellar populations in the \MgbFe\ plane.

%%%%%%%%%%%%%%%%%%%%%%%%%%%%%%%%%%%%%%%%%%%%%%%%%%%%%%%%%%%%%%%%%%%%%%%%%%%%%%%
\section*{Acknowledgements}
%%%%%%%%%%%%%%%%%%%%%%%%%%%%%%%%%%%%%%%%%%%%%%%%%%%%%%%%%%%%%%%%%%%%%%%%%%%%%%%

L.C.~acknowledges financial support from the European Community’s
Seventh Framework Program (/FP7/2007-2013/) under grant agreement No.
229517.  We would like to thank Dave Wilman for fruitful discussions on
the implications of a disk regrowth scenario. We wish to thank Peter
Erwin for making his \texttt{Imfit} code available to us.  We wish to
thank Eric Emsellem for making the Python version of the Voronoi binning
code available to us. This research made use of APLpy, an open-source
plotting package for Python hosted at http://aplpy.github.com.  This
research has made use of the NASA/IPAC Extragalactic Database (NED)
which is operated by the Jet Propulsion Laboratory, California Institute
of Technology, under contract with the National Aeronautics and Space
Administration. Some of the data presented in this paper were obtained
from the Mikulski Archive for Space Telescopes (MAST). STScI is operated
by the Association of Universities for Research in Astronomy, Inc.,
under NASA contract NAS5-26555. Support for MAST for non-HST data is
provided by the NASA Office of Space Science via grant NNX13AC07G and by
other grants and contracts. This work is based in part on archival data
obtained with the Spitzer Space Telescope, which is operated by the Jet
Propulsion Laboratory, California Institute of Technology under a
contract with NASA.\ Support for this work was provided by an award
issued by JPL/Caltech. Funding for the SDSS and SDSS-II has been
provided by the Alfred P. Sloan Foundation, the Participating
Institutions, the National Science Foundation, the U.S. Department of
Energy, the National Aeronautics and Space Administration, the Japanese
Monbukagakusho, the Max Planck Society, and the Higher Education Funding
Council for England. The SDSS Web Site is http://www.sdss.org/.  The
SDSS is managed by the Astrophysical Research Consortium for the
Participating Institutions. The Participating Institutions are the
American Museum of Natural History, Astrophysical Institute Potsdam,
University of Basel, University of Cambridge, Case Western Reserve
University, University of Chicago, Drexel University, Fermilab, the
Institute for Advanced Study, the Japan Participation Group, Johns
Hopkins University, the Joint Institute for Nuclear Astrophysics, the
Kavli Institute for Particle Astrophysics and Cosmology, the Korean
Scientist Group, the Chinese Academy of Sciences (LAMOST), Los Alamos
National Laboratory, the Max-Planck-Institute for Astronomy (MPIA), the
Max-Planck-Institute for Astrophysics (MPA), New Mexico State
University, Ohio State University, University of Pittsburgh, University
of Portsmouth, Princeton University, the United States Naval
Observatory, and the University of Washington.

%%%%%%%%%%%%%%%%%%%%%%%%%%%%%%%%%%%%%%%%%%%%%%%%%%%%%%%%%%%%%%%%%%%%%%%%%%%%%%%
\appendix
\section{On the choice of regularisation for the kinematic extraction} 
\label{apx:smoothing}

The Maximum Penalized Likelihood method for the derivation of
non-parametric LOSVDs introduces the smoothing parameter as described in
\refsec{sec:kin_extract}. It penalizes the sum of the squared second
derivative of the LOSVD.\ If the smoothing factor is set to zero no
penalization occurs and the derived LOSVDs tend to show strong
noise-induced fluctuations. Non-zero smoothing factors dampen the
fluctuations at the cost of the potential introduction of biases.
Similar regularisation schemes are also part of other algorithms for
the derivation of LOSVDs such as pPXF \citep{Cappellari2004} and the
actual degree of penalization is typically chosen through Monte Carlo
simulations: a sample of synthetic spectra with artificial noise
according to specific signal-to-noise value is generated. The degree of
regularisation is then chosen such that it does not bias the moments of
the derived LOSVDs significantly \citep[e.g.][]{Nowak2008}. 

In this work we specifically wish to study the shape of the full LOSVD.\
We therefore cannot base our choice on the moments of the LOSVD.\ We
decide instead to use a smoothing factor that is small enough that it
results in a non-significant increase of the RMS value of the residuals
between the recorded spectra and the preferred broadened model. This
very conservative approach results in LOSVDs which still show a fair
degree of oscillations (see \reffig{fig:losvds}). These oscillations
have amplitudes that are generally compatible with the statistical error
in each of the velocity channels. They are due to the fact that the
smoothing acts on a typical velocity scale, the size of which is
controlled by the smoothing factor: the penalization of the second
derivative necessarily introduces a correlation between neighbouring
channels. Any noise spike will pull up the channels next to it. An
increase of the smoothing will act over larger velocity separations and
therefore increase the velocity scale that the oscillations occur on.

\begin{figure}
\begin{center}
\includegraphics[width=0.49\textwidth]{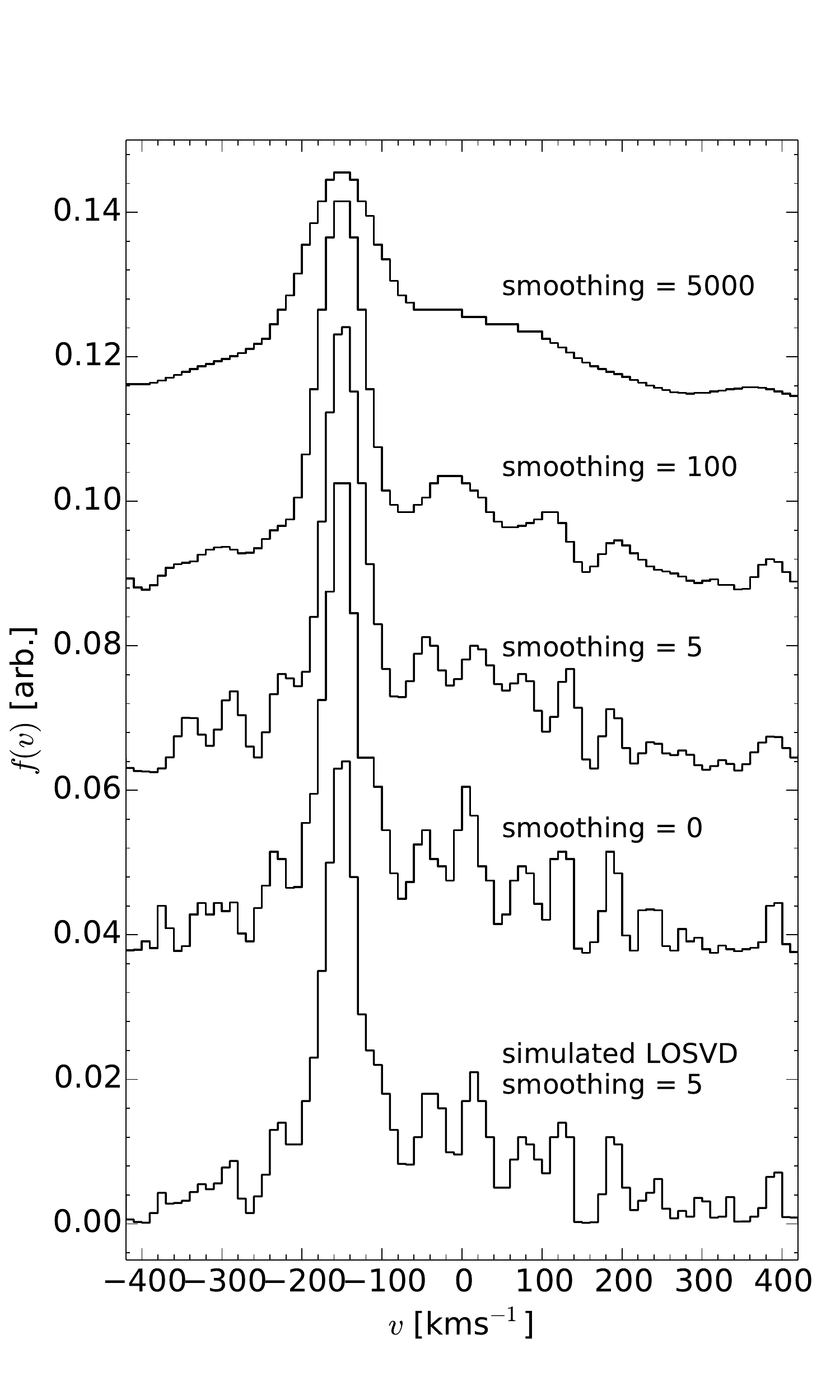}
% plotSmoothingTest2.py
\end{center}
\caption{The best-fitting non-parametric LOSVDs for different choices of
    the smoothing value. All but the bottom-most curves represent LOSVDs
    derived from the actual spectrum of bin 130, as shown in the upper
    panel of \reffig{fig:losvds}. Stronger penalization of the second
    derivative results in smoother LOSVDs at the cost of artificially
    broadening the narrow dispersion component at -150\,\kms.
    The bottom-most distribution was derived from a simulated spectrum
    that was broadened with a double-Gaussian kernel with the same
    parameters that we derive in the decomposition for bin 130.
}
\label{fig:smoothing}
\end{figure}

The effect of different degrees of regularisation is shown for the
example of a single bin in \reffig{fig:smoothing}. The upper four curves
show the LOSVDs that we derive from the spectrum of bin 130. With
stronger smoothing, oscillations occur on increasingly longer scales
until they are completely dampened out by a smoothing of 5000. The
genuine narrow dispersion component at about -150\,\kms\ does not
disappear, but it does get significantly broadened. The underlying broad
component remains nearly unaffected. The bottom-most distribution in
\reffig{fig:smoothing} was derived for a simulated spectrum that was
created by broadening a template spectrum with a two-component Gaussian
kernel with the same parameters that we derived for bin 130. The level
of noise-induced oscillations is very similar to the one seen in the
LOSVDs that we derived with the same smoothing from the actual data. In
the case of a single component LOSVD that is well-approximated by a
single Gaussian distribution, the smoothing can be set large enough that
over that scale of the LOSVD any oscillations are dampened out. For the
case of the two component LOSVDs of NGC\,7217 this is not possible, but
the choice of our small smoothing values guarantees that our derived
dispersions are not biased toward large values.

%%%%%%%%%%%%%%%%%%%%%%%%%%%%%%%%%%%%%%%%%%%%%%%%%%%%%%%%%%%%%%%%%%%%%%%%%%%%%%%
\bibliographystyle{mn2e}
\bibliography{Fabricius}
%\clearpage

\appendix

\label{lastpage}

\end{document}